\documentclass[12pt]{article}

\usepackage{lineno,hyperref,amsmath,amsfonts,amssymb,amsthm,diagbox,graphicx}
\usepackage{mathtools}
\usepackage{caption, subcaption, color, natbib}
\usepackage{setspace}
\usepackage[left=1in,top=1in,right=1in,bottom=1in,nohead,paperwidth=8.5in, paperheight=11in]{geometry} 
\usepackage{authblk}
\usepackage{float}
\usepackage{comment}
\usepackage{mleftright}
\mleftright

\theoremstyle{thmstyleone}%
\newtheorem{theorem}{Theorem}
\theoremstyle{thmstyleone}%
\newtheorem{cor}{Corollary}
\newtheorem{lemma}{Lemma}

\theoremstyle{thmstylethree}%

\newcommand{\norm}[1]{\left\lVert#1\right\rVert}
\DeclareMathOperator*{\argmax}{arg\,max}

\title{\bf Generalized Spherical Principal Component Analysis}
\author[a]{Sarah Leyder\thanks{Corresponding author. Email: sarah.leyder@uantwerpen.be}}
\author[b]{Jakob Raymaekers}
\author[a,c]{Tim Verdonck}
\affil[a]{Department of Mathematics, University of Antwerp, Belgium}
\affil[b]{Department of Quantitative Economics, Maastricht University, The Netherlands}
\affil[c]{IDLab, Imec Antwerp, Belgium}
\date{\today}

\begin{document}
\onehalfspacing
\maketitle

\begin{abstract}
Outliers contaminating data sets are a challenge to statistical estimators. Even a small fraction of outlying observations can heavily influence most classical statistical methods. In this paper we propose generalized spherical principal component analysis, a new robust version of principal component analysis that is based on the generalized spatial sign covariance matrix. Supporting theoretical properties of the proposed method including influence functions, breakdown values and asymptotic efficiencies are studied, and a simulation study is conducted to compare our new method to existing methods. We also propose an adjustment of the generalized spatial sign covariance matrix to achieve better Fisher consistency properties. We illustrate that generalized spherical principal component analysis, depending on a chosen radial function, has both great robustness and efficiency properties in addition to a low computational cost.
\bigskip


\end{abstract}
\doublespacing

\newpage

\section{Introduction}\label{sec1}

A well-known and frequently used technique to analyze the structure of data sets is principal component analysis (PCA). The objective of this technique is to obtain a new set of uncorrelated variables while preserving as much as possible of the variation present in the original data. These new variables are constructed as linear combinations of the original variables. Equivalently, they are obtained by projecting the original data on the PCA loading vectors of the data set, i.e. the directions in which the data has the greatest variability. PCA is a key building block in statistical data analysis and is widely used as a first step in clustering, discriminant analysis and regression.

In classical PCA (CPCA), the principal components can be calculated through a spectral decomposition of the covariance matrix. However, it is widely known that this matrix is very sensitive to outliers and potentially heavily influenced by anomalous observations. As a result, the directions of greatest variability are easily attracted towards these outliers, distorting the output of PCA. In order to avoid this, robust PCA methods have been developed which are resistant to such outlying observations.
There exist many different approaches to robust principal component analysis.

One approach is to use a spectral decomposition of a robust estimate of the covariance matrix.
For affine equivariant covariance matrices, this approach was studied by \cite{campbell1980robust} and \cite {boente1987asymptotic} who used M-estimators,
which unfortunately cannot withstand many outliers. 
\cite{Croux2000} revisited this approach, suggesting instead to use high-breakdown estimators of location and scatter such as 
S-estimators \citep{davies1987asymptotic,rousseeuw1984robust} or the MCD \citep{rousseeuw1984least}. In particular, they derive general expressions for the influence functions and efficiencies of the resulting eigenvector and eigenvalue estimates. A drawback of the approach is that many highly robust covariance estimators are computationally demanding. 
Additionally, some of them can only be computed when the number of samples is (substantially) larger than the dimension.

A second approach works incrementally starting from the principal component corresponding with the largest projected variance. Each new principal component is then estimated as a maximizer of the projected
robust variance conditional on being orthogonal to the already estimated components. This approach is
detailed in \cite{li1985projection,Croux2005,croux2007algorithms}. The approach works well when a relatively small number of principal components is required. However, it can also be computationally demanding as the number of projections needed should increase rapidly with the dimension of the data to guarantee a stable performance. A combination of the two approaches mentioned above was used by \cite{ROBPCA} to develop the ROBPCA algorithm. ROBPCA often outperforms the projection-pursuit based methods as well as the covariance-based methods. It remains fairly slow to compute, especially on larger data sets.

To mitigate the computational burden of the previously mentioned approaches to robust principal component analysis, one elegant and popular approach for robust PCA is  \textit{spherical principal component analysis} (SPCA), introduced independently by \citet{marden1999some} and \citet{Locantore}. Spherical PCA starts by projecting the centered data onto a unit sphere before performing classical principal component analysis on this transformed data set. It was studied by, among others, \cite{visuri2001subspace,taskinen2012robustifying,croux2002sign}.
This procedure is equivalent to performing PCA on the \textit{spatial sign covariance matrix} (SSCM): 
\begin{equation*}
        \Sigma_{\text{SSCM}}(X) = \mathbb{E}\left[ \frac{(X - \mu)(X- \mu)^T}{\norm{X- \mu}^2}\right]
\end{equation*}
with $X$ a $p$-variate random variable, $\mu$ the location of the distribution of $X$ and $\norm{.}$ the Euclidean norm. Under mild assumptions on the underlying distribution, the SSCM is a Fisher consistent estimator of the eigenvectors and it preserves the order of the eigenvalues. The SSCM was studied in detail in \cite{magyar2014asymptotic, durre2014spatial, durre2016eigenvalues,boente2019spatial}.

In \citet{GSSCM} a generalisation to the SSCM was introduced, namely the \textit{generalized spatial sign covariance matrix} (GSSCM). They found that the SSCM is part of a large class of orthogonally equivariant scatter estimates, namely the generalized spatial sign covariance matrices. Whereas in the SSCM all centered data points $x_i$ are given the weight $1/\norm{x_i}$, the generalized SSCM assigns different weights,  depending on the distribution of the random variable:
\begin{align}
    \label{defGSSCM}
    \Sigma_{g_X}(X) &= \mathbb{E}\left[ g_X(X -\mu) g_X(X-\mu)^T \right]\\
    &\text{with}  \quad  g_X(t) = t \ \xi_X(\norm{t}) \nonumber
\end{align}
where $\xi_X: \mathbb{R}^{+} \to \mathbb{R}^{+}$ is the \textit{radial function}. By using the Euclidean norm, the GSSCM becomes an orthogonally equivariant scatter estimator. In \citet{GSSCM}, it is shown that the GSSCM inherits the consistency properties of the SSCM in that it is a Fisher consistent estimator of the eigenvectors and preserves the ranks of the eigenvalues under the same assumptions.

A new robust method for principal component analysis emerges when we combine the idea of spherical PCA with the GSSCM. Instead of computing the principal components from the SSCM, we can compute them from the generalized SSCM. We refer to this method as \textit{generalized spherical principal component analysis} (GSPCA).

In this paper, we introduce and investigate the GSPCA method from a robustness perspective. The rest of the paper is organized as follows.
Section \ref{sec:meth} formally introduces GSPCA. Section \ref{SectionTheoreticalProp} covers several theoretical properties of the new method, such as breakdown values, influence functions and asymptotic variances. Additionally, we further improve our approach by adapting the GSSCM estimator to get Fisher consistency for the eigenvalues. In Section \ref{SectionSimulationStudy} an extensive simulation study compares our method to SPCA \citep{Locantore} and ROBPCA \citep{ROBPCA}, the state-of-the-art method. Section \ref{sec:dataexamples} finalizes by applying GSPCA to real life data sets.

\section{Generalized spherical principal component analysis}\label{sec:meth}

As discussed in the introduction, the GSPCA method computes the principal component directions as the eigenvectors of the GSSCM given in Equation \eqref{defGSSCM}. For a finite, $p$-variate data set $\mathbf{X}$ containing $\{x_1,\dots,x_n\}$ the sample GSSCM becomes
\begin{align*}
    S&_{g_{\mathbf{X}}}(\mathbf{X})= \frac{1}{n} \sum_{i=1}^n \xi_{\mathbf{X}}^2(\norm{x_i - T(\mathbf{X})}) \ (x_i - T(\mathbf{X})) (x_i - T(\mathbf{X}))^T= \frac{1}{n} \ g_{\mathbf{X}}(\mathbf{X})^T g_{\mathbf{X}}(\mathbf{X})
\end{align*}
where $T$ is a (orthogonally equivariant) location estimator for the center of the data matrix $\mathbf{X} \in \mathbb{R}^{n \times p}$. In \citet{GSSCM}, it is suggested to use the \textit{$k$-step least trimmed squares (LTS) estimator} as a robust location. This estimator starts from the spatial median, but adds some extra successive steps to improve robustness against outliers. The $k$-step LTS estimator has a breakdown value of $\left \lfloor{(n+1)/2}\right \rfloor /n$.\\
In this paper we will consider the following five radial functions suggested in \citet{GSSCM}, apart from the evident radial functions $\xi(r) = 1$ and $\xi(r) = 1/r$, respectively corresponding to the classical covariance matrix and the SSCM.
\begin{enumerate}
    \item Winsorizing (Winsor):
        \begin{flalign}
        \label{defWinsor}
            \qquad \xi_X(r) &=
              \begin{cases}
                1       & \quad \text{if }  r \leq Q_2\\
                Q_2/r   & \quad \text{if } Q_2 < r
              \end{cases}&&
        \end{flalign}
    \item Quadratic Winsor (Quad):
        \begin{flalign}
            \qquad \xi_X(r) &=
              \begin{cases}
                1           & \quad \text{if }  r \leq Q_2\\
                Q_2^2/r^2   & \quad \text{if } Q_2 < r
              \end{cases}&&
        \end{flalign}
    \item Ball:
        \begin{flalign}
        \label{defBall}
            \qquad \xi_X(r) &=
              \begin{cases}
                1   & \quad \text{if }  r \leq Q_2\\
                0   & \quad \text{if } Q_2 < r
              \end{cases}&&
        \end{flalign}
    \item Shell
        \begin{flalign}
        \label{defShell}
            \qquad \xi_X(r) &=
              \begin{cases}
                0   & \quad \text{if }  r < Q_1\\
                1   & \quad \text{if } Q_1 \leq r \leq Q_3\\
                0   & \quad \text{if } Q_3 < r
              \end{cases}&&
        \end{flalign}
    \item Linearly Redescending (LR):
        \begin{flalign}
        \label{defLR}
            \qquad \xi_X(r) &=
              \begin{cases}
                1   & \ \text{if }  r \leq Q_2\\
                \frac{Q_3^* - r}{Q_3^* - Q_2}   & \ \text{if } Q_2 < r \leq Q_3^*\\
                0   & \ \text{if } Q_3^* < r
              \end{cases}&&
        \end{flalign}
\end{enumerate}

The cutoffs $Q_1,Q_2,Q_3 \text{ and } Q_3^*$ are robust estimates of the first, second and third quartile of the distribution of the Euclidean distances. We will come back to these in the next section.

For a given data set $\mathbf{X}$ we can now use the spectral decomposition of the sample GSSCM to obtain the GSPCA loading vectors:
\begin{equation*}
    S_{g_{\mathbf{X}}}(\mathbf{X}) = V_g \Lambda_g V_g^T
\end{equation*}
Here the matrix $\Lambda_g$ is the diagonal matrix containing the eigenvalues $\lambda_{g,i}$ of $S_{g_{\mathbf{X}}}(\mathbf{X})$ in descending order ($\lambda_{g,1} > \dots > \lambda_{g,p})$ and the columns of the matrix $V_g$ consist of the corresponding eigenvectors $v_{g,i}$. The GSPCA loading vectors then coincide with these eigenvectors and the principal components correspond to the data projected onto these vectors.

Alternatively we can also state that the principal components are the uncorrelated vectors maximizing the variance of the transformed data. In particular, the $i$-th loading vector $a_i$ can be defined as  
\begin{align*}
  &a_i = \argmax_{a \in A} \left\{ \text{var}(g_{\mathbf{X}}(\mathbf{X})a) \right\}\\
  \text{ where } A &=\{a \in \mathbb{R}^p \  \vert \norm{a}_2 = 1 \text{ and } a \cdot a_j = 0 \text{ for all } j= 1, \ldots, {i-1}\}
\end{align*} 
It should be clear from the definition that GSPCA generally has a fairly low computational cost.
The only uncertain factor is the computation and evaluation of the radial functions $g$.  
In the cases considered above however, it requires the computation of the Euclidean norms of the observations followed by robust estimates of the quantiles of these norms. The first can be done in $\mathcal{O}(np)$ time, and the latter in $\mathcal{O}(n)$. Therefore, GSPCA has a computational complexity given by $\mathcal{O}(n + np + np^2+p^3) = \mathcal{O}(np^2+p^3)$, the same as classical PCA.

Additionally, like CPCA, GSPCA has the property that it can be computed through the singular value decomposition (SVD) as well as the spectral composition of the GSSCM. In order to use the SVD for GSPCA, we first need to transform the observations using the function $g_X$ of Equation \eqref{defGSSCM} after which we can apply SVD to the transformed data $g_{\mathbf{X}}(\mathbf{X})$. More precisely, we first compute the singular value decomposition 
\begin{equation*}
    g_{\mathbf{X}}(\mathbf{X}) = UDV^T
\end{equation*}
where the orthogonal matrices $U$ and $V$ respectively contain the left and right singular vectors of $g_{\mathbf{X}}(\mathbf{X})$ and the diagonal matrix $D$ contains the corresponding singular values $\delta_{g,i}$. Next it can easily be shown that the columns $v_{g,i}$ of $V$ correspond to the GSPCA loading vectors and the values $\delta_{g,i}^2/n$ correspond to the GSPCA eigenvalues:
\begin{align*}
    S_{g_{\mathbf{X}}}(\mathbf{X}) &= \frac{1}{n} \  g_{\mathbf{X}}(\mathbf{X})^T g_{\mathbf{X}}(\mathbf{X})
    = \frac{1}{n} \ V D U^T U D V^T \\
    &= \frac{1}{n} \ V D D V^T \\ &= V \text{diag}(\delta_{g,1}^2/n,\dots, \delta_{g,p}^2/n) V^T = V_g \Lambda_g V_g^T
\end{align*} The SVD procedure has the advantage that the computational cost of GSPCA can be lowered if the required number of principal components is known in advance. In this scenario we can use truncated singular value decomposition on the GSSCM to calculate only the first $k$ singular vectors, which correspond with the first $k$ eigenvectors of the matrix $S_{g_{\mathbf{X}}}(\mathbf{X})$. This not only avoids the calculation of the remaining principal components, but also does not require the explicit construction of the $p\times p$ covariance estimate $S_{g_{\mathbf{X}}}(\mathbf{X})$. 

\section{Theoretical properties}
\label{SectionTheoreticalProp}
In this section we study the robustness and efficiency of GSPCA. To evaluate robustness, the breakdown value of our method is calculated and influence functions are derived. The latter are then used to evaluate the efficiency through the computation of the asymptotic variances. We use the obtained theoretical properties to objectively compare different radial functions and explain their statistical behavior.

\subsection{Breakdown value}
\label{subsectionBDV}
The breakdown value is a global measure for robustness defined by Hampel based on an idea of Hodges \citep{Hampel}. For a given estimator, it is defined as the smallest fraction of observations in the data set that needs to be changed to carry the estimate arbitrarily far. More specifically, for a scatter estimator $S$, it is defined as the minimal amount of contamination required to make the largest eigenvalue $\lambda_1$ arbitrarily large (explosion) or the smallest eigenvalue $\lambda_p$ arbitrarily close to zero (implosion):
{\small
\begin{align*}
    \varepsilon (S,\mathbf{X}) = \min& \left\{ \frac{m}{n} : \sup_{\mathbf{X}_m^*} \ \max [ \lambda_1(S(\mathbf{X}_m^*)), \lambda_p^{-1}(S(\mathbf{X}_m^*)) ] = \infty \right\}
\end{align*}}

with $\mathbf{X}_m^*$ the data set $\mathbf{X}$ where $m$ observations have been replaced.

While in many applications, implosion of the covariance matrix is undesirable, in the context of PCA, it is the explosion breakdown that is more relevant. After all, if the dimension is equal to or larger than the sample size, we will have that the smallest eigenvalue is zero and would thus have ``breakdown'' if we were to use the definition above. In \citet{GSSCM} it is shown that the breakdown value of the GSSCM is $\left \lfloor{(n-p+1)/2}\right \rfloor /n$. This seems unsatisfactory, since it is in contrast with the SSCM which has a higher breakdown value, namely $\left \lfloor{(n+1)/2}\right \rfloor /n$. Fortunately, the lower breakdown value of the GSSCM is due to the implosion breakdown value of the ``hard redescending'' radial functions such as Ball, Shell and LR, in combination with the estimation of the cutoffs $Q_1,Q_2,Q_3 \text{ and } Q_3^*$ used in these radial functions. 

If we are no longer concerned with the implosion breakdown value, we can take a slightly different approach. First, we estimate the cutoffs using the estimators
\begin{align*}
     Q_1 &= \Big[\text{med}_i \left(\norm{x_i - T(\mathbf{X})}^{2/3}\right) - \text{MAD}_i \left(\norm{x_i - T(\mathbf{X})}^{2/3}\right)   \Big]^{3/2} \\
   Q_2 &= \text{med}_i (\norm{x_i - T(\mathbf{X})}) \\
    Q_3 &= \Big[\text{med}_i \left(\norm{x_i - T(\mathbf{X})}^{2/3}\right)   + \text{MAD}_i \left(\norm{x_i - T(\mathbf{X})}^{2/3}\right)   \Big]^{3/2} \\
    Q_3^* &= \Big[\text{med}_i \left(\norm{x_i - T(\mathbf{X})}^{2/3}\right)   + 1.4826 \cdot \text{MAD}_i \left(\norm{x_i - T(\mathbf{X})}^{2/3}\right)   \Big]^{3/2}
\end{align*}

In these definitions med and MAD are the median and the median absolute deviation. Unlike in \cite{GSSCM}, these estimators for the cutoffs no longer depend on the order statistic $h=\left \lfloor{(n+p+1)/2}\right \rfloor$, which was required to prove the implosion breakdown value of the GSSCM. The dependence on the $h$-th order statistic is undesirable in the context of PCA, as the value of $p$ is often close to $n$ or even considerably larger than $n$ which would lead to $h>n$, making the cutoffs meaningless.

The theorem below states that for our adapted GSSCM, with cutoffs based on the median and MAD, we obtain an explosion breakdown value independent of $p$, equal to that of the SSCM. The proof can be found in the Supplementary material.

\begin{theorem}[Explosion breakdown value]\label{thm:breakdown}
Given $\mathbf{X} = \{x_1,...,x_n\}$ an $n \times p$ dimensional data set and a location estimator $T(\mathbf{X})$ with a breakdown value of at least $\left \lfloor{(n+1)/2}\right \rfloor /n$. Suppose that
\begin{enumerate}
    \item The radial function takes values in $[ 0,1 ]$
    \item $\forall \text{ data sets }\mathbf{X}: \#\{x_i: \xi(\norm{x_i - T(\mathbf{X})}) = 1 \} \geq \left \lfloor{(n+1)/2}\right \rfloor $
    \item $\forall t: \ \norm{g(t)} = \norm{t} \xi(\norm{t}) \leq \text{med}_i (d_i) + 1.4826 \cdot \text{MAD}_i(d_i)$
\end{enumerate}
Then the explosion breakdown value $\varepsilon$ of the GSSCM is $\left \lfloor{(n+1)/2}\right \rfloor /n$.
\end{theorem}

Note that the conditions in the theorem are all satisfied for the radial functions in Equations \eqref{defWinsor} -- \eqref{defLR}. In addition, the k-step LTS estimator has a breakdown value of $\left \lfloor{(n+1)/2}\right \rfloor /n$. Hence, we conclude that GSPCA based on the GSSCM is robust up to $\frac{\left \lfloor{(n+1)/2}\right \rfloor}{n} \approx 50 \%$ contamination.

\subsection{Influence functions of the loading vectors}
\label{subsectionIFs}
We now consider the influence functions relevant for GSPCA. The influence function for an estimator $T$ at distribution $F$ is defined as (see \citet{Hampel}):
\begin{align}
\label{eq:IF}
    \text{IF}(x,T,F) &= \lim_{\varepsilon \to 0} \frac{T(F_{\varepsilon,x}) - T(F)}{\varepsilon}= \frac{\partial}{\partial \varepsilon} T(F_{\varepsilon,x}) \Big\vert_{\varepsilon=0}
\end{align}

with $F_{\varepsilon,x} = (1-\varepsilon)F + \varepsilon \Delta_x$ the distribution contaminated by $x$ where $\Delta_x$ is the distribution putting all its mass in $x$.\par 
In contrast to the breakdown value, the influence function is a local measure of robustness instead of a global one. Informally, it can be interpreted as a measure for the effect that an infinitesimal small amount of contamination has on the estimate. Therefore it is a complementary measure and describes a different aspect of our method.\par

In the context of PCA, the relevant functionals $T$ in Equation \eqref{eq:IF} are the eigenvector and eigenvalue functionals, which we denote by $v_{g,j}$ and $\lambda_{g,j}$ for $j=1,\dots,p$. In this section specifically, we will study the eigenvector functionals, therefore the relevant influence function is given by
\begin{equation*}\label{eq:IF2}
    \text{IF}(x,v_{g, j},F) = \lim_{\varepsilon \to 0} \frac{v_{g, j}(F_{\varepsilon,x}) - v_{g, j}(F)}{\varepsilon}
\end{equation*}
In what follows, we assume that $F$ is a distribution with the center at the origin and covariance matrix $\Sigma$ which has distinct eigenvalues. Furthermore, we denote the spectral decomposition of $\Sigma$ by $\Sigma = V \Lambda V^T$ where $V$ is a matrix with the eigenvectors $v_i$ in its columns and $\Lambda$ is a diagonal matrix with the eigenvalues $\lambda_1 > \ldots > \lambda_p$ on its diagonal.

Note that the functionals $v_{g, j}$ inherit the Fisher consistency properties of the GSSCM in case of elliptically symmetric distributions, see \cite{GSSCM}. We say that a distribution $F_X$ has an elliptically symmetric density if the density can be written as
\[ f_X(x) = det(\Sigma)^{-1/2} \cdot h((x-\mu)^T \Sigma^{-1} (x-\mu))  
\]

\noindent
In this case one thus has that $v_{g,j}(F) = v_j$. \\
Additionally, $v_{g, j}$ is orthogonally equivariant as the GSSCM is orthogonally equivariant:
\begin{align*}
  \forall \ U &\in \mathbb{R}^{p \times p} \text{ orthogonal matrix:} v_{g, j}(F_U) = U v_{g, j}(F) U^T
\end{align*}
\noindent
where $F_U$ is the distribution of $UX$ if $X \sim F$.

We now state the expression for the influence functions corresponding to the loading vectors of GSPCA, for which the proof can be found in Section \ref{app:thm2} of the Supplementary material:

\begin{theorem}[Influence functions of the eigenvectors of the GSSCM]\label{thm:IF1}
For the influence functions of the eigenvectors resulting from GSPCA we have the following analytical expression:
\begin{align}
\label{IFeigenvectorsGSSCM}
     \text{IF}(x,v_{g,j}, F) = \sum_{k=1, k \neq j}^{p} \frac{1}{\lambda_{g,j} - \lambda_{g,k}} \Big[&v_{g,k}^T \int   \left( dge(X) g(X)^T \right. +  \left. g(X) dge(X)^T  \right)  dF(X)\ v_{g,j}\nonumber\\
     &+ (v_{g,k}^T g(x)) (v_{g,j}^T g(x)) \Big]   v_{g,k}
\end{align}
where $ dge(x) = \frac{\partial}{\partial \varepsilon}  g_{\varepsilon}(X) \Big\vert_{\varepsilon=0}$
\end{theorem}

While the above expression looks rather involved, we can simplify it further without (much) loss of generality. First, note that we can assume that $F$ has a diagonal covariance matrix since the GSSCM is orthogonally equivariant. In that case, the eigenvectors of the covariance matrix are the standard unit vectors, i.e. $v_j = e_j$. Additionally, we can assume that $F$ is elliptically symmetric. This assumption is very natural since it is required for the Fisher consistency of the eigenvector functionals $v_{g, j}$ of the GSSCM, yielding $v_{g,j}(F) = v_j$. We then obtain the result below, the proof of which is in Section \ref{app:col1} of the Supplementary material.

\begin{cor}\label{col:IF1}
Assume that $F$ is a centered elliptically symmetric distribution, that it has a density function and a diagonal covariance matrix $\Sigma$, then the following holds:
\begin{equation*}
\label{corr2IFeigenvectors}
    \text{IF}(x,v_{g,j}, F) = \sum_{k=1, k \neq j}^{p} \frac{1}{\lambda_{g,j} - \lambda_{g,k}} \big( g(x)_k g(x)_j \big)   v_k
\end{equation*}
\end{cor}

We thus see that, under reasonable assumptions, the influence function of the eigenvectors of GSPCA reduces to a fairly simple expression. In particular, if we plug in the identity function for $g$, we obtain the influence function of the loadings of classical PCA. In particular, we obtain 
$\sum_{k=1, k \neq j}^{p} \frac{x_k x_j}{\lambda_{j} - \lambda_{k}}   v_k$, corresponding to the result obtained in \cite{Croux2000}.

\begin{figure*}[ht]
\centering
\begin{subfigure}[t]{.24\linewidth}
\includegraphics[width=\textwidth]{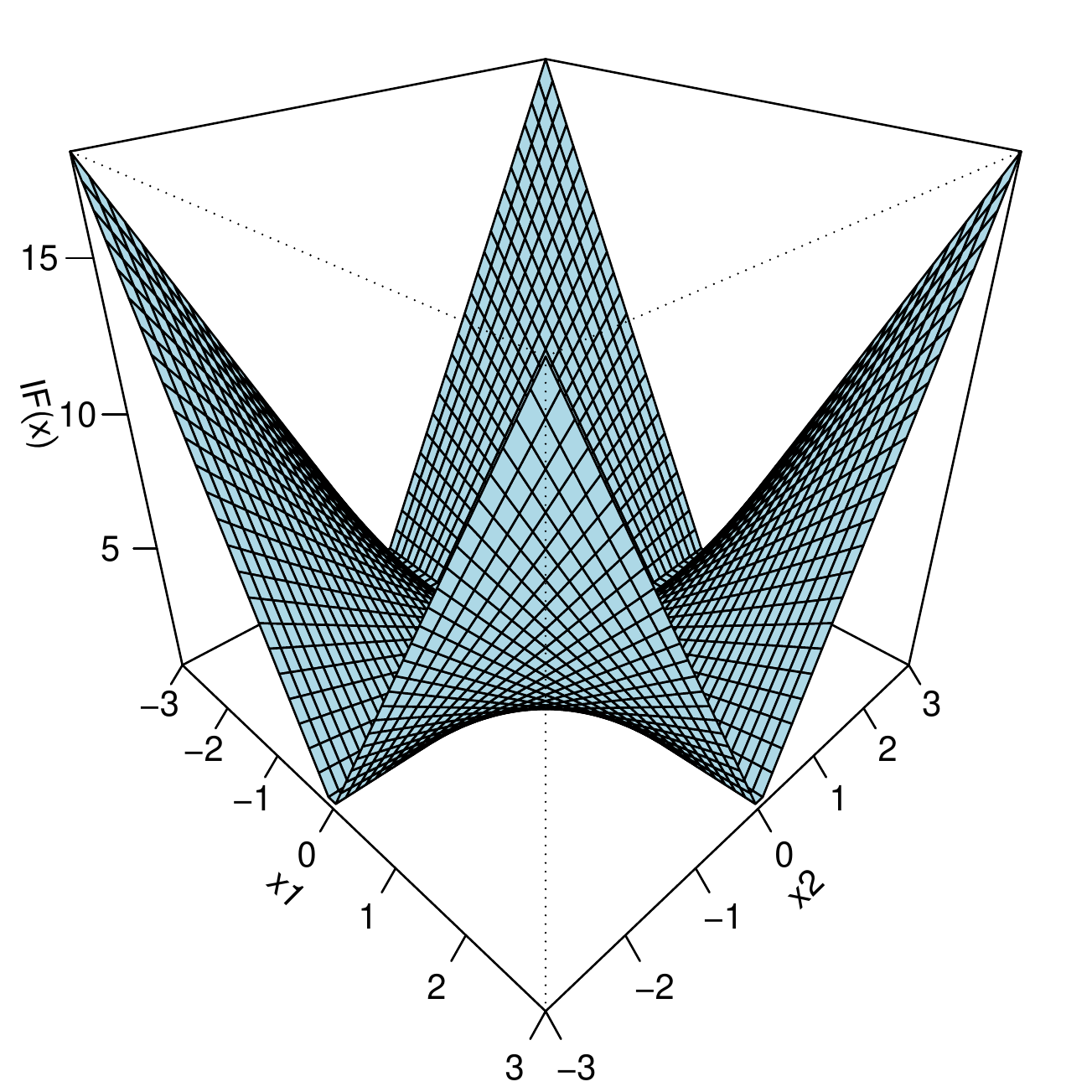}
\caption{classical covariance matrix}
\end{subfigure}
\begin{subfigure}[t]{.24\linewidth}
\includegraphics[width=\textwidth]{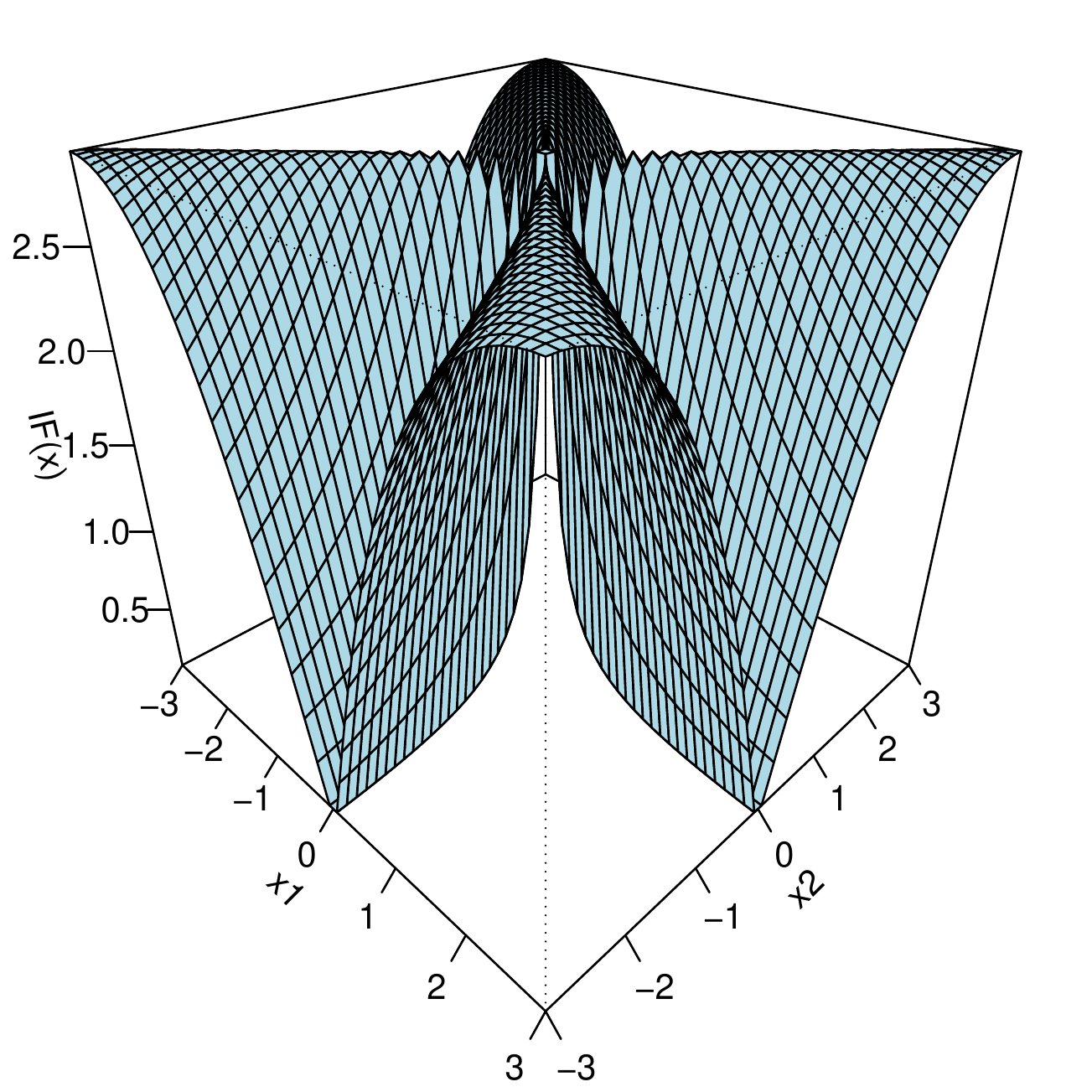}
\caption{SSCM}
\end{subfigure}
\begin{subfigure}[t]{.24\linewidth}
\includegraphics[width=\textwidth]{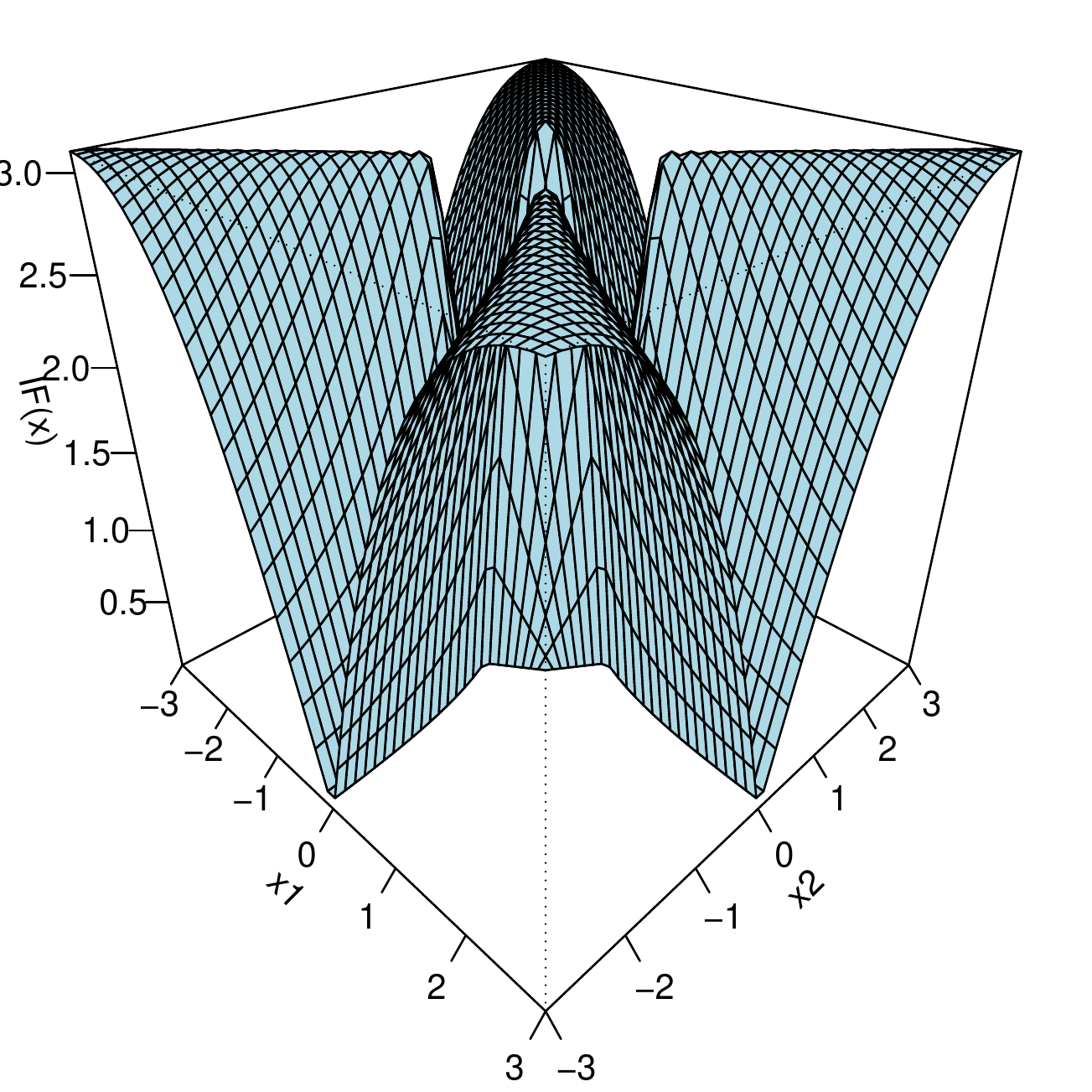}
\caption{Winsor}
\end{subfigure}
\begin{subfigure}[t]{.24\linewidth}
\includegraphics[width=\textwidth]{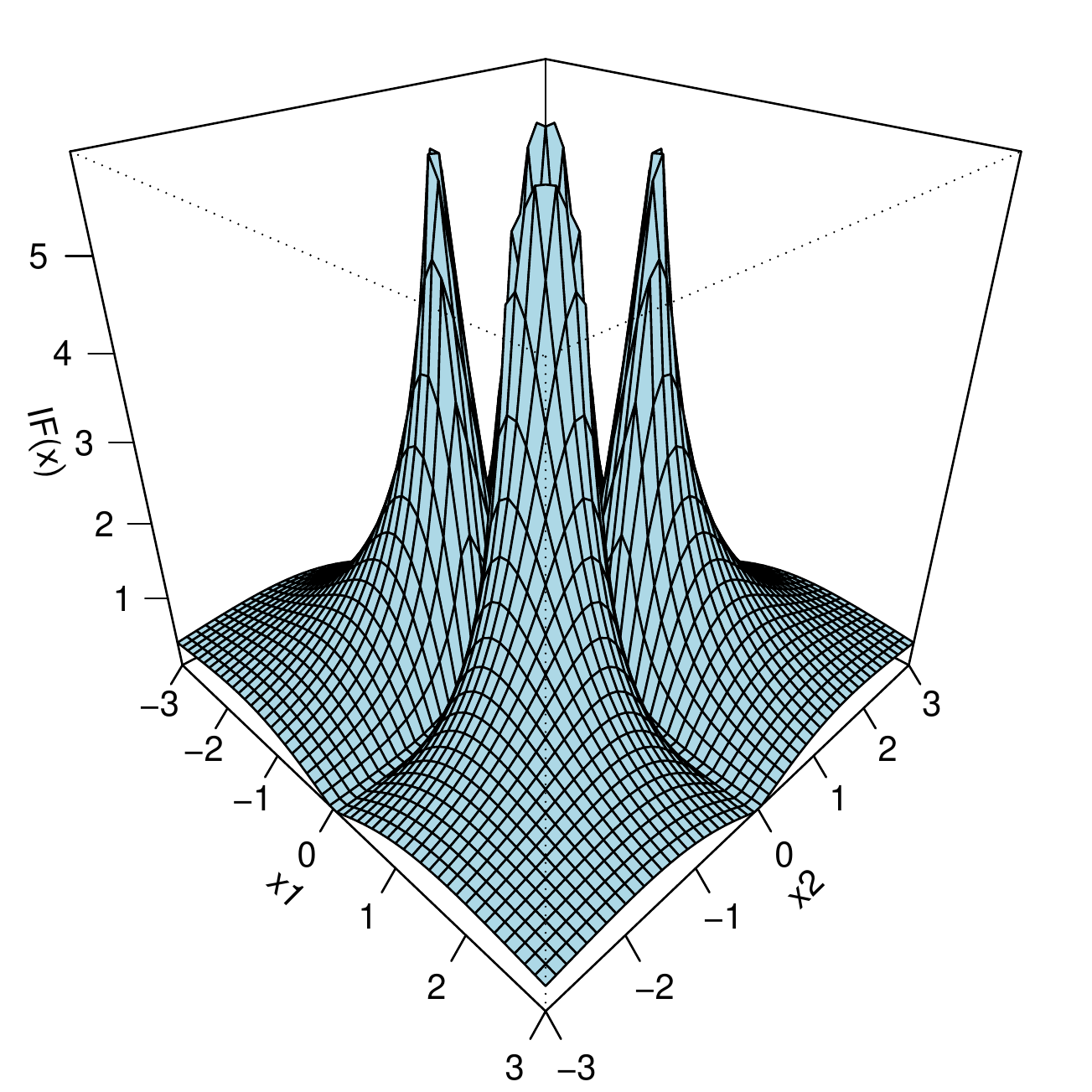}
\caption{Quad}
\end{subfigure}
\begin{subfigure}[t]{.24\linewidth}
\includegraphics[width=\textwidth]{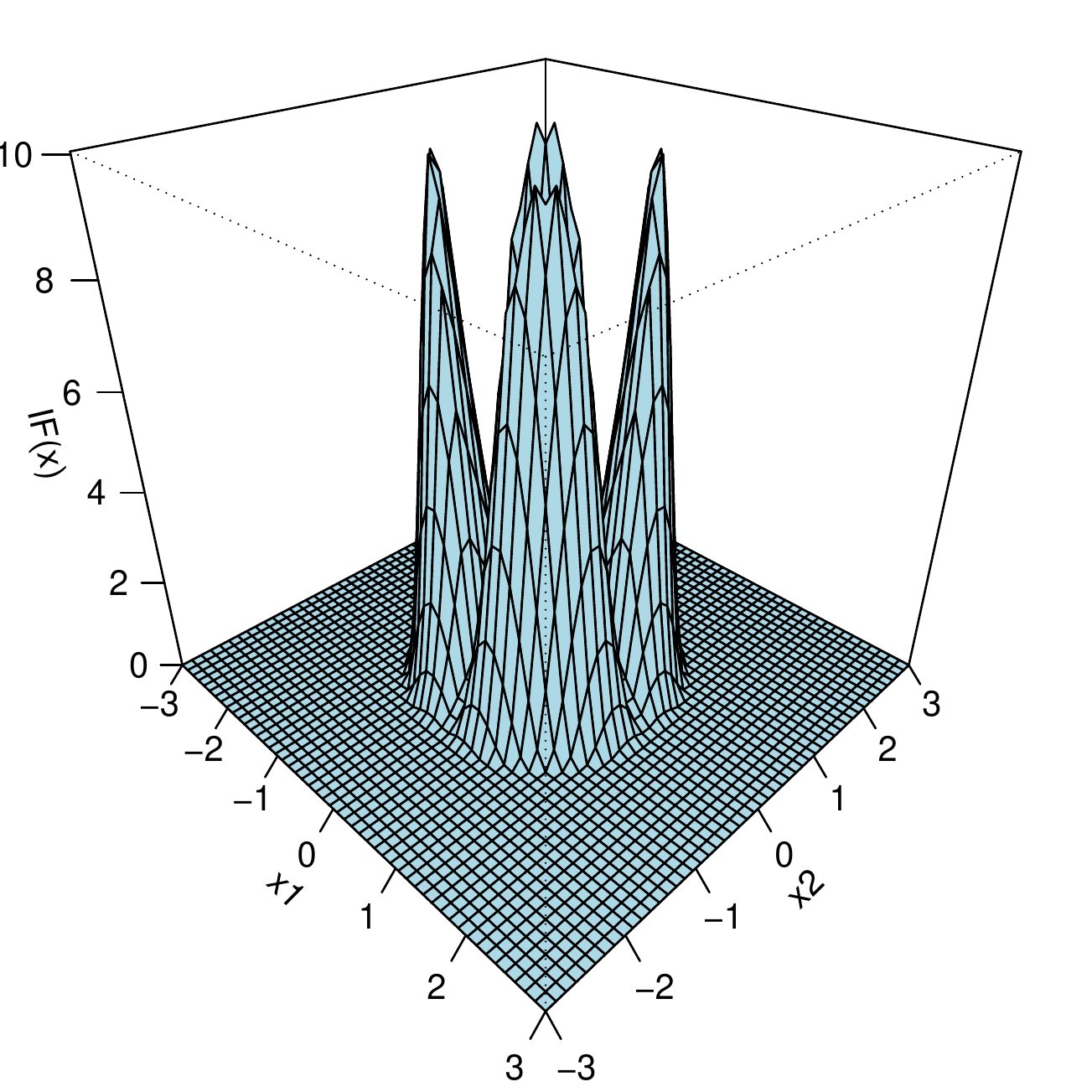}
\caption{LR}
\end{subfigure}
\begin{subfigure}[t]{.24\linewidth}
\includegraphics[width=\textwidth]{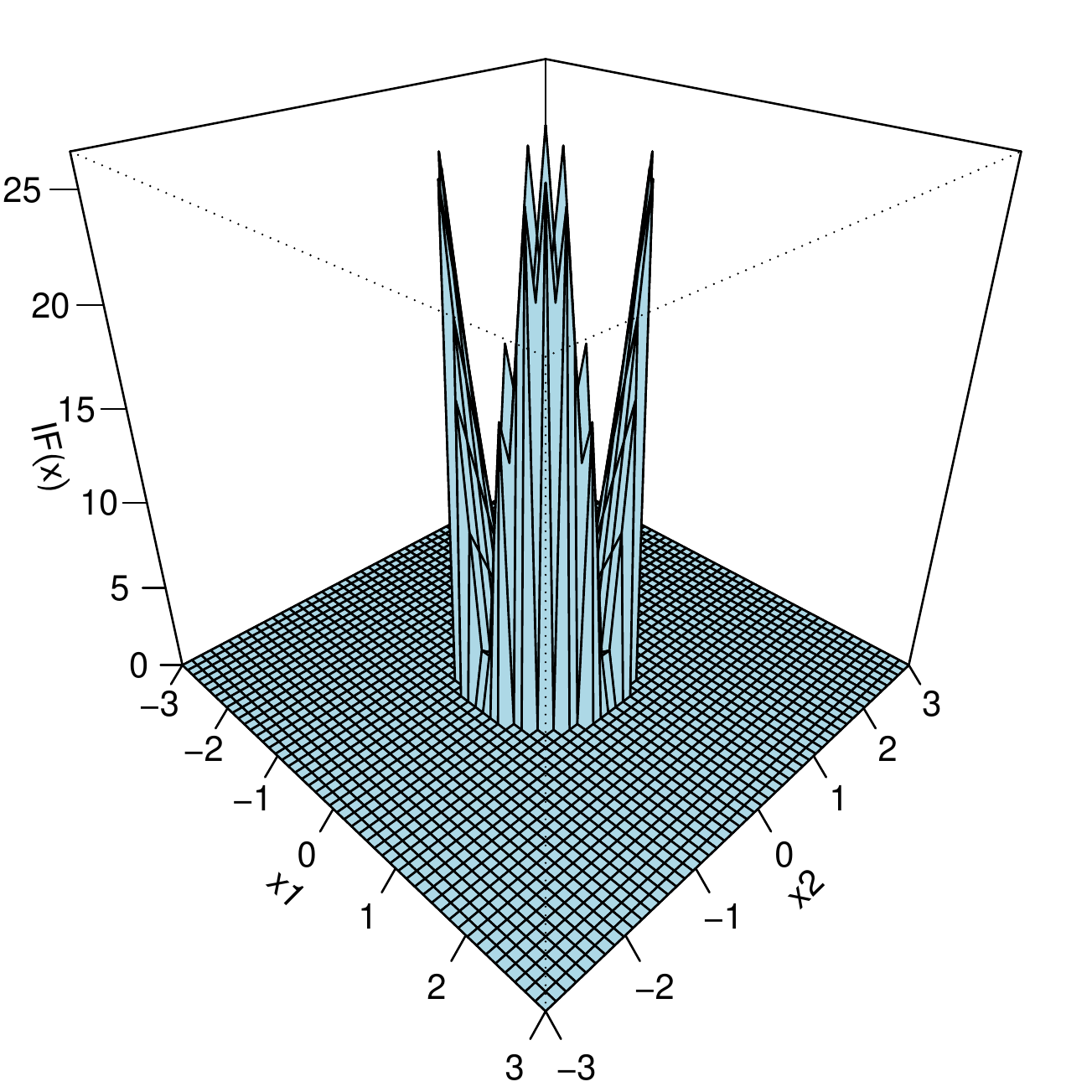}
\caption{Ball}
\end{subfigure}
\begin{subfigure}[t]{.24\linewidth}
\includegraphics[width=\textwidth]{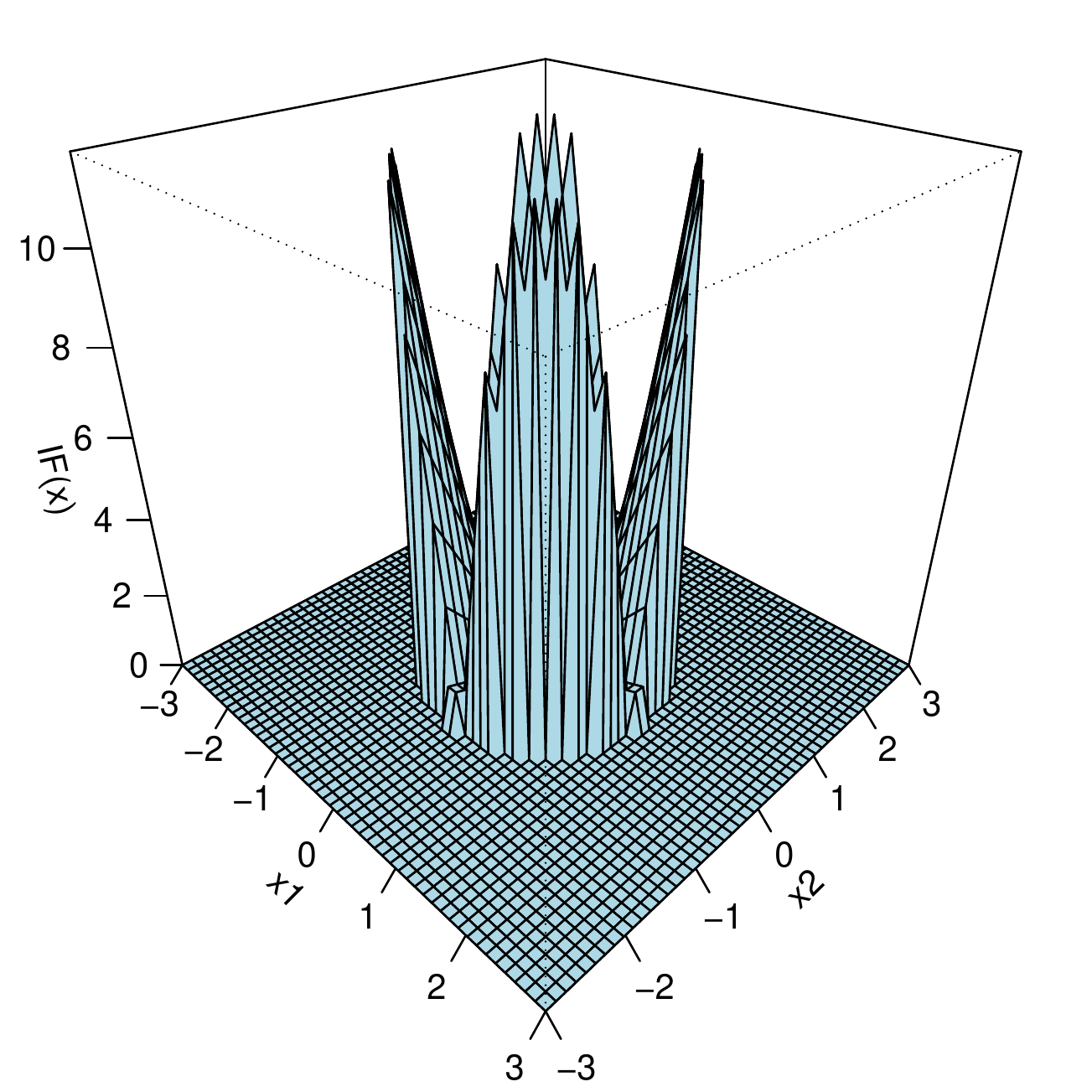}
\caption{Shell}
\end{subfigure}
\caption{\centering  \small Norm of the influence function of the largest eigenvector at $\mathcal{N}(0_2,\text{diag}(1,0.5))$ for different radial functions}
\label{normiflargestEV}
\end{figure*}

Having obtained analytical forms for the influence functions of the loading vectors resulting from GSPCA, we will now visualize them to compare the different radial functions in Equations \eqref{defWinsor} -- \eqref{defLR} and illustrate their robustness. We consider a bivariate normal distribution, $F =  \mathcal{N}(0_2, \Sigma)$ with $\Sigma = \text{diag}(1,0.5)$, and plot the norm of the influence function of the largest eigenvector for the different radial functions resulting in the 3D-plots shown in Figure \ref{normiflargestEV}.

First, we observe that the eigenvectors of the classical covariance matrix are the only ones with an unbounded influence function, all others are bounded. Second, all GSSCM radial functions are redescending to zero, except for Winsor, whose influence function looks like a smoothed version of the influence function of the SSCM, suggesting that Winsor will attain higher efficiency than the SSCM. Further, we see that the norms of the influence functions of LR, Ball and Shell look quite similar, all three demonstrate four large spikes whereafter each influence function becomes zero. This is due to their cutoffs at the second or third quantile. Quad only descends to zero in the limit, making it more robust than Winsor, but not quite as robust as the redescending radial functions. Overall these results suggest high robustness for the LR, Ball and Shell radial functions.

Additionally, we study a robustness measure derived from the influence function, being the gross-error sensitivity (GES). It is defined as follows for an estimator $T$ at distribution $F$:
\begin{equation*}
    \gamma^*(T,F) = \sup_x \vert \text{IF}(x,T,F) \vert
\end{equation*}

The gross-error sensitivity measures the maximal influence an infinitesimally small amount of contamination can have on an estimator $T$. Therefore it should be finite for robust estimators and preferably small.

For the loading vectors of GSPCA, we now compute the gross-error sensitivities per radial function numerically to compare robustness. We again consider the bivariate normal distribution $\mathcal{N}(0_2,\text{diag}(1,0.5))$ and study the first loading vector. We compute the gross-error sensitivity as the supremum of the norm of the corresponding influence function. This results in Table \ref{GrossErrorSensitivity}.\\
\begin{table}[h!]
    \centering
    \begin{tabular}{|c|cccc|}
    \hline
         & classical & SSCM & Winsor & Quad  \\
         \hline
         $\gamma^*$ & $\infty$ & 2.914213 & 3.100523  & 6.569927 \\
         \hline
          & LR & Ball & Shell &\\
         \hline
         & 10.74909 & 29.27686 & 12.42133&\\
         \hline
    \end{tabular}
    \captionsetup{width=\textwidth}
    \caption{\centering \small Gross-error sensitivities of 
    the largest eigenvector at $\mathcal{N}(0_2,\text{diag}(1,0.5))$}
    \label{GrossErrorSensitivity}
\end{table}
Once again, we observe that CPCA is not robust, its GES is unbounded. In contrast, GSPCA has bounded GES for each radial function, implying robustness. The smallest values are attained by SSCM and Winsor. While this may suggest these are the superior methods, the plots of the influence functions do show that the Quad, LR, Ball and Shell methods will have a smaller influence function for most values of the contamination $x$.

\subsection{Asymptotic variances and asymptotic relative efficiencies of the loading vectors}
\label{subsectionASV}

We will now use the results on the influence functions to study the asymptotic variances and efficiencies of GSPCA. This will allow us to compare precision among the different radial functions. 
Following \citet{Hampel}, we obtain that for well-behaved functionals $T$, the corresponding estimator $T_n = T(F_n)$ is asymptotically normal:
$\sqrt{n}\left(T_n - T(F)\right)\xrightarrow[]{D} \mathcal{N}\left(0, \text{ASV}(T, F)\right)$ where $$\text{ASV}(T, F) = \mathbb{E}_F\left[\mbox{IF}(x, T, F)\mbox{IF}(x, T, F)^T\right].$$

In order to simplify the exposition, we consider the case of a bivariate normal distribution $\mathcal{N}(0_2,\text{diag}(1,\gamma))$. Since we are interested in eigenvectors for PCA, we first calculate the asymptotic variance (ASV) of the second element of the largest eigenvector, given that it is important that we estimate this second element close to zero to get a good estimation of the first eigenvector. This is similar to the approach taken in \cite{Croux2010}. We denote the asymptotic variance of the second element of the largest eigenvector simply as $\text{ASV}_g$ per radial function.

In Section \ref{subsectionIFs} we found analytical expressions for the influence functions of the eigenvectors. Hence, we can use them to calculate the asymptotic variances.

For the second component of the influence function of the first eigenvector, which we denote by $v_{g,(1, 2)}$, we obtained the following expression:
\begin{equation*}
    \text{IF}(x,v_{g,(1, 2)},F) = \frac{1}{\lambda_{g,1}-\lambda_{g,2}} g(x)_1 g(x)_2
\end{equation*}
Using this, we obtain the asymptotic variance:
\begin{align}
    \label{eq:formule_ASV}
    \text{ASV}_g(F) = \frac{1}{(\lambda_{g,1}-\lambda_{g,2})^2} \int \left(g(X)_1 g(X)_2 \right)^2 \ dF(X)
\end{align}

\begin{figure}[h!]
        \centering
        \begin{subfigure}[t]{.45\linewidth}
            \includegraphics[trim={0 1.5cm 0 1.8cm},clip,width=\textwidth]{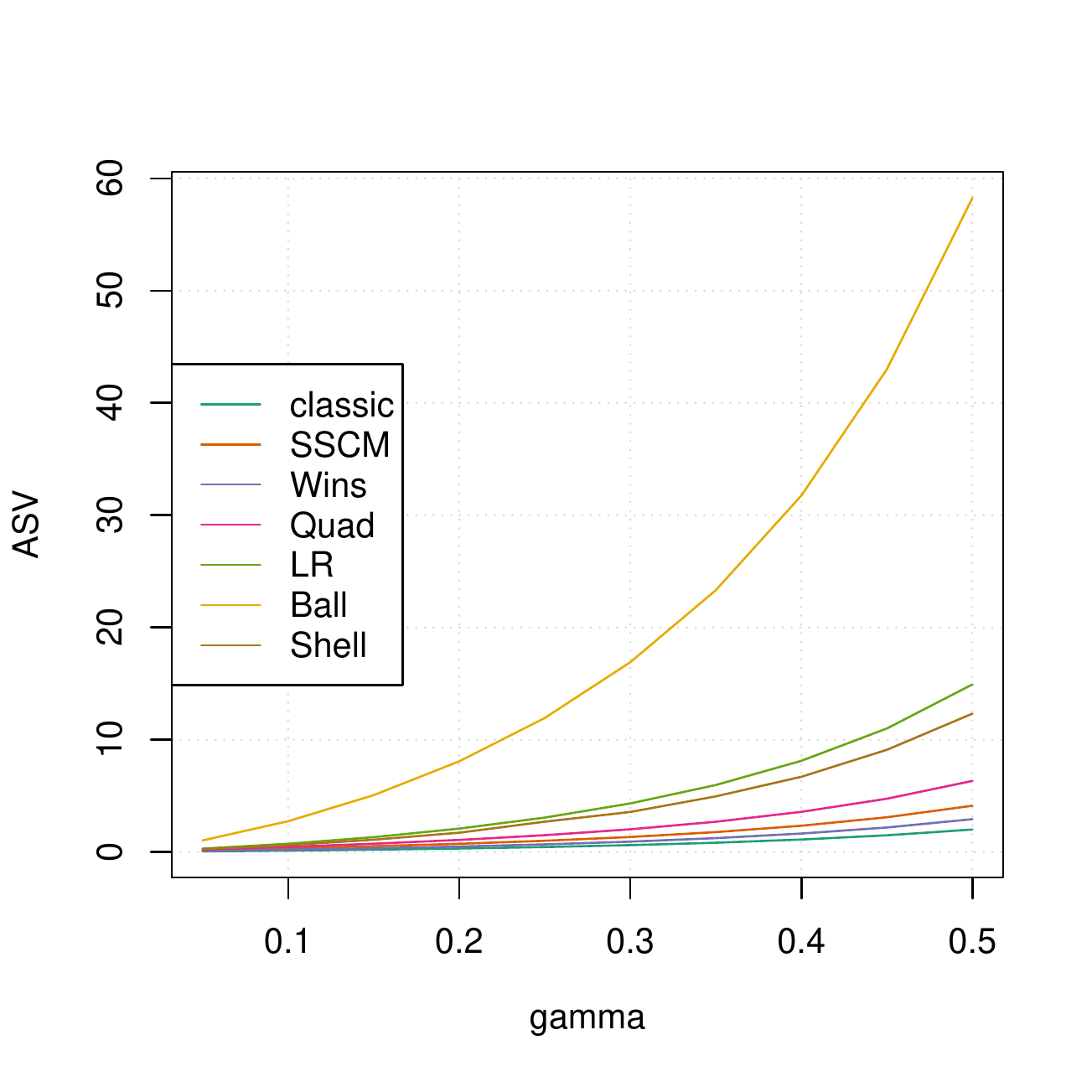} 
            \caption{$0 \leq \gamma \leq 0.5 $}
        \end{subfigure}
        \begin{subfigure}[t]{.45\linewidth}
            \includegraphics[trim={0 1.5cm 0 1.8cm},clip,width=\textwidth]{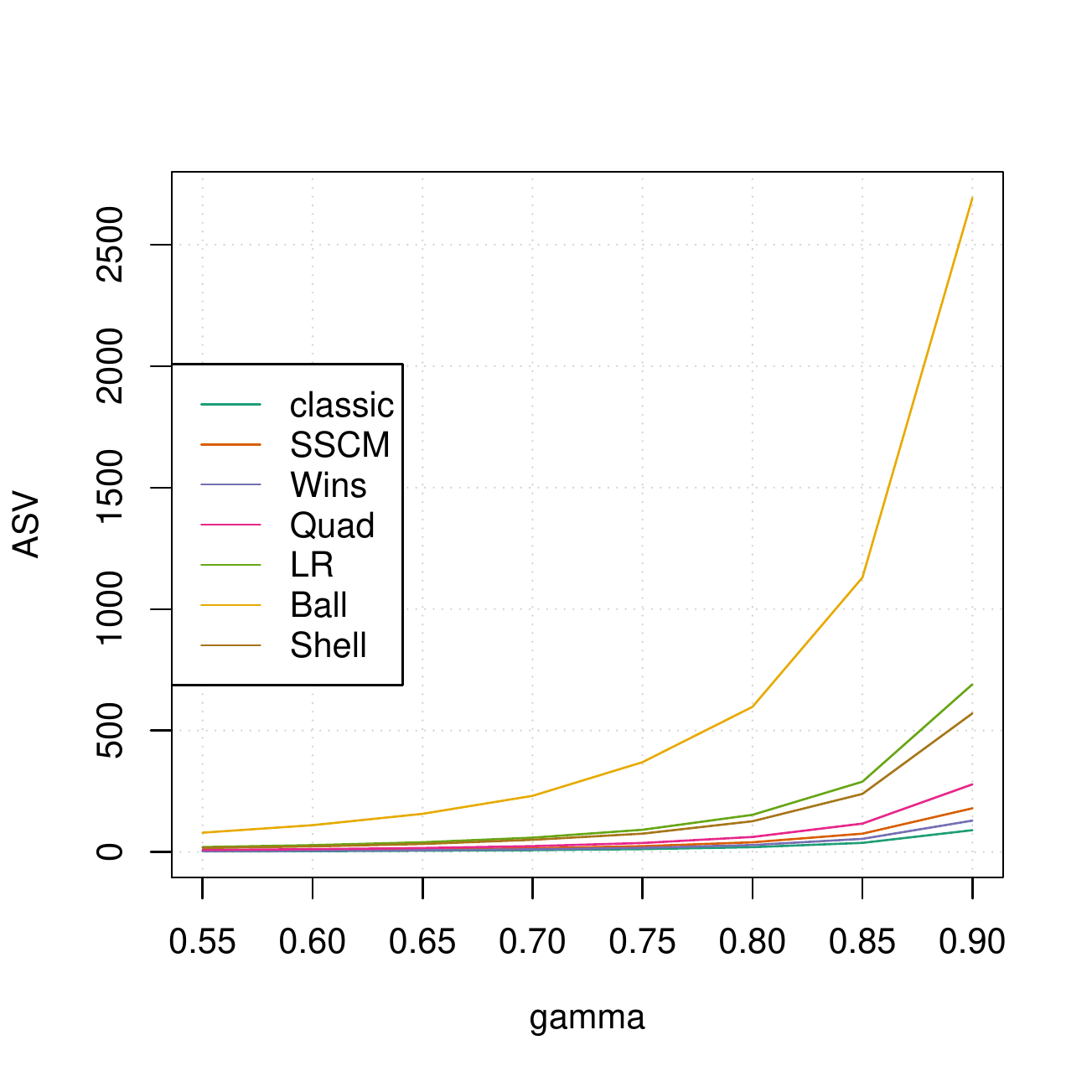} 
            \caption{$0.55 \leq \gamma \leq 0.90 $}
        \end{subfigure}
        \caption{\centering \small Asymptotic variance of the second component of the largest eigenvector for different values of $\gamma$ at bivariate normal distribution $\mathcal{N}(0_2,\text{diag}(1,\gamma))$}
        \label{fig:figASVs}
    \end{figure}

When we assume that $F$ is distributed as $\mathcal{N}(0_2,\text{diag}(1,\gamma))$, we can evaluate the asymptotic variance of Equation \eqref{eq:formule_ASV} for different values of $\gamma$. The results of this computation are presented in Figure \ref{fig:figASVs}.

From the plots in Figure \ref{fig:figASVs}, it is clear that Ball performs the poorest of all radial functions by a large margin. LR and Shell perform somewhat average. The lowest ASV is evidently obtained by classical PCA. However, Winsor's ASV is quite close to it, followed by SSCM. Quad also has a low asymptotic variance.

\begin{figure}[h!]
       \centering
        \begin{subfigure}[t]{.45\linewidth}
            \includegraphics[trim={0 1.5cm 0 1.8cm},clip,width=\textwidth]{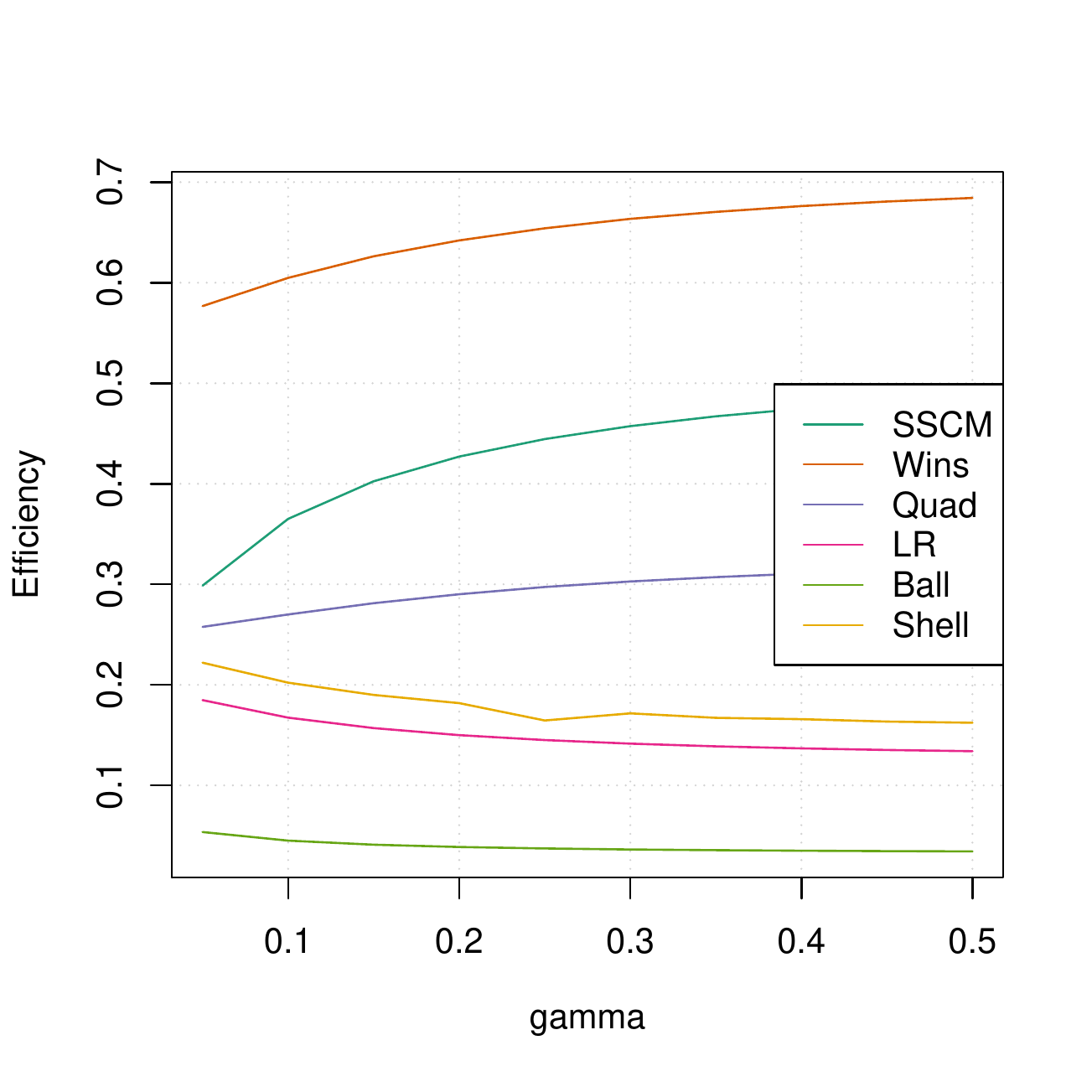} 
            \caption{$0 \leq \gamma \leq 0.5 $}
        \end{subfigure}
        \begin{subfigure}[t]{.45\linewidth}
            \includegraphics[trim={0 1.5cm 0 1.8cm},clip,width=\textwidth]{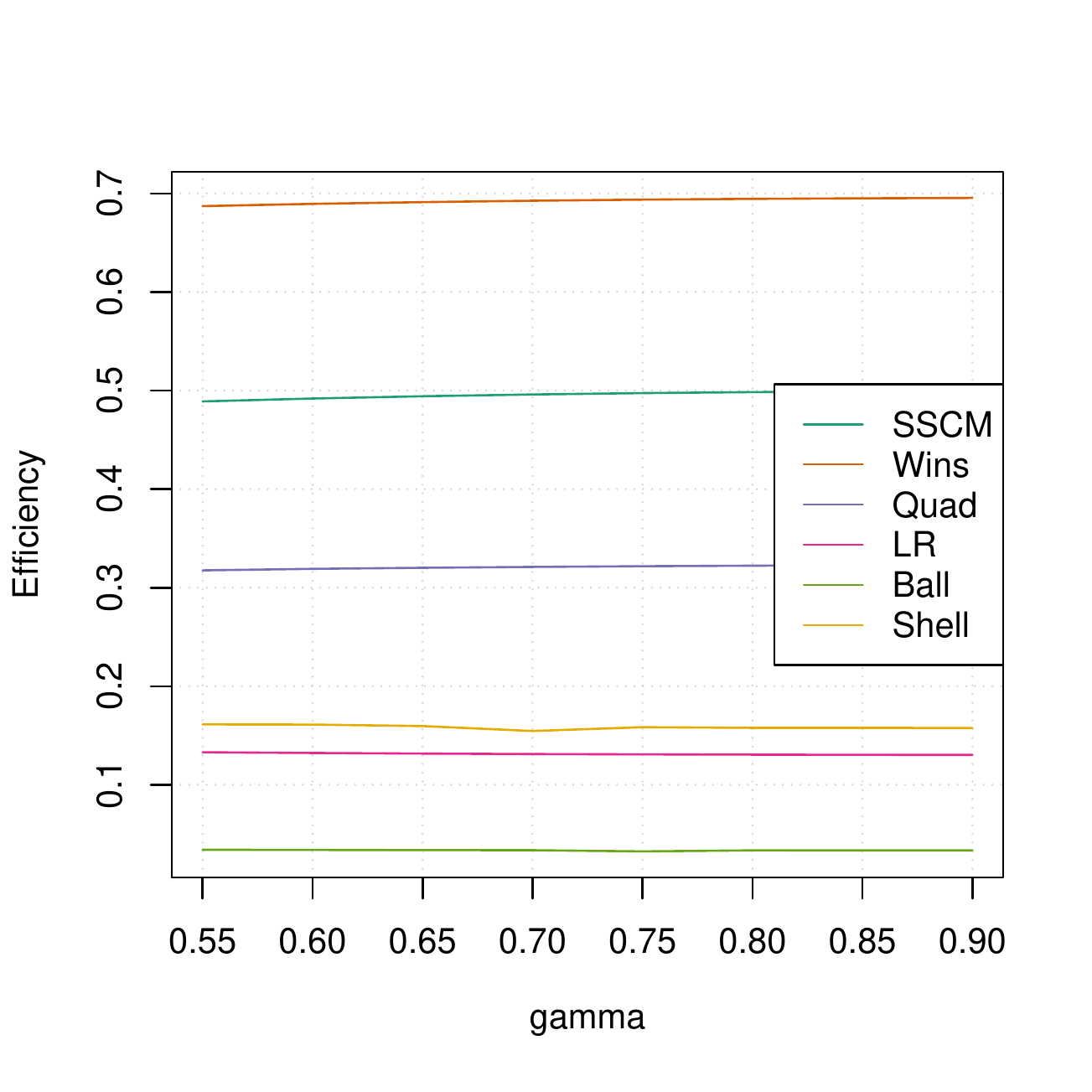} 
            \caption{$0.55 \leq \gamma \leq 0.90 $}
        \end{subfigure}
        \caption{\centering \small Asymptotic relative efficiencies of the second component of the largest eigenvector for different values of $\gamma$ at bivariate normal distribution $\mathcal{N}(0_2,\text{diag}(1,\gamma))$}
        \label{figureEfficiencies}
    \end{figure}

We now use these asymptotic variances to calculate the asymptotic relative efficiency in comparison to the classical method as follows:
\begin{equation*}
    \text{Eff}_g(F) = \frac{\text{ASV}_\text{class}(F)}{\text{ASV}_g(F)}
\end{equation*}
where $g$ refers to the chosen radial function. The closer this value is to one, the more efficient is our method based on the corresponding radial function $g$.
This yields Figure \ref{figureEfficiencies}.
It is immediately clear that the highest efficiency is obtained by Winsor, it seems to converge to 0.7 for higher values of $\gamma$. SSCM is inferior, but still achieves values around 0.5. This is interesting, since it suggests that we can improve on the efficiency of the popular SSCM without sacrificing robustness (in terms of having a bounded influence function).
Next we observe Quad, attaining values around 0.3. Shell, LR and Ball have low efficiencies at bivariate normal distributions. Overall Winsor clearly outperforms the other radial functions in terms of relative efficiency.

The high variance of Ball translates into the lowest efficiency. 
This does not come as a surprise, as Ball bears similarities to the affine equivariant robust covariance estimators which use only half of the data such as the minimum covariance determinant  and minimum volume ellipsoid estimators \citep{rousseeuw1984least}. These are known to have fairly low efficiencies on Gaussian data but strong robustness properties \citep{davies1992asymptotics,butler1993asymptotics, croux1999influence,cator2012central}. Ball is also similar to the BACON algorithm \citep{billor2000bacon}, which also has strong robustness but weak efficiency properties.
    
\subsection{Influence functions of the eigenvalues}
\label{subsectionIFsEVS}
In addition to the eigenvectors, we also study the influence functions of the eigenvalues obtained by GSPCA. Denoting these functionals 
by $\lambda_{g,j}$ for $j = 1, \ldots, p$, we could be interested in 
\begin{equation*}
    \text{IF}(x,\lambda_{g, j},F) = \lim_{\varepsilon \to 0} \frac{\lambda_{g, j}(F_{\varepsilon,x}) - \lambda_{g, j}(F)}{\varepsilon}.
\end{equation*}

The problem with using the expression above however, is that the eigenvalue functionals are not Fisher consistent for elliptical distributions unlike their eigenvector counterparts. Otherwise stated, in general we have $\lambda_{g, j}(F) \neq \lambda_{j}(F)$, as well as $\lambda_{g_1, j}(F) \neq \lambda_{g_2, j}(F)$ for two different functions $g_1$ and $g_2$. As a result, we cannot directly compare these influence functions for different functions $g$ which makes them less interesting.\par
In this section, we will correct the estimated eigenvalues to make them Fisher consistent. We will then calculate the influence functions of these corrected eigenvalue estimators, which now can be compared for different radial functions. Finally, we briefly discuss the covariance matrix obtained by combining the eigenvectors with the re-estimated eigenvalues.

\subsubsection{Modification of GSPCA for Fisher consistent eigenvalues}
\label{SectionModification}

To adapt our method for Fisher consistency, we will use the approach suggested in \citet{Croux2005}. The idea proposed there is that we can correct the eigenvalues by first projecting the data onto the eigenvectors before calculating the eigenvalues. More specifically we proceed as follows:
\begin{enumerate}
    \item First, we calculate the eigenvectors $v_{g,j}$ through GSPCA as done before. We know that these are Fisher consistent provided that we have an elliptical distribution.
    \item Second, we project the data onto these eigenvectors after which we can use a robust, equivariant scale estimator $S$ to calculate the eigenvalues:
    \begin{equation*}
        \lambda_{S,j} = S^2(v_{g,j}^T X)
    \end{equation*}
    and for a finite sample:
    \begin{equation*}
        \lambda_{S_n,j} = S_n^2(v_{g,j}^T x_1,...,v_{g,j}^T x_n)
    \end{equation*}
    Provided that the estimator $S$ itself is a Fisher consistent estimator, this procedure yields Fisher consistent eigenvalue estimates at elliptically symmetric distributions (see, e.g., \cite{Croux2005}).

    \item Lastly, should we be interested in the covariance matrix, we can combine the two steps above to obtain a new estimator $\Sigma_\text{comb}$ for the covariance matrix:
    \begin{align*}
        \Sigma_{\text{comb}}(F) &= \sum_{k=1}^p \lambda_{S,k}(F) v_{g,k}(F) v_{g,k}(F)^T = \sum_{k=1}^p \lambda_{k}(F) v_{k}(F) v_k(F)^T
    \end{align*}
    Here the last equality holds if $F$ is an elliptically symmetric distribution, then we have Fisher consistency for $\Sigma$.
\end{enumerate}
In summary, we can correct our method to obtain Fisher consistency for the eigenvalues and the resulting covariance estimate at elliptically symmetric distributions. Next, we will take a look at the corresponding influence functions and asymptotic variances to evaluate the robustness and efficiency of the adapted GSPCA.

\subsubsection{Influence functions of the new eigenvalues}
In order to obtain the influence function for the corrected eigenvalue functionals, we need a general property of projections of elliptically symmetric distributions. Denote the distribution of the random vector projected on $a$ as $H^a$, i.e., if $X \sim H$ then $a^T X \sim H^a$. We will make use of following lemma from \citet{Croux2005} on the distribution of this projection $H^a$.
\begin{lemma}[Projected distributions]\label{lem:proj}
Assume $H$ is an elliptically symmetric distribution with location parameter $\mu$ and covariance matrix $\Sigma$. Then there exists a univariate symmetric distribution $F_0$ such that:
\begin{equation*}
    H^a(z) = F_0 \left( \frac{z-\mu^T a}{\sqrt{a^T \Sigma a}} \right)
\end{equation*}
\end{lemma}
Lemma \ref{lem:proj} ensures that all projections of a random vector with an elliptically symmetric distribution follow the same symmetric distribution $F_0$ after proper scaling and centering.\par
We can now obtain an expression for the influence functions of the corrected eigenvalue functionals. The proof can be found in Section \ref{app:thm3} of the Supplementary material. 
\begin{theorem}[Influence functions of the new eigenvalues]\label{thm:IF2}
For our modified eigenvalues $\lambda_{S,k}$ with $S$ an equivariant scale functional, we have following expression for the influence function at elliptically symmetric distributions:
\begin{equation}
    \label{IF_new_eigenvalue}
    \text{IF}\left(x,\lambda_{S,k},H\right)  = 2 \lambda_k \ \text{IF}\left(\frac{x^T v_k}{\sqrt{\lambda_k}},S,F_0\right) 
\end{equation}
\end{theorem}
This is the same expression as the one obtained by \citet{Croux2005} in the context of robust PCA based on projection-pursuit. In particular, note that the expression does not depend on the influence function of any of the eigenvectors.

From the expression of the influence function of the eigenvalues, we readily obtain the asymptotic variance of the new eigenvalues at elliptically symmetric distribution:
\begin{align*}
     \text{ASV}(\lambda_{S,k},H) &= \mathbb{E} [\text{IF}(X,\lambda_{S,k},H)^2]
    = \mathbb{E} \left[4 \lambda_k^2 \ \text{IF}\left( \frac{v_k^T X}{\sqrt{\lambda_k}}, S,F_0 \right)^2 \right]
    = 4 \lambda_k^2 \ \text{ASV}(S, F_0)
\end{align*}

Note that the expression for the influence function of the eigenvalues is proportional to the influence function of the scale functional used. This means that known optimality properties of equivariant scale functionals carry over to the estimation of these eigenvalues. In particular, we can leverage the theory on M-estimation to find optimal M-estimators of scale for estimating the eigenvalues. This is stated in Corollary \ref{cor:MestScale}, which follows from Theorem \ref{thm:IF2} and \cite{Hampel} (Section 2.5e),   and leads us to the use of the the median absolute deviation (MAD) defined by 
\begin{align*}
        \text{MAD}(X) &= c \cdot \text{med} \vert X - \text{med}(X) \vert
\end{align*}
Here $c = 1/\Phi^{-1}\left(\frac{3}{4}\right) \approx 1.4826$ is a consistency factor at normal distributions, hence we obtain Fisher consistency for the estimated eigenvalues.

\begin{cor}\label{cor:MestScale}
If $F_0 = \mathcal{N}(0,1)$, the median absolute deviation is the most B-robust M-estimator of the eigenvalues, i.e. it has the lowest gross error sensitivity among all (well-behaved) Fisher consistent M-estimators of scale. Similarly, Huber's M-estimator is the optimal B-robust M-estimator of the eigenvalues as it has the highest efficiency for a given bound on the gross error sensitivity.
\end{cor}

 We visualize the influence function of the new eigenvalue for the multivariate normal distribution below. Suppose $H\sim \mathcal{N}(0_2, \mbox{diag}(1, \rho))$ and thus $F_0 \sim \mathcal{N}(0, 1) = \Phi$.
 The influence function of the MAD is given by (\citet{Hampel}, p. 107):
\begin{equation*}
    \text{IF}(x,\text{MAD},\Phi) = \frac{\text{sign}\left(\vert x \vert -\Phi^{-1}\left(\frac{3}{4}\right)\right)}{4 \Phi^{-1}(\frac{3}{4}) \phi(\Phi^{-1}\left(\frac{3}{4}\right))}.
\end{equation*}

\begin{figure}[H]
    \centering
    \includegraphics[trim={0 0.6cm 0 2cm},clip,scale=0.45]{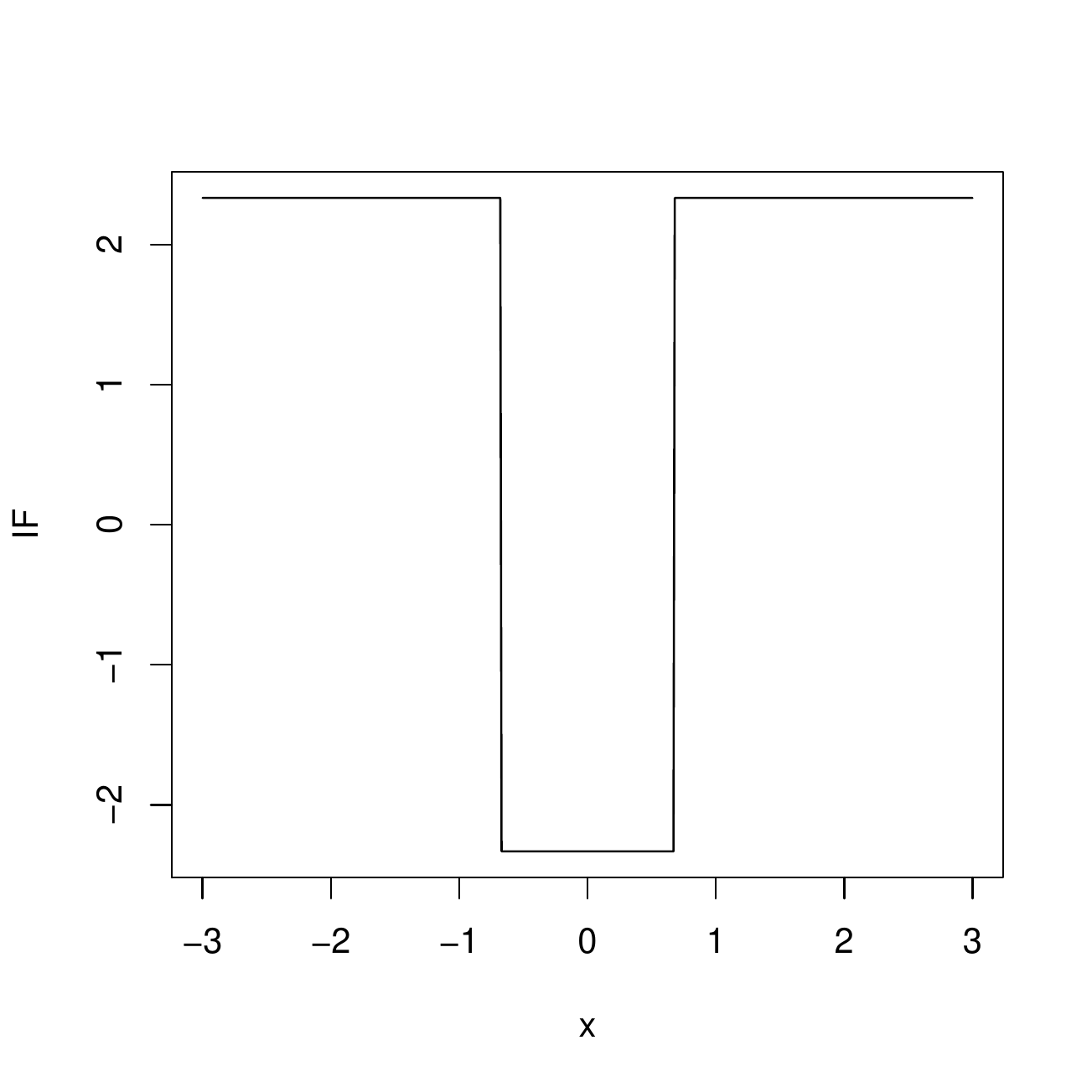}
    \caption{\centering \small Influence function of the largest (new) eigenvalue for $x \in [-3,3]$}
\label{fig:IFmad}
\end{figure}

The influence function of the largest eigenvalue $\lambda_1 = 1$ then becomes $2 \ \text{IF}(x_1,\text{MAD},\Phi)$, which is shown in Figure \ref{fig:IFmad}.

\subsubsection{Influence function of the new covariance estimator}

\begin{figure}[h!]
        \centering
         \begin{subfigure}[t]{.45\linewidth}
            \includegraphics[trim={0 1.5cm 0 1.8cm},clip,width=\textwidth]{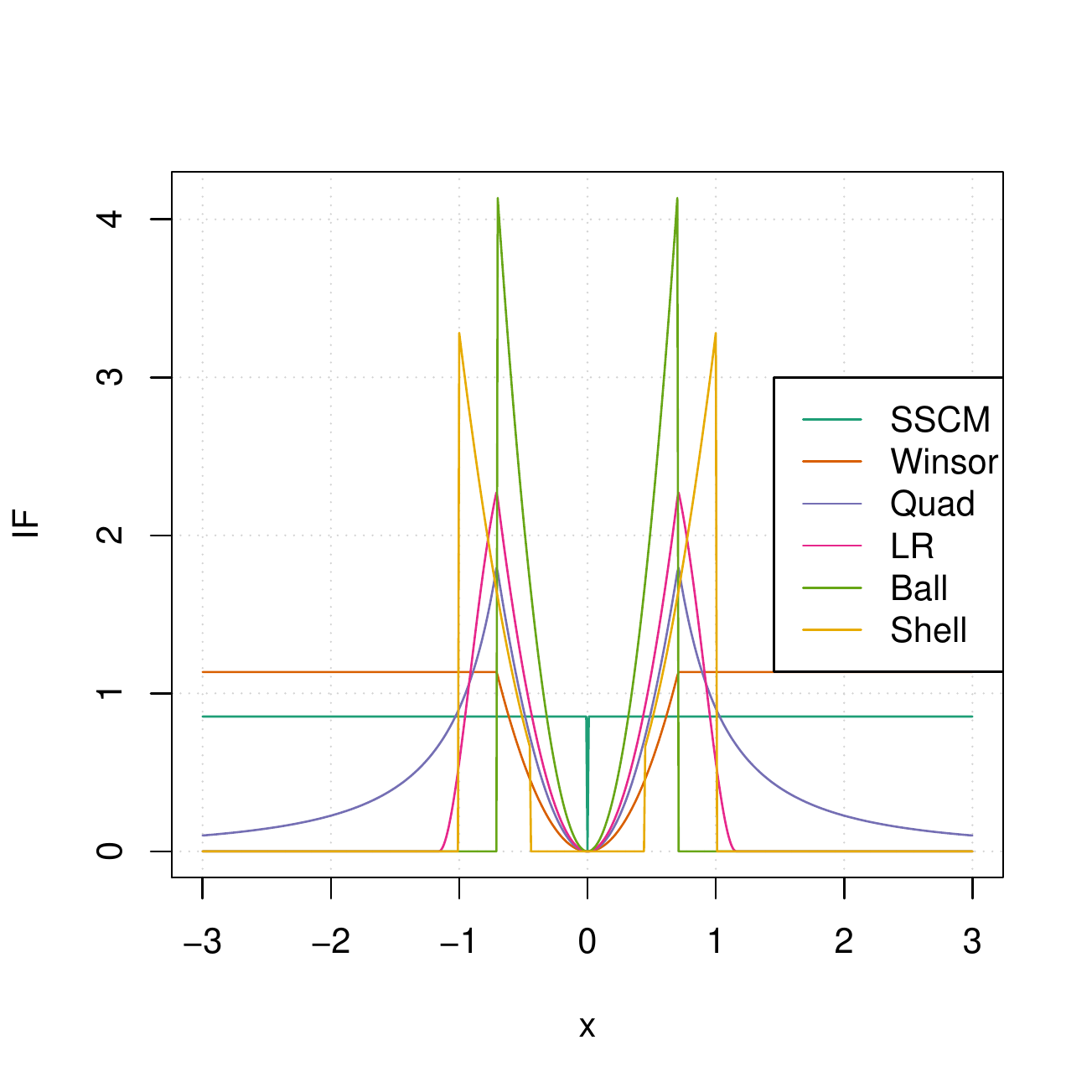} 
        \end{subfigure}
        \begin{subfigure}[t]{.45\linewidth}
            \includegraphics[trim={0 1.5cm 0 1.8cm},clip,width=\textwidth]{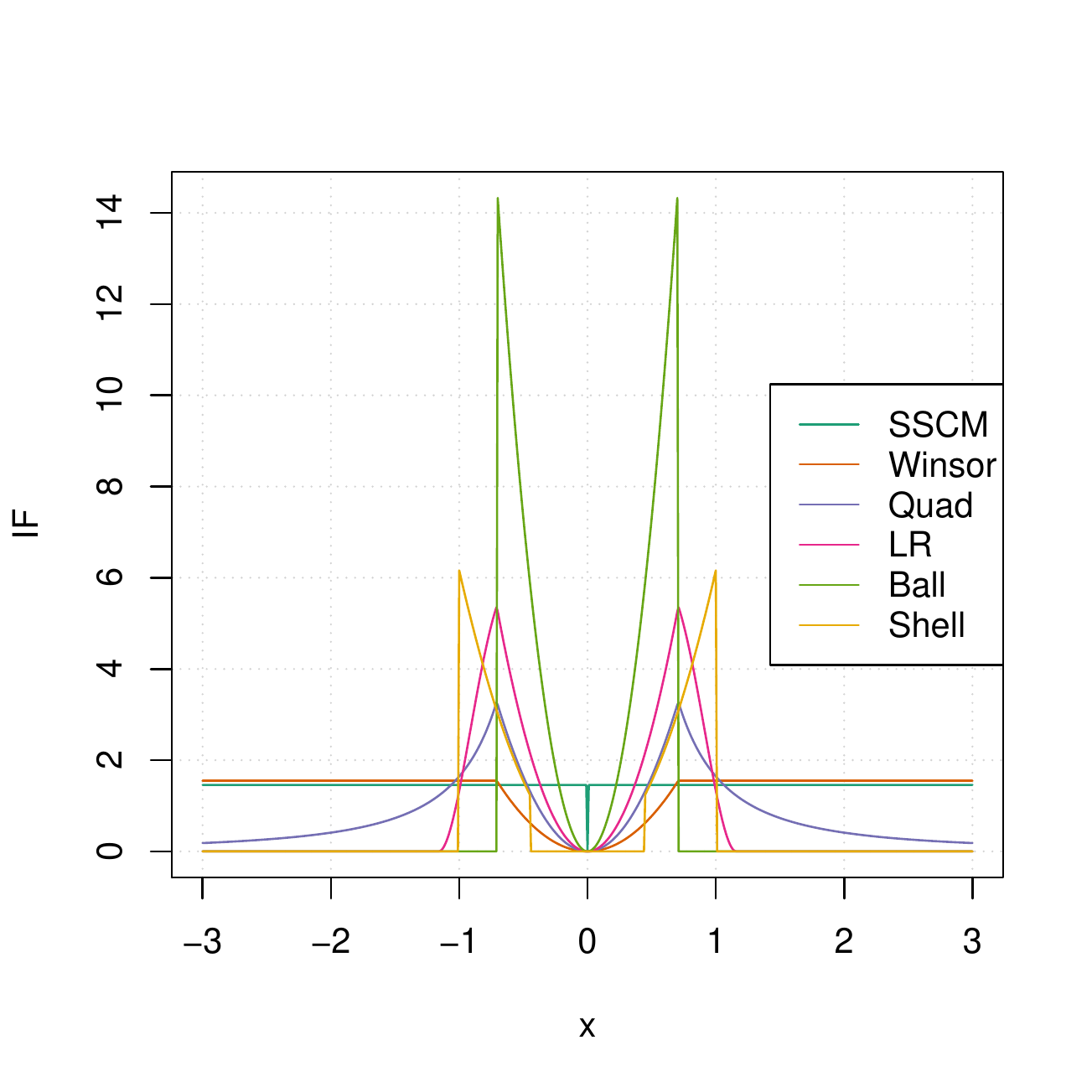} 
        \end{subfigure}
        \caption{\centering \small Influence function of the uncorrected GSSCM $\Sigma$ (left) and the corrected GSSCM $\Sigma_{\text{comb}}$ (right) for contamination in the direction of $(x,x)^T$: the off-diagonal element}
        \label{IFcorrectedGSSCM}
    \end{figure}

Finally, as an additional result, we compute the influence function of the new estimate for the covariance matrix $\Sigma_{\text{comb}}$ (see Section \ref{app:thm4} of the Supplementary Material for the proof).
\begin{theorem}[Influence function of the new covariance matrix $\Sigma_{\text{comb}}$]\label{thm:IF3}
Let $H$ be an elliptically symmetric distribution with a diagonal covariance matrix and $S$ an equivariant scale estimator. Then
\begin{align*}
    &\text{IF}(x,\Sigma_{\text{comb}},H) = 2  \sum_{k=1}^p \lambda_k \ \text{IF} \left( \frac{v_k^T x}{\sqrt{\lambda_k}}, S, F_0 \right) v_k v_k^T  \nonumber \\ &+ \sum_{k=1}^p \lambda_k 
    \sum_{j=1, j \neq k}^{p} \frac{1}{\lambda_{g,k} - \lambda_{g,j}} (g(x)_j g(x)_k) (v_j v_k^T + v_k v_j^T)
\end{align*}
\end{theorem}
Next we compare the influence functions of the new covariance estimate using MAD as scale estimator $S$ with the common GSSCM. For this we plot the influence functions for $H \sim \mathcal{N}(0_2,\text{diag}(1,\rho))$ and focus on the off-diagonal element. For that, we obtain:
\begin{equation*}
    \text{IF}(x,\Sigma_{\text{comb}},H)_{1,2} = \frac{1 - \rho}{\lambda_{g,1} - \lambda_{g,2}} g(x)_1 g(x)_2
\end{equation*}
This corresponds to a rescaling of the influence function of the common GSSCM. For $\rho = 0.5$ and contamination in the direction of $(x,x)^T$ this yields Figure \ref{IFcorrectedGSSCM}.

\noindent

As we can see from the comparison of the raw GSSCM and our proposed correction, the relative performances among different radial functions remain relatively stable. However, the influence function of Winsor is now similar to that of SSCM except in the center where it is lower. This again speaks in favor of Winsor instead of the popular SSCM. Contamination in the direction of $(x,0)^T$ results in a zero influence function for each radial function since $g(x)_2 = 0$.

\section{Simulation study}
\label{SectionSimulationStudy}

In this section we conduct a simulation study on synthetically sampled data to investigate the performance of generalized spherical PCA using the five radial functions defined in Equations \eqref{defWinsor} -- \eqref{defLR}. We will compare them with classical PCA, spherical PCA based on the SSCM and ROBPCA, the state-of-the-art method for robust PCA.

The ROBPCA method was published by \citet{ROBPCA}. It is based on the projection pursuit and the minimum covariance determinant estimator. Just as GSPCA, ROBPCA has the favorable property of being orthogonally equivariant. For our simulation, we will largely base ourselves on the methodology of the study conducted in the prior mentioned ROBPCA paper \citep{ROBPCA}.
\begin{table*}[h!]
\small
\centering
\scalebox{0.85}{
\begin{tabular}{|c | c c | c c c c c c c c| }
\hline
 & $n$ & $p$ & CPCA & ROBPCA & SSCM & Ball & LR & Quad & Shell & Winsor\\ 
\hline
Normal & 100 & 4 & \textbf{.100} & .138 & .137 & .257 & .163 & .132 & .149 & .113\\
 & 50 & 100 & \textbf{.216} & .245 & .275 & .332 & .272 & .243 & .311 & .227\\
\hline
$t_5$ & 100 & 4 & .134 & .138 & .131 & .243 & .163 & .137 & .170 & \textbf{.113}\\
 & 50 & 100 & .303 & .281 & .274 & .395 & .313 & .277 & .378 & \textbf{.254}\\
\hline
$t_3$ & 100 & 4 & .185 & .140 & .122 & .222 & .159 & .135 & .174 & \textbf{.113} \\
 & 50 & 100 & .419 & .305 & .272 & .416 & .332 & .294 & .407 & \textbf{.267} \\
\hline
$t_2$ & 100 & 4 & .249 & .156 & .129 & .235 & .162 & .142 & .186 & \textbf{.119} \\
 & 50 & 100 & .554 & .354 & \textbf{.280} & .455 & .358 & .318 & .445 & .285 \\
\hline
$t_1$ & 100 & 4 & .403 & .183 & .120 & .236 & .173 & .155 & .204 & .\textbf{117} \\
(Cauchy) & 50 & 100 & .710 & .495 & \textbf{.277} & .494 & .403 & .366 & .497 & .304 \\
\hline
\end{tabular}
}
\captionof{table}{\small \centering Maxsub measure at uncontaminated data}
\label{MaxsubNoContamination}
\end{table*}
\subsection{Setting}
\label{subsectionSimSetting}
In the simulation, we will look at uncontaminated data ($\varepsilon = 0 $) and data with different levels of contamination ($\varepsilon = 0.1 , \ \varepsilon = 0.2$) obtained from the following $p$-variate Gaussian distribution and the  $p$-variate Student's $t$-distribution with five degrees of freedom:
\begin{align*}
    (1 - \varepsilon) \ \mathcal{N}_p(0, \Sigma) + \varepsilon  \ \mathcal{N}_p(\tilde{\mu}, \tilde{\Sigma})\\
    (1 - \varepsilon) \ t_5(0, \Sigma) + \varepsilon  \ t_5(\tilde{\mu}, \tilde{\Sigma})
\end{align*}
From these distributions, we will repeatedly generate 500 data samples of size $n$. As the GSSCM scatter estimator is orthogonally equivariant, we only need to look at diagonal covariance matrices $\Sigma$. More in specific, we consider the following two situations, just as in the ROBPCA paper \citep{ROBPCA}:
\begin{enumerate}
    \item \underline{Low-dimensional data}: \\
    $n=100$, $p=4$, $\Sigma = \text{diag}(8,4,2,1)$, $\tilde{\mu} = f_1 \cdot e_4 = (0,0,0,f_1)^T$, $\tilde{\Sigma}=\Sigma / f_2$
    \item \underline{High-dimensional data}:\\
    $n=50$, $p=100$, $\Sigma = \text{diag}(17,13.5,8,3,1,0.095,...,0.002,0.001)$, $\tilde{\mu} = f_1 \cdot e_6$, $\tilde{\Sigma}=\Sigma / f_2$
\end{enumerate}
In the low-dimensional case we compute $k=3$ principal components, whereas in the high-dimensional case we compute $k=5$ components, as we want our principal component analysis to explain at least 90\% of the total variance. Parameter $f_1 \in\{ 6,8,10,\dots,20\}$ determines the location shift for the contaminated data in the direction of the $k+1^{\text{th}}$ principal component, being $e_4$ or $e_6$. Parameter $f_2 \in \{1,15\}$ specifies the concentration of the contaminated data.

\subsection{Maxsub measure}
\label{maxsubsection}
To evaluate performance in our simulation, we will compute the maxsub measure, which calculates the maximal angle between the estimated PCA subspace and the space spanned by $\{ e_1, e_2, ... , e_k\}$. This can be computed as follows \citep{ROBPCA}:

\begin{equation*}
    \text{maxsub} = \text{arccos}(\sqrt{\lambda_k})/(\pi/2)
\end{equation*}
\noindent
where $\lambda_k$ represents the smallest eigenvalue of $I_{p,k}^T V_{p,k} V_{p,k}^T I_{p,k}$ with $I_{p,k} = (e_1 \ e_2 \dots e_k)$ and $V_{p,k} = (v_1 \ v_2 \dots v_k)$. We divide by $\pi/2$ to standardize the value of the angle. The ideal value of the maxsub measure is zero, the worst value is one.

\subsubsection{Uncontaminated data}
In a first step, we simulate uncontaminated data ($\varepsilon = 0$). As discussed, we sample from the multivariate normal distribution and the Student's $t$-distribution. We however do not limit ourselves here to five degrees of freedom for the $t$-distribution,  we also consider $t_3, \ t_2 \ \text{and} \ t_1$. The mean of the maxsub measure for 500 samples is shown in Table \ref{MaxsubNoContamination}.

When there is no contamination, we see that for the normal and for the $t_5$-distribution the best values are attained by classical PCA (evidently) and GSPCA with the Winsor radial function. This result is in line with the computed efficiencies in Section \ref{subsectionASV}, where Winsor attained the highest efficiency of all radial functions. ROBPCA, SPCA and Quad perform almost as good. The worst results are those of GSPCA with the Ball function, making this method least efficient at no contamination. Shell also performs poorly, especially in the high dimensional case. The low efficiency at no contamination for Ball and Shell is due to the fact that a lot of data points are given a zero weight, see Equation \eqref{defBall} and \eqref{defShell}.

For the $t_1$-, $t_2$- and $t_3$-distribution the results are different. These distributions are heavy tailed, meaning that some sampled data points are located far away from the sample mean. These points at the tails typically have a significant effect on the classical covariance matrix, which is clear from Table \ref{MaxsubNoContamination}. As the degrees of freedom in the $t$-distribution decrease, making it more heavy tailed, CPCA attains high values for the maxsub measure, meaning that it fails. GSPCA with Winsor and SPCA on the other hand, are not affected by the heavy tailed distribution. Hence in this scenario, the use of Winsor or the SSCM is suggested over the use of CPCA, even if there is no contamination.

\begin{figure*}[h!]
\centering
\begin{tabular}{cc}
\includegraphics[trim={0.2cm 0.7cm 1.05cm 0.7cm},clip,width=0.43\textwidth]{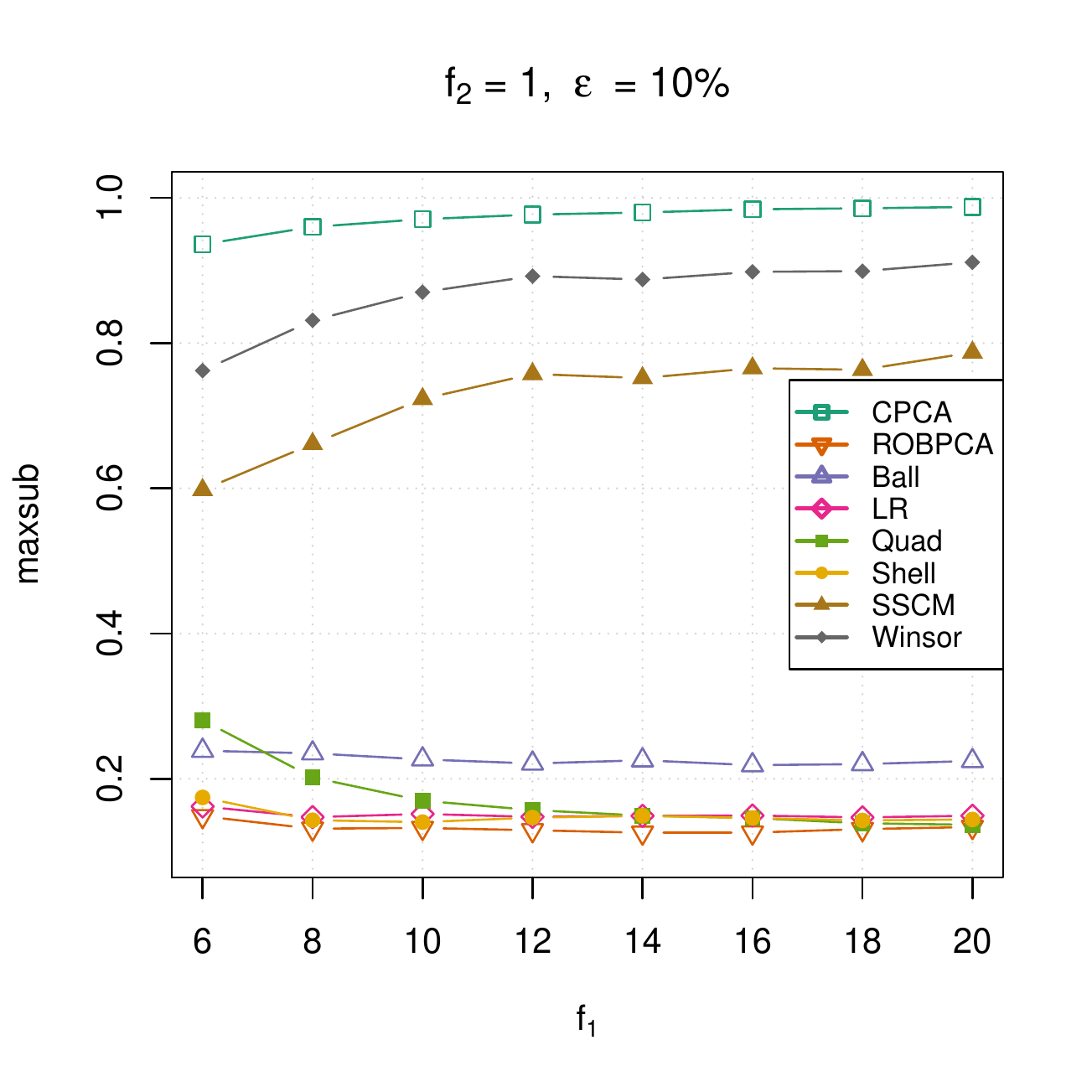} &
\includegraphics[trim={0.2cm 0.7cm 1.05cm 0.7cm},clip,width=0.43\textwidth]{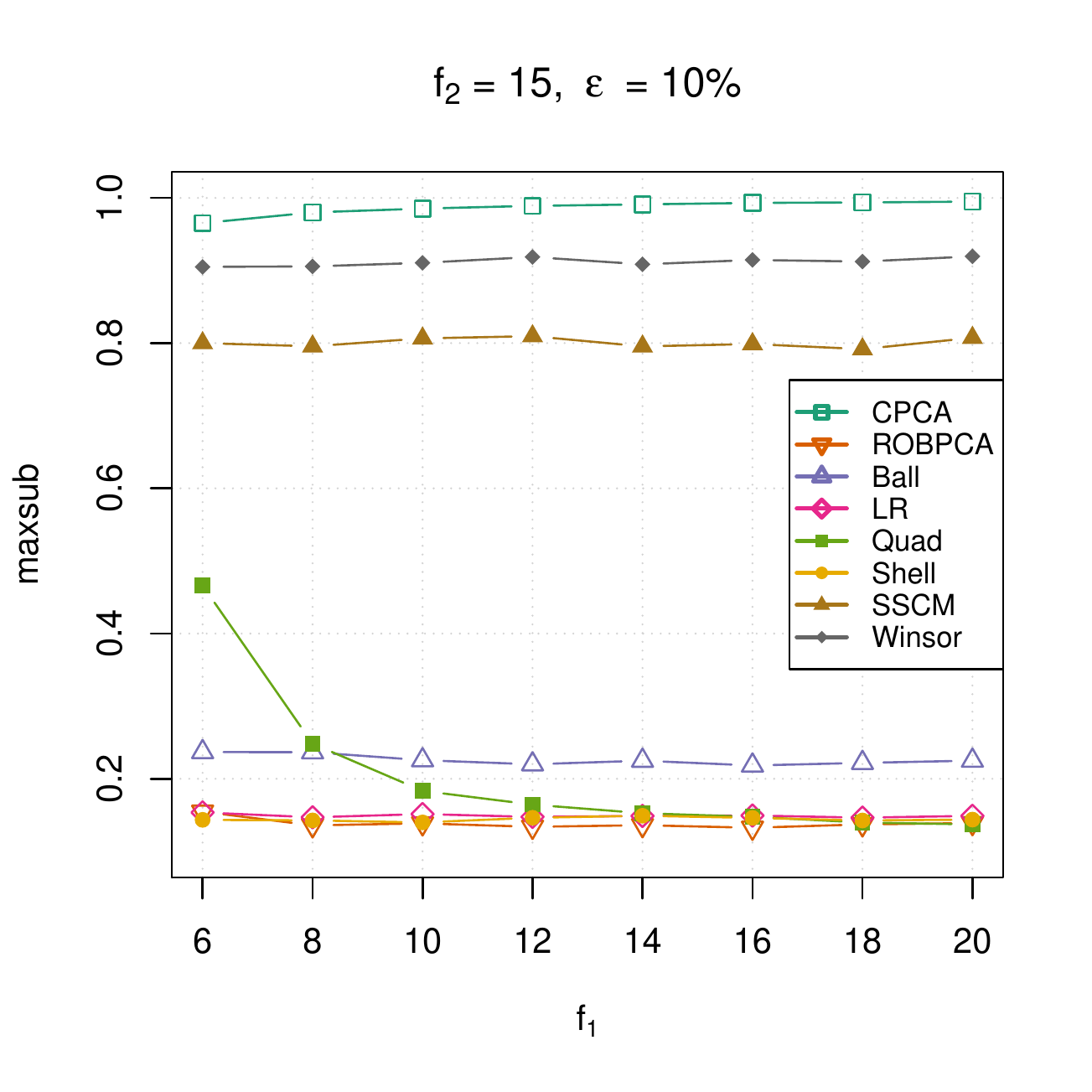} \\
\end{tabular}
\begin{tabular}{cc}
\includegraphics[trim={0.2cm 0.7cm 1.05cm 0.7cm},clip,width=0.43\textwidth]{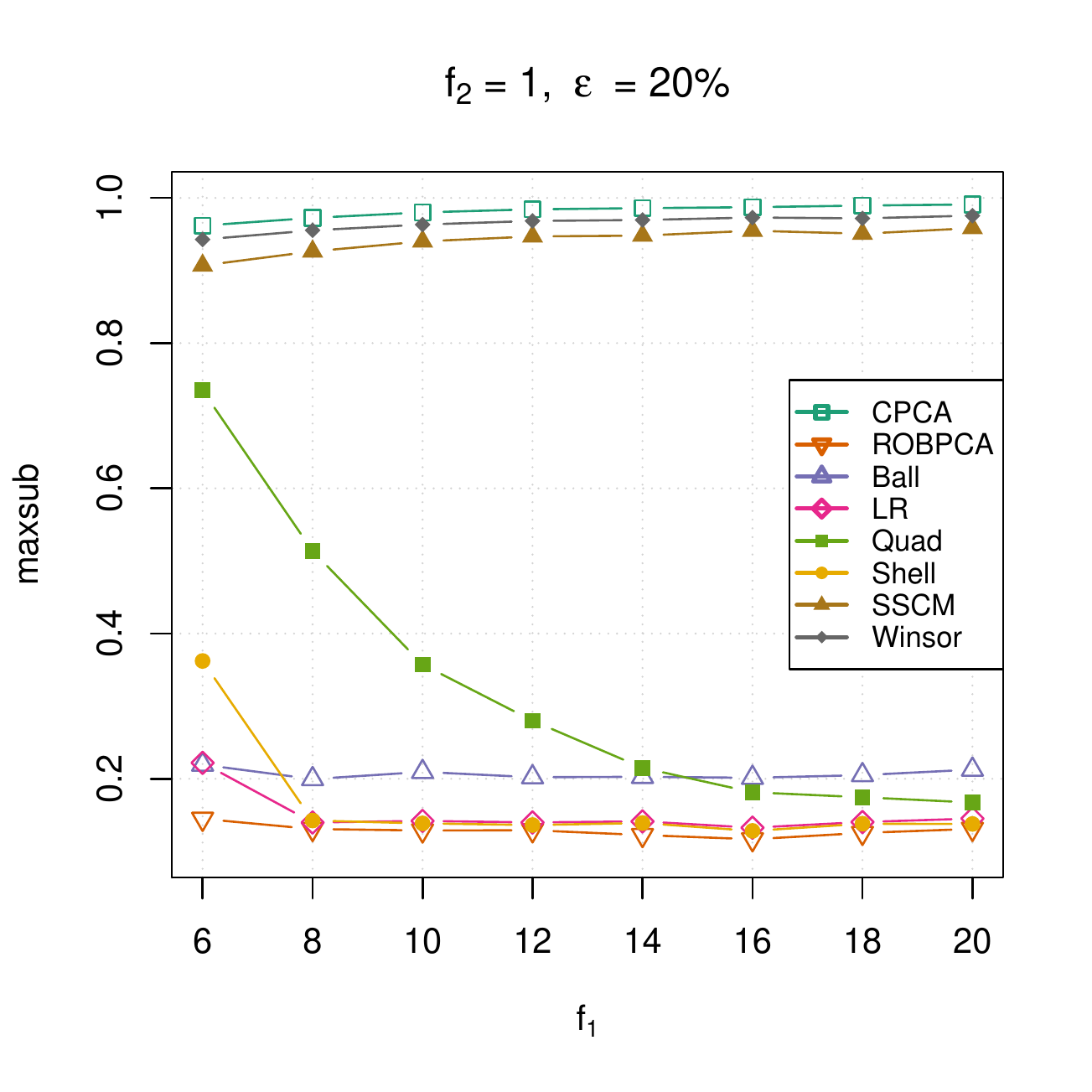} &
\includegraphics[trim={0.2cm 0.7cm 1.05cm 0.7cm},clip,width=0.43\textwidth]{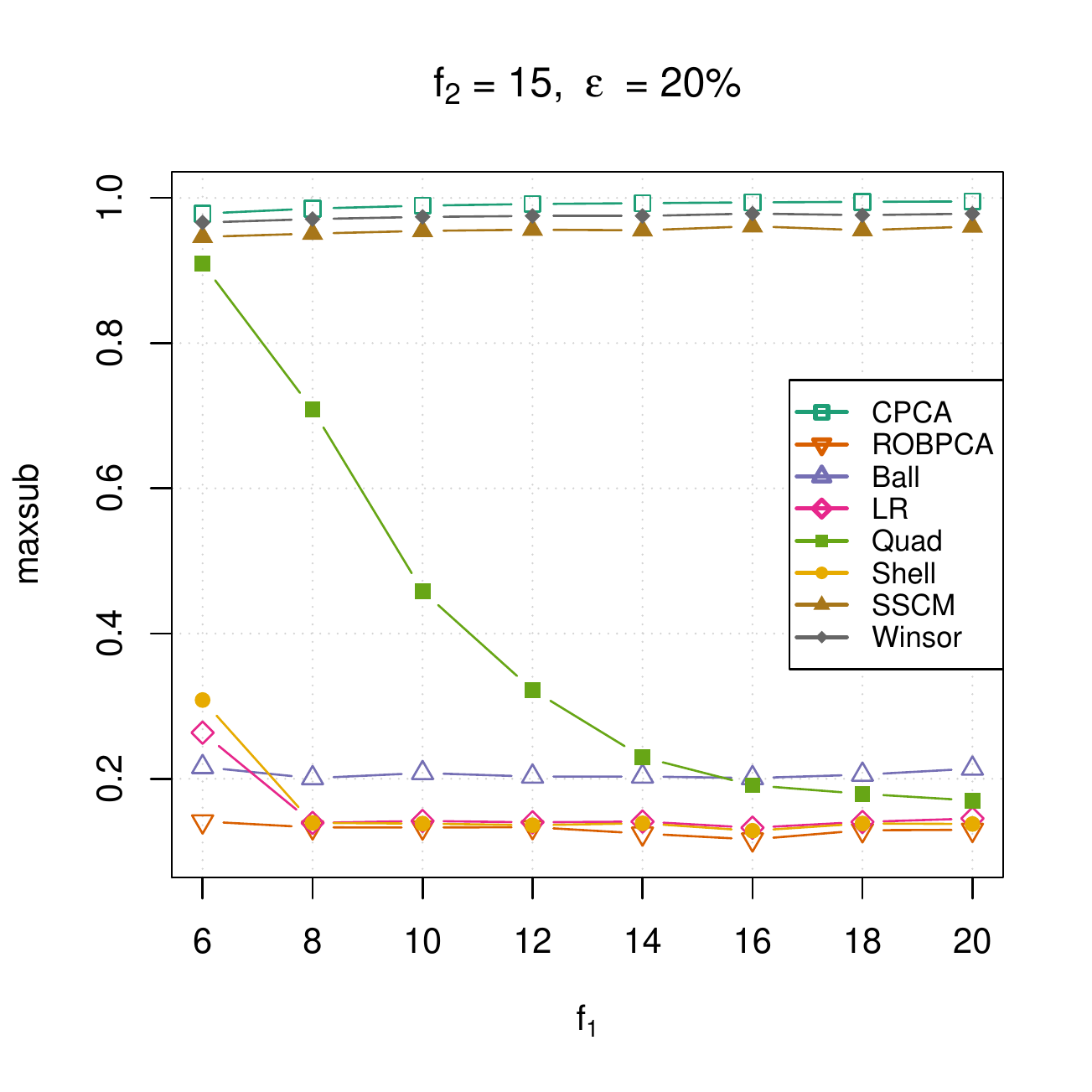} \\
\end{tabular}
\captionsetup{width=0.9\textwidth}
\caption{\centering \small Maxsub measure for low-dimensional, multivariate normal data}
\label{lowdimnormal}
\end{figure*}

\subsubsection{Contaminated data}
\begin{figure*}[h!]
\centering
\begin{tabular}{cc}
\includegraphics[trim={0.2cm 0.7cm 1.05cm 0.7cm},clip,width=0.43\textwidth]{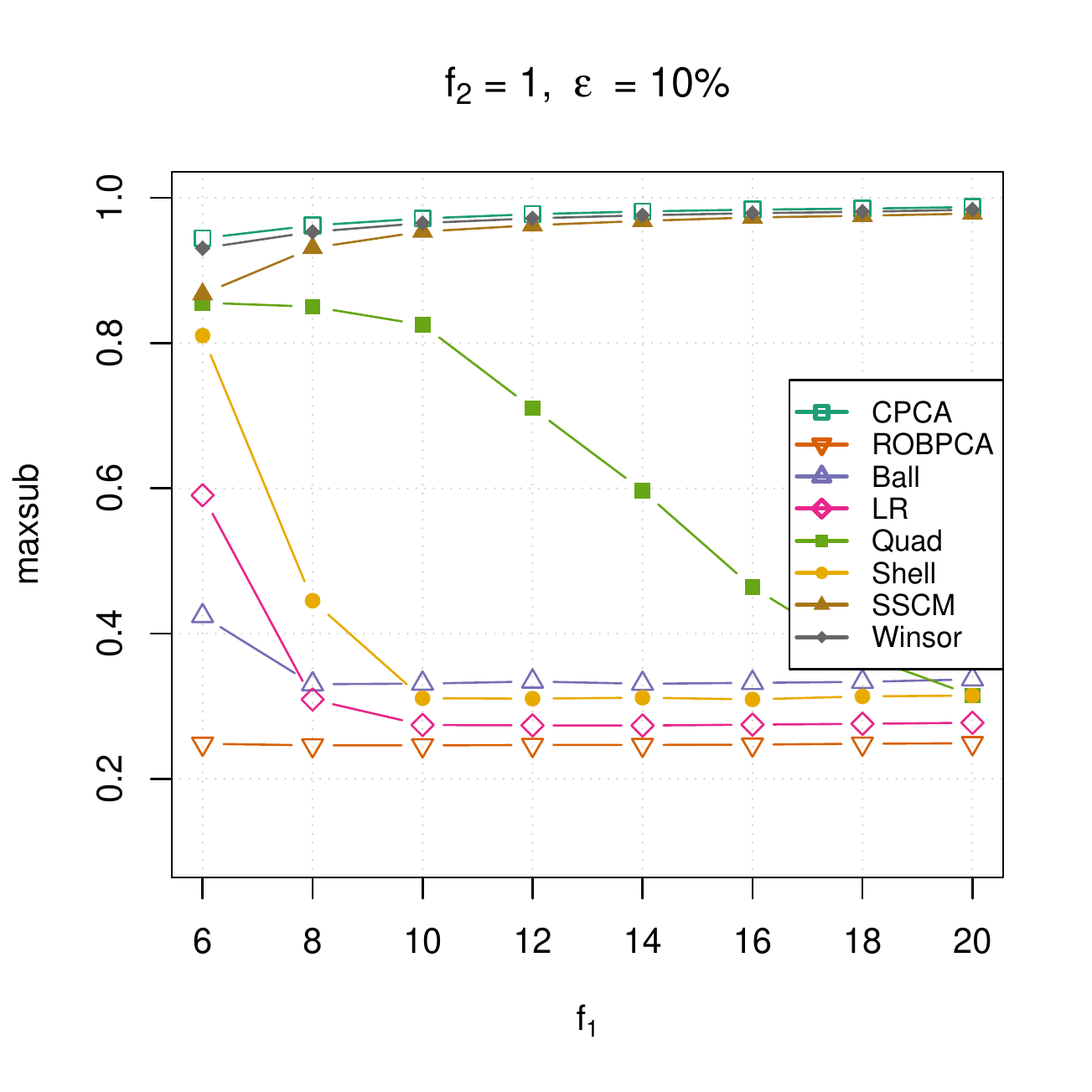} &
\includegraphics[trim={0.2cm 0.7cm 1.05cm 0.7cm},clip,width=0.43\textwidth]{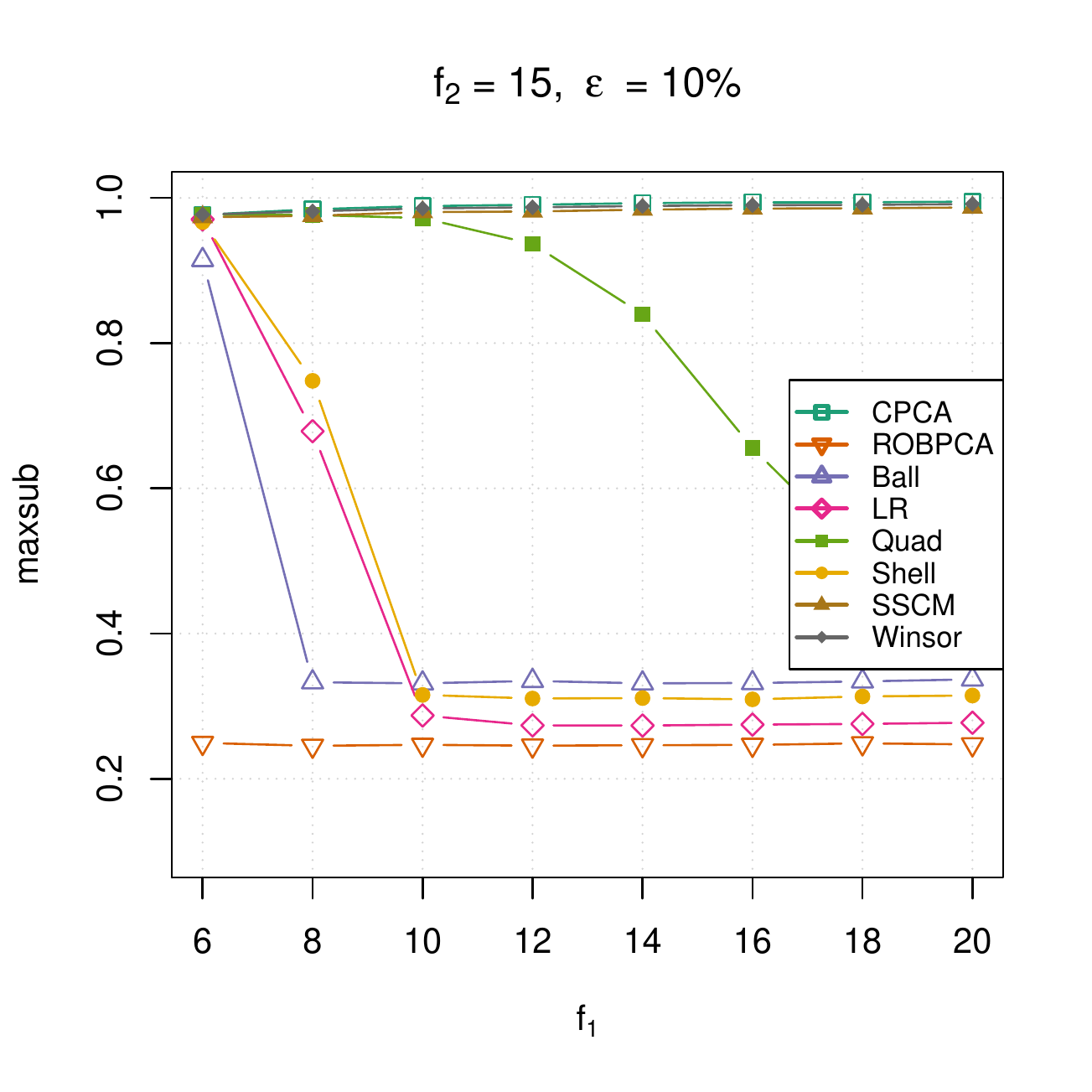} \\
\end{tabular}
\begin{tabular}{cc}
\includegraphics[trim={0.2cm 0.7cm 1.05cm 0.7cm},clip,width=0.43\textwidth]{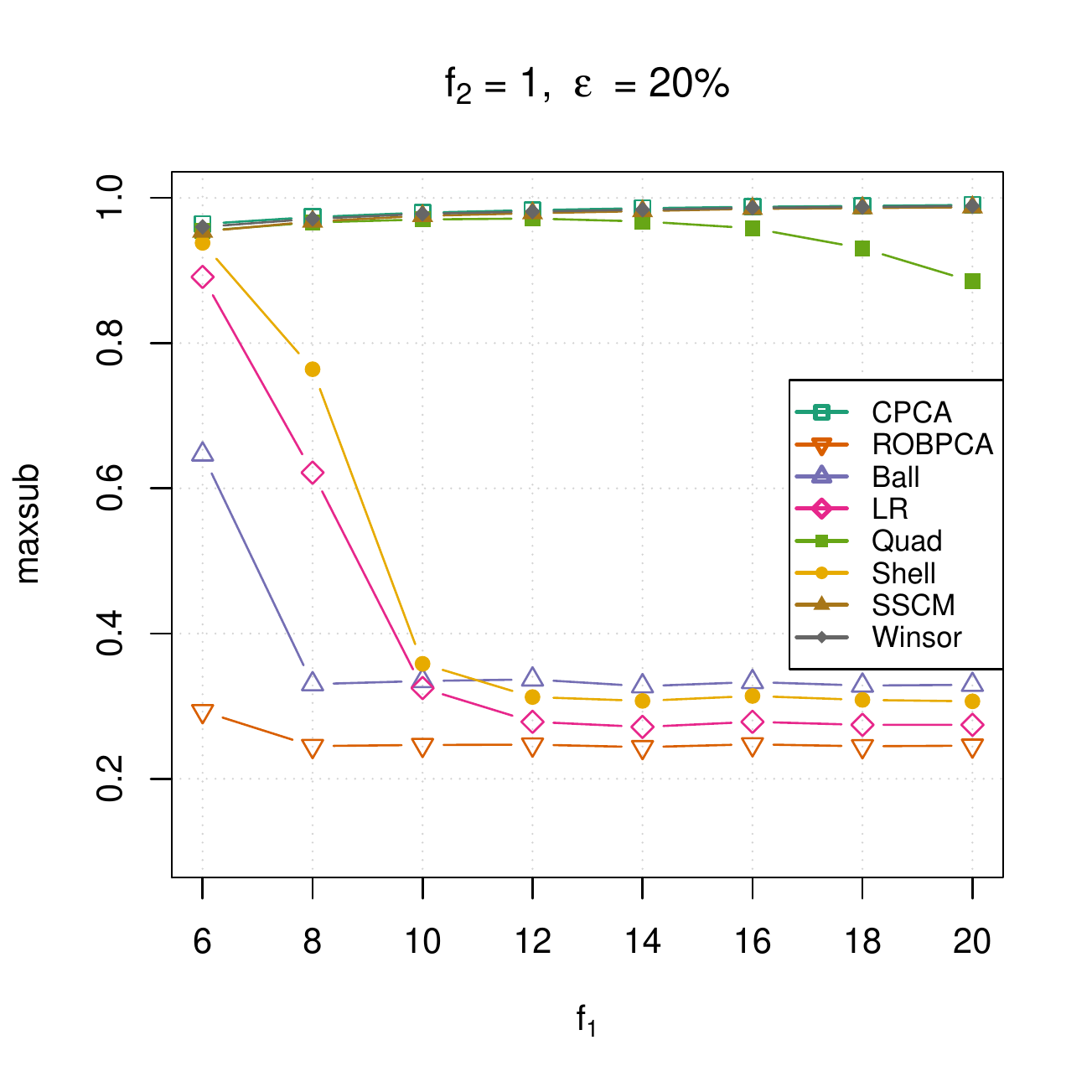} &
\includegraphics[trim={0.2cm 0.7cm 1.05cm 0.7cm},clip,width=0.43\textwidth]{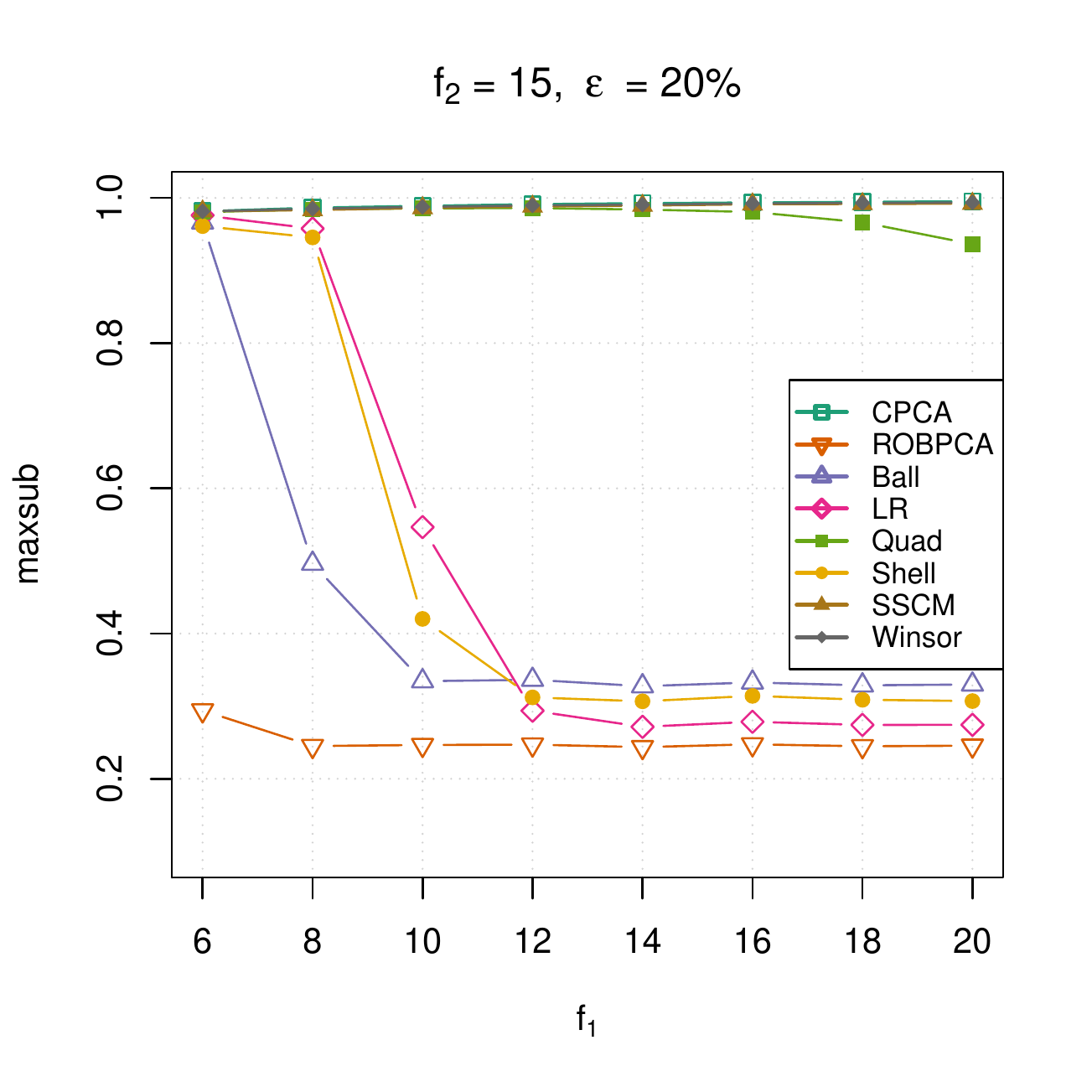} \\
\end{tabular}
\captionsetup{width=0.9\textwidth}
\caption{\centering \small Maxsub measure for high-dimensional, multivariate normal data}
\label{highdimnormal}
\end{figure*}

\begin{figure*}[h!]
\centering
\begin{tabular}{cc}
\includegraphics[trim={0.2cm 0.7cm 1.05cm 0.7cm},clip,width=0.43\textwidth]{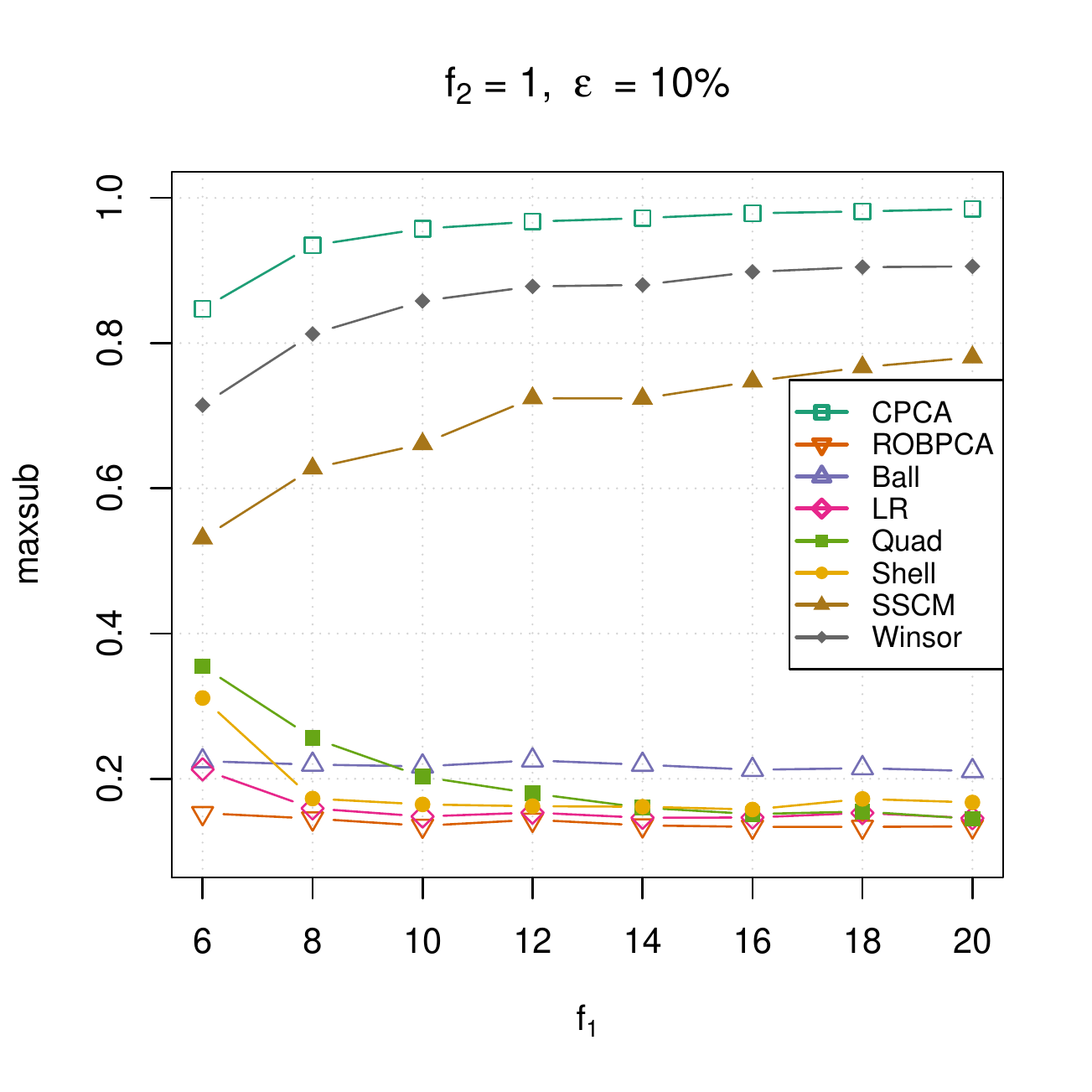} &
\includegraphics[trim={0.2cm 0.7cm 1.05cm 0.7cm},clip,width=0.43\textwidth]{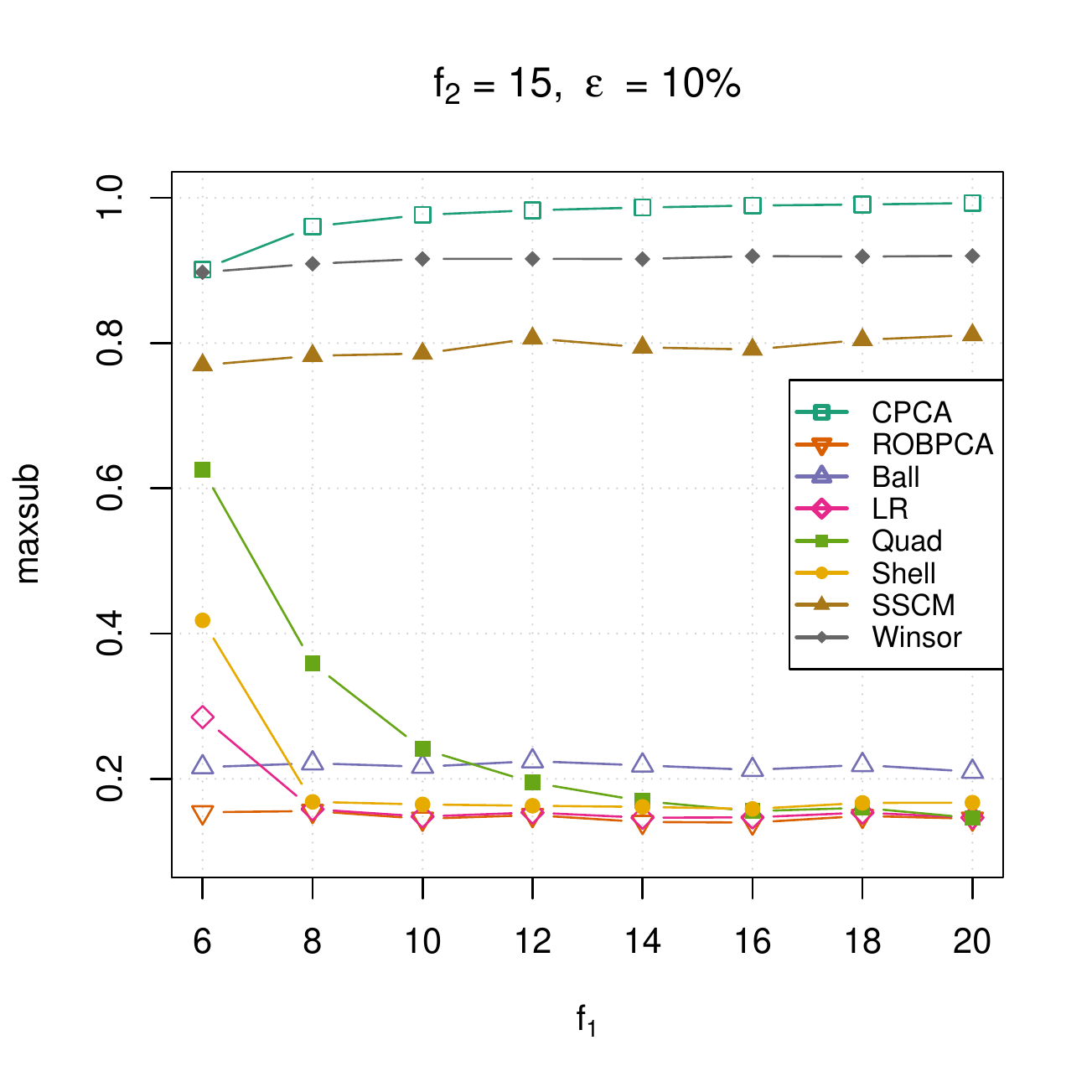} \\
\end{tabular}
\begin{tabular}{cc}
\includegraphics[trim={0.2cm 0.7cm 1.05cm 0.7cm},clip,width=0.43\textwidth]{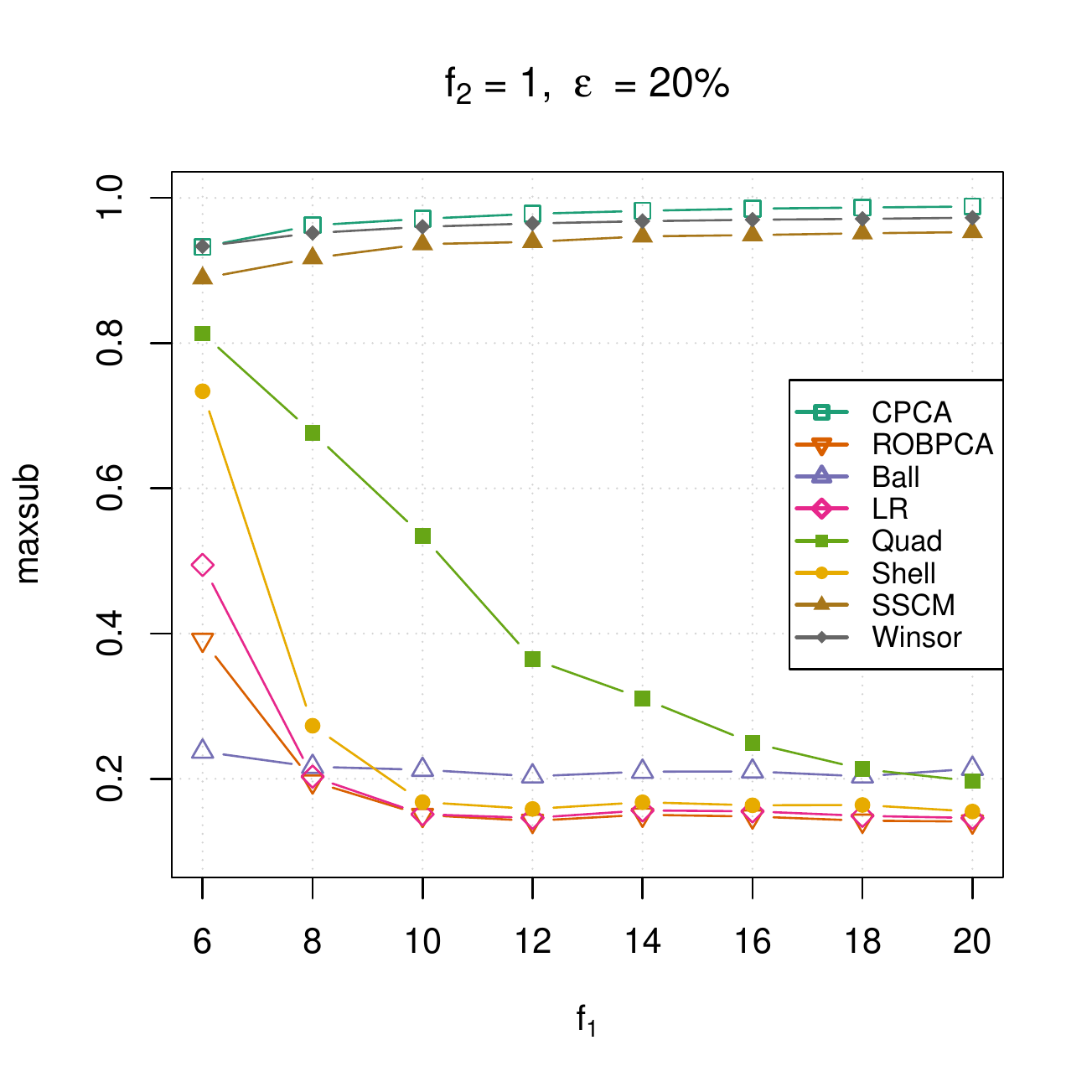} &
\includegraphics[trim={0.2cm 0.7cm 1.05cm 0.7cm},clip,width=0.43\textwidth]{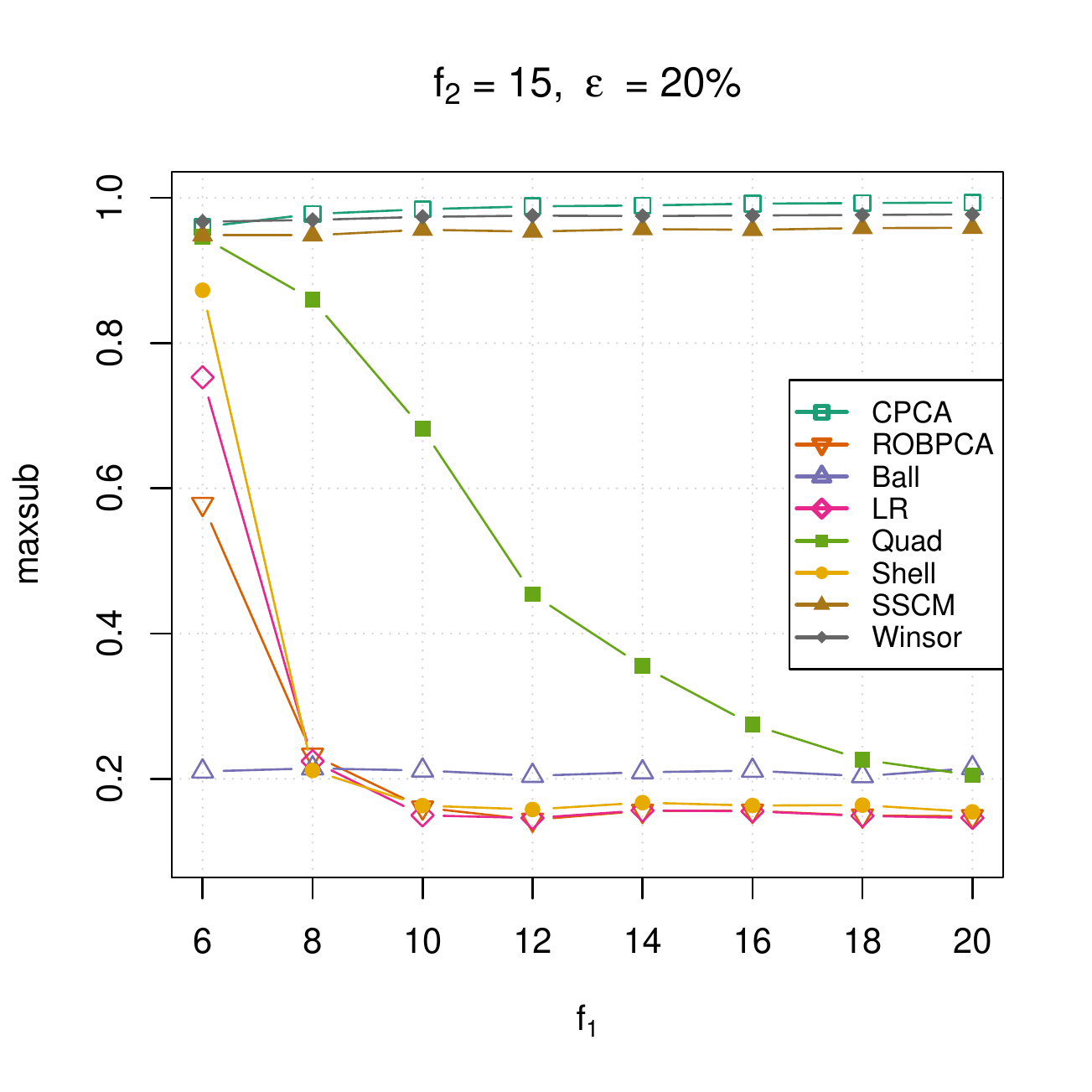} \\
\end{tabular}
\captionsetup{width=0.9\textwidth}
\caption{\centering \small Maxsub measure for low-dimensional, multivariate $t_5$-data}
\label{lowdimstud}
\end{figure*}

\begin{figure*}[h!]
\centering
\begin{tabular}{cc}
\includegraphics[trim={0.2cm 0.7cm 1.05cm 0.7cm},clip,width=0.43\textwidth]{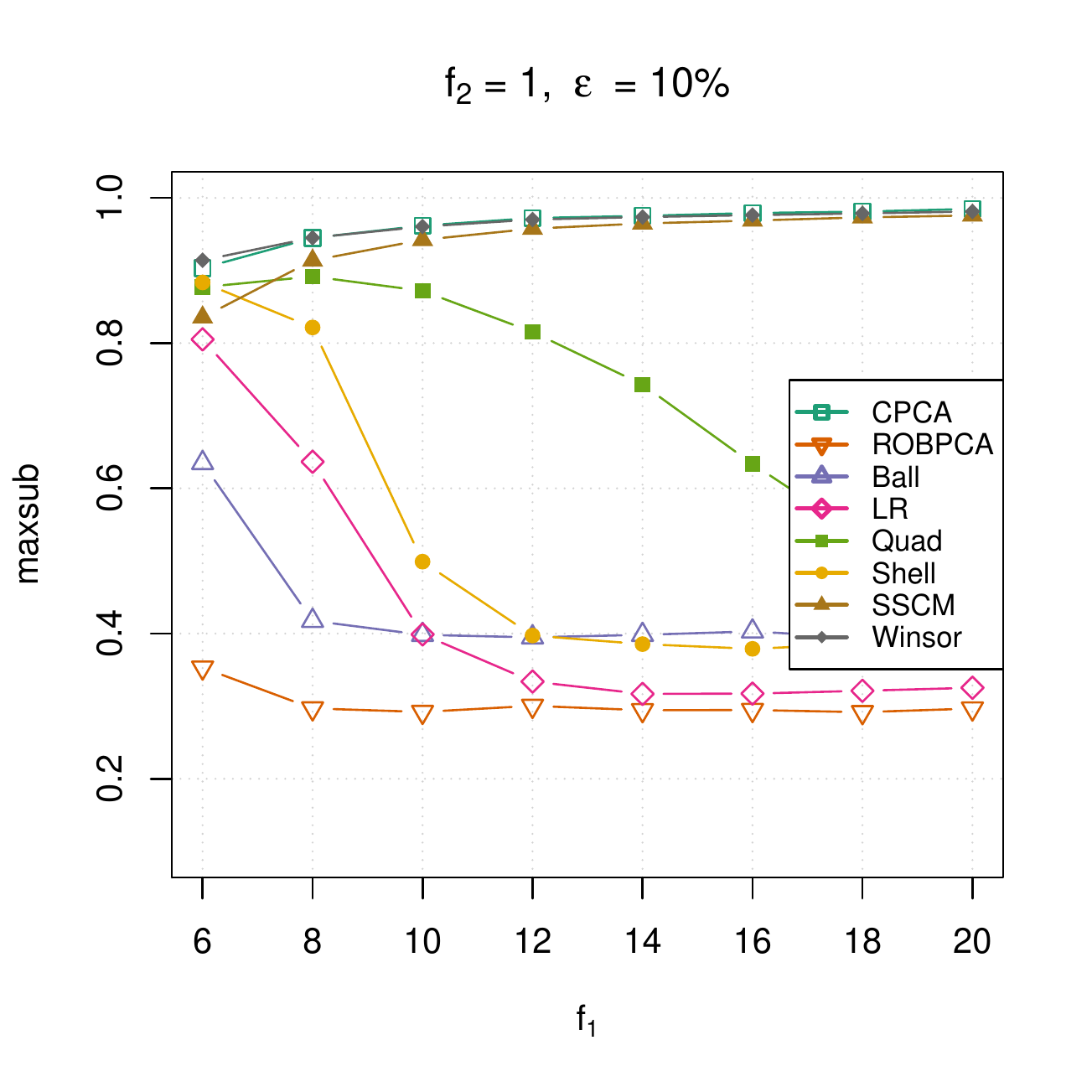} &
\includegraphics[trim={0.2cm 0.7cm 1.05cm 0.7cm},clip,width=0.43\textwidth]{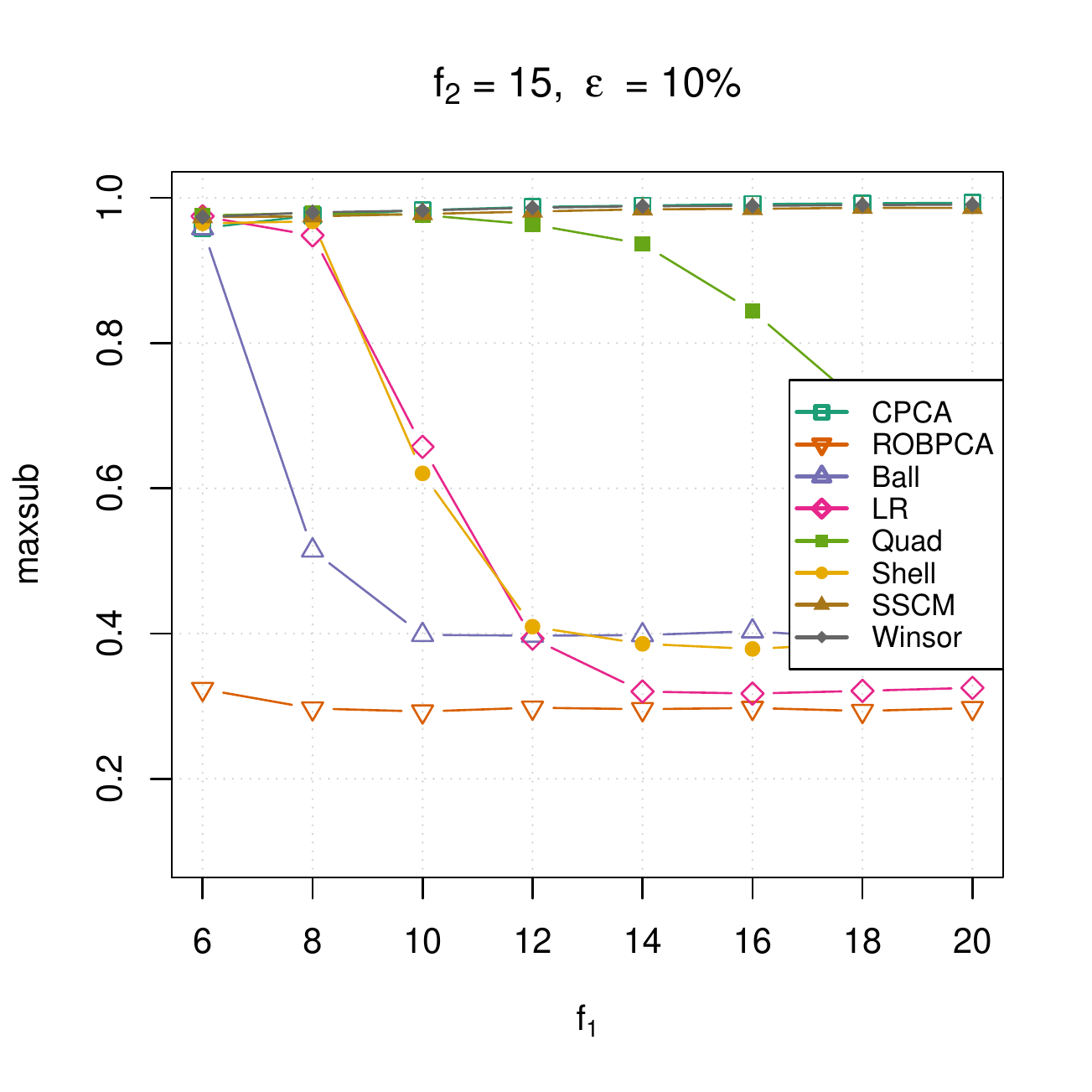} \\
\end{tabular}
\begin{tabular}{cc}
\includegraphics[trim={0.2cm 0.7cm 1.05cm 0.7cm},clip,width=0.43\textwidth]{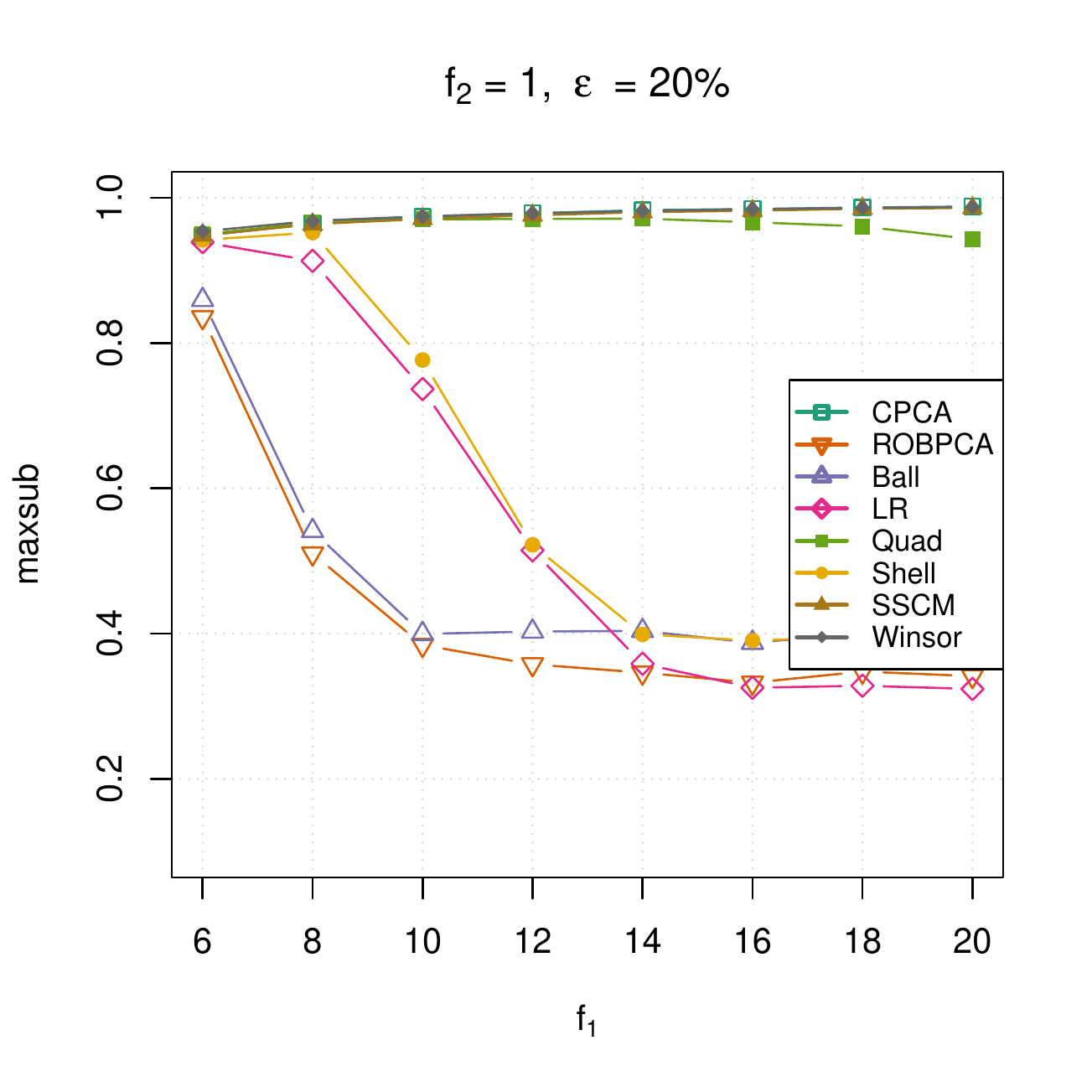} &
\includegraphics[trim={0.2cm 0.7cm 1.05cm 0.7cm},clip,width=0.43\textwidth]{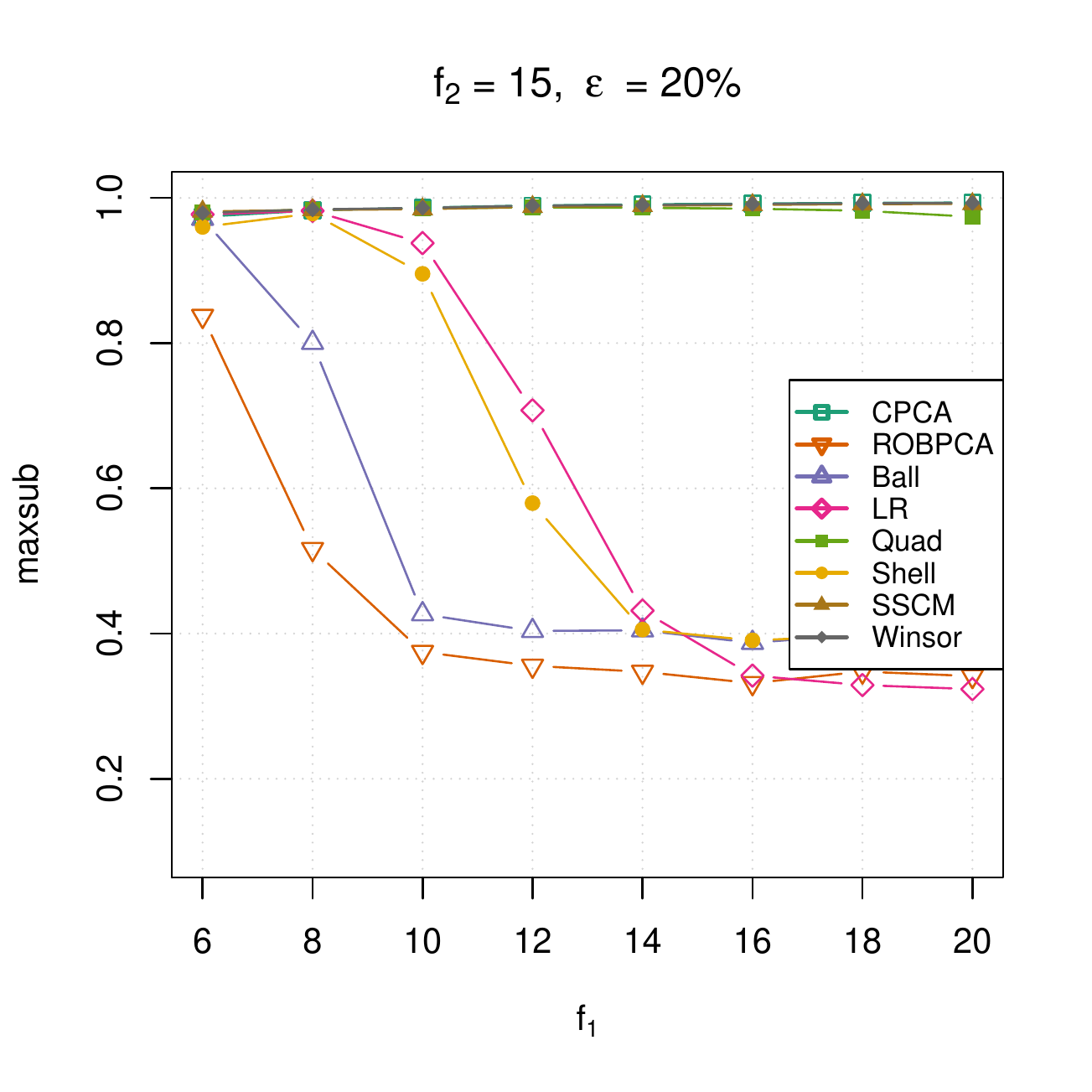} \\
\end{tabular}
\captionsetup{width=0.9\textwidth}
\caption{\centering \small Maxsub measure for high-dimensional, multivariate $t_5$-data}
\label{highdimstud}
\end{figure*}

In a second step, we look at the mean of the maxsub measure for 500 samples of data with different levels of contamination ($\varepsilon = 0.1$ and $\varepsilon=0.2$). The results for multivariate normal and $t_5$-data, for the low- and high-dimensional case, are shown in Figures \ref{lowdimnormal} to \ref{highdimstud}.

\textbf{Low-dimensional, normal data:}\\
For the low-dimensional, multivariate normal data (Figure \ref{lowdimnormal}) we can clearly observe that CPCA, SPCA and Winsor fail. Their maxsub measure is close to one, implying that outliers influenced the estimated PCA subspace to the extent that one of the estimated principal components is orthogonal to $\text{span}\{e_1,\dots,e_k\}$.

The poor result for these radial functions can be explained by their influence functions in Section \ref{subsectionIFs}, where we saw that the influence functions of the eigenvectors of the classical covariance matrix, the SSCM and the GSSCM of Winsor did not redescend to zero in contrast with the others. The maxsub measure of Quad also attains some high values, especially for $\varepsilon=0.2$ and with $f_1$ rather small. When $f_1$ is larger, indicating that the outliers are shifted far enough from the regular observations, Quad performs better. In contrast, we attain very good maxsub measure results for LR, Shell and Ball, whose values are as good as the values for ROBPCA. These three radial functions had influence functions that became zero after a certain value.\\

\textbf{High-dimensional, normal data:}\\
In the high-dimensional normal case (Figure \ref{highdimnormal}), Quad performs worse and fails, just as CPCA, SPCA and Winsor. LR, Shell and Ball are able to distinguish the outliers when $f_1 \geq 10$, in which case their results are comparable to ROBPCA. However, for $f_1 <10$, only ROBPCA achieves low values. 

\textbf{Low-dimensional, $t_5$-data:}\\
For $t_5$-data in the low-dimensional case (Figure \ref{lowdimstud}), the results are similar to low-dimensional normal data. The only difference is that for $\varepsilon = 0.2$ and $f_1 = 6$, all methods perform poorly, except for Ball who still attains a low value for the maxsub measure. So in this situation, when the outliers are very close to the regular observations, Ball is the only one who can distinguish them.

\textbf{High-dimensional, $t_5$-data:}\\
Lastly, for the high-dimensional multivariate $t_5$-data (Figure \ref{highdimstud}), we observe that once again CPCA, SPCA, Winsor and Quad fail. However, Ball and ROBPCA perform well once $f_1$ is greater than or equal to 10. LR and Shell attain higher values, they only achieve good results when $f_1$ exceeds 12 or 14. So for the high-dimensional $t_5$ case, the performance of the methods depends on whether the outliers are separated enough from the regular observations.\\
In summary, we can conclude that for the multivariate normal data GSPCA with radial functions LR, Shell and Ball performs comparable to ROBPCA. For the $t_5$-data, Ball achieves the same results as ROBPCA. LR and Shell perform somewhat worse, but can still be considered as good alternatives. The fact that GSPCA with the Ball radial function achieves such great results is somewhat surprising, since from its definition in Equation \eqref{defBall} it follows that half of the data points get weight zero and therefore do not contribute anymore. However, this cutoff makes Ball very robust, as can be seen by the influence functions of its eigenvectors in Section \ref{subsectionIFs}, Ball's influence function is the first one that becomes zero among all radial functions.

\subsection{Computational time}
One of the biggest advantages of our new PCA method is its computational speed. It is considered to be as fast as classical PCA, while ROBPCA is a significantly slower method. To illustrate this, we will compare the computational time of CPCA and ROBPCA with GSPCA using the radial functions Ball and LR.

We will use the same setting as previously and sample multivariate normal data, from the low- and high-dimensional case, with the mean vector and covariance matrix as specified in Section \ref{subsectionSimSetting}. We set $\varepsilon = 0$ (no contamination) and measure the total computational time for 100 runs. The computations were done on a 2.50 GHz core i5 processor ($7^{th}$ gen) and results are shown in Table \ref{CompTimeLowDimNorm} and \ref{CompTimeHighDimNorm}.

From both tables it is clear that GSPCA outperforms ROBPCA when we consider the computational time. The time required for GSPCA is comparable to that of classical PCA for both the low- and high-dimensional situation. However, the time needed for ROBPCA increases significantly when the sample size $n$ becomes larger. We conclude that a lot of time can be saved by using GSPCA instead of ROBPCA.

\begin{table*}[h!]
\centering
\begin{tabular}{|c | c c c c c c c c c c| }
\hline
 $n$ & 50 & 100 & 150 & 200 & 250 & 300 & 350 & 400 & 450 & 500\\ 
\hline
CPCA &0.38	&0.25	&0.36	&0.32	&0.33	&0.35	&0.34	&0.37	&0.47	&0.41
\\
ROBPCA &1.66	&4.74	&13.18	&28.61	&54.52	&92.37	&146.24 & & & 			
\\
GSPCA Ball &0.23	&0.23	&0.38	&0.37	&0.41	&0.45	&0.52	&0.58	&0.62	&0.71
\\
GSPCA LR &0.25	&0.22	&0.5	&0.47	&0.46	&0.6	&0.55	&0.58	&0.61	&0.73
\\
\hline
\end{tabular}
\caption{\centering \small Total computational time for 100 runs (in seconds) for low-dimensional, multivariate normal data ($p=4$, $k=3$) for various values of $n$}
\label{CompTimeLowDimNorm}
\end{table*}

\begin{table*}[h!]
\centering
\begin{tabular}{|c | c c c c c c c c c c| }
\hline
 $n$ & 50 & 100 & 150 & 200 & 250 & 300 & 350 & 400 & 450 & 500\\ 
\hline
CPCA & 0.89	&1.34	&1.88	&2.28	&2.86	&2.93	&3.58	&3.88	&4.18	&4.6
\\
ROBPCA &2.78	&9.34	&23.78	&50.18	&95.04	&154.86	&240.79	 & & &		
\\
GSPCA Ball &1.14	&1.4	&1.58	&2	&2.19	&2.67	&3.11	&3.45	&4.11	&4.35
\\
GSPCA LR &1.11	&1.38	&1.65	&2	&2.3	&2.56	&3.41	&3.66	&3.76	&4.56
\\
\hline
\end{tabular}
\caption{\centering \small Total computational time for 100 runs (in seconds) for high-dimensional, multivariate normal data ($p=100$, $k=5$) for various values of $n$}
\label{CompTimeHighDimNorm}
\end{table*}

\section{Data Examples} \label{sec:dataexamples}
In this section we illustrate the performance of GSPCA on two real data sets: the Top Gear car data set and a surveillance video of a beach, previously studied in a different context in \cite{DirOut}.

\subsection{Top Gear data}
\begin{figure}[h!]
    \centering
    \subfloat{{\includegraphics[trim={0 0.5cm 0 0.8cm},clip,width=6.1cm]{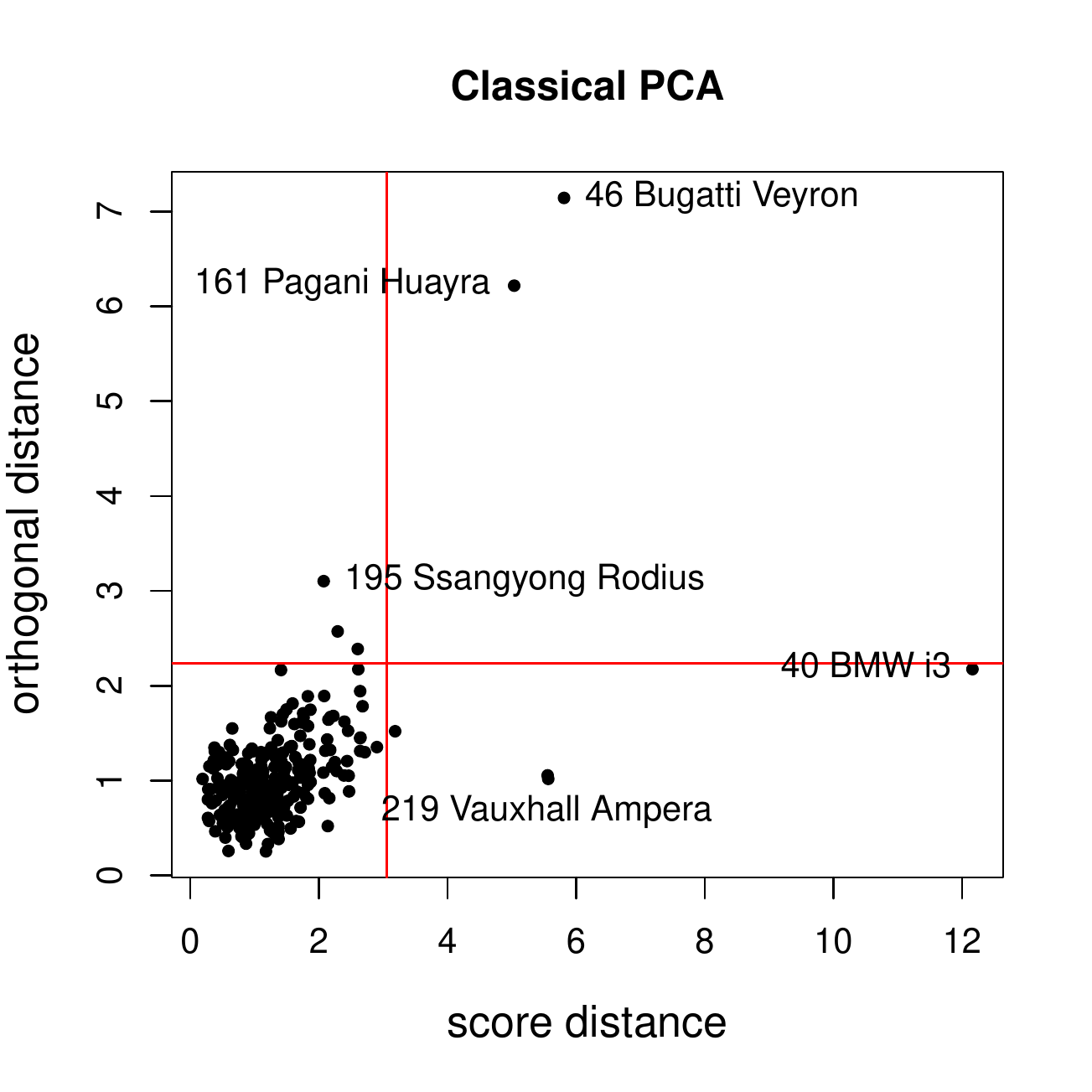} }}
    \vspace{0.5cm}
    \subfloat{{\includegraphics[trim={0 0.5cm 0 0.8cm},clip,width=6.1cm]{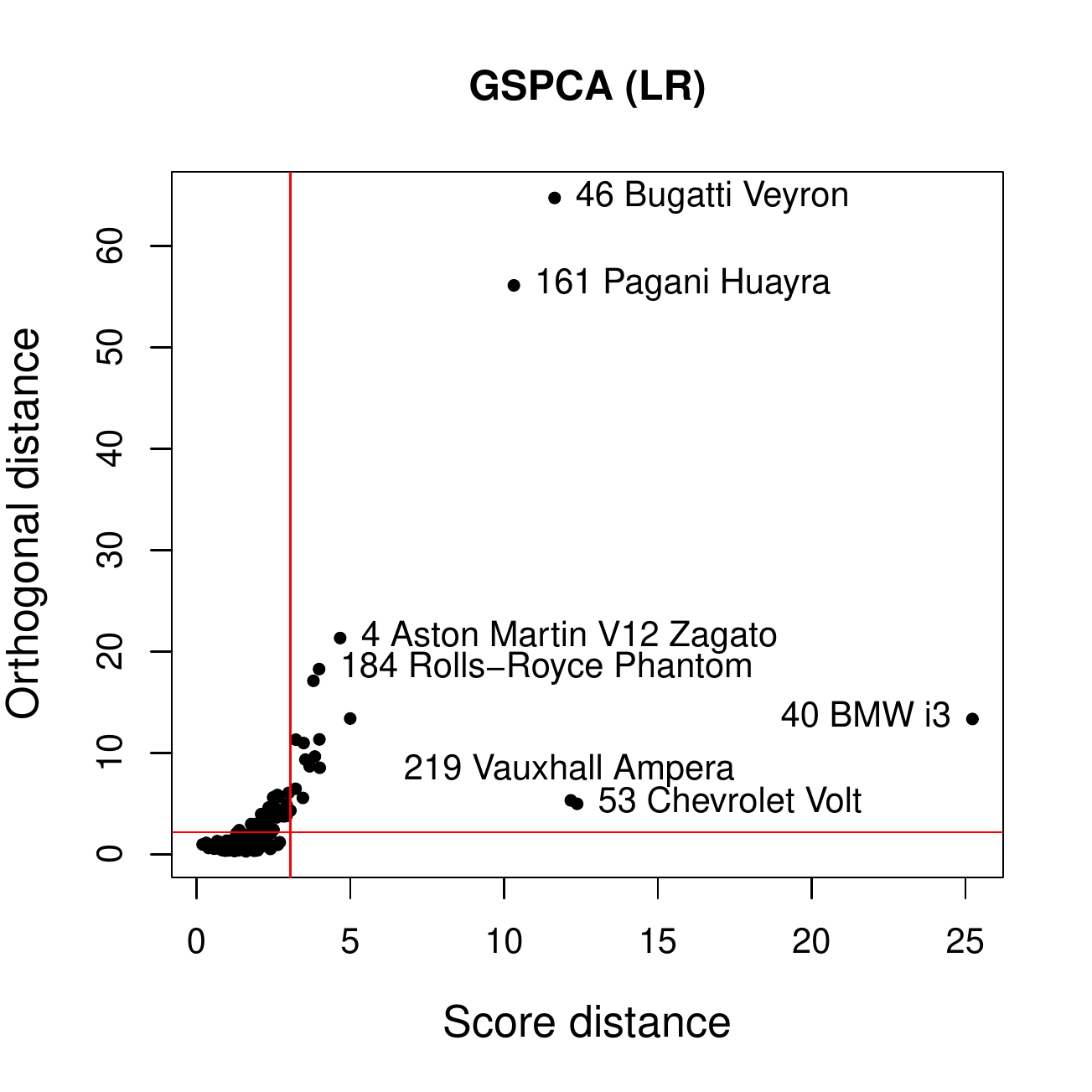} }}
    \captionsetup{width=0.5\textwidth}
    \caption{\centering \small Diagnostic plots for CPCA and GSPCA on the Top Gear data}
    \label{fig:topgear}
\end{figure}

The first data set comprises the Top Gear car data set from the R package RobustHD, which has been studied frequently in the context of robustness and PCA. We consider the numeric variables Price, Displacement, BHP, Torque, Acceleration, TopSpeed, MPG, Weight, Length, Width and Height and remove incomplete observations, retaining 245 observations. Next we scale the data and perform CPCA and GSPCA combined with the LR radial function, retaining 3 principal components to explain at least 85\% of the total variance.
To illustrate the robustness of GSPCA, we use the diagnostic plot defined by \cite{ROBPCA} to classify the flagged outliers, resulting in Figure \ref{fig:topgear}.

We observe that GSPCA identifies many bad leverage points, outlying in score distance and orthogonal distance,  while CPCA only flags 2 of them as bad leverage points. This suggests that CPCA was heavily influenced by the bad leverage points identified by GSPCA. We also ran the state-of-the-art method ROBPCA on the data, its results were very similar to GSPCA. 

\subsection{Video data}

\begin{figure*}[h!]
\centering
\begin{tabular}{cccc}
\includegraphics[trim={2.8cm 1cm 2.8cm 1cm},clip,width = 0.225\textwidth]{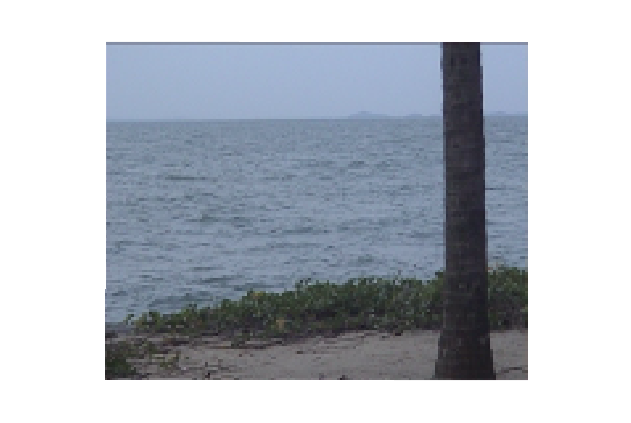}
&
\includegraphics[trim={2.8cm 1cm 2.8cm 1cm},clip,width = 0.225\textwidth]{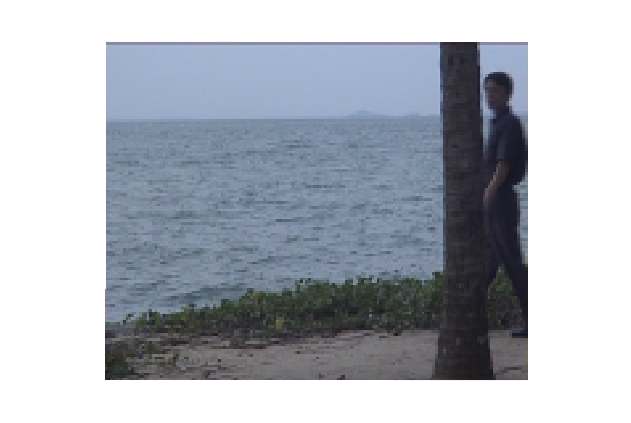}
&
\includegraphics[trim={2.8cm 1cm 2.8cm 1cm},clip,width = 0.225\textwidth]{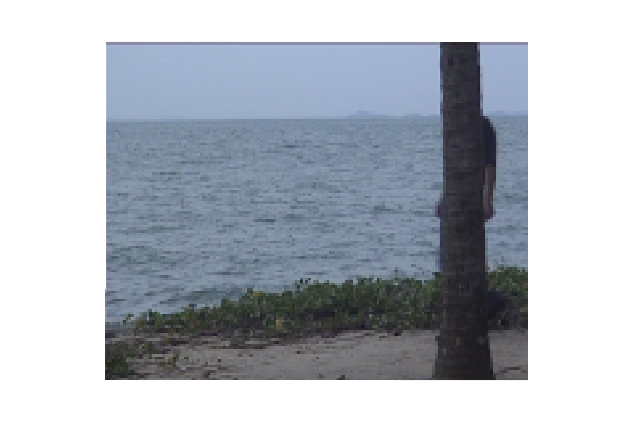}
&
\includegraphics[trim={2.8cm 1cm 2.8cm 1cm},clip,width = 0.225\textwidth]{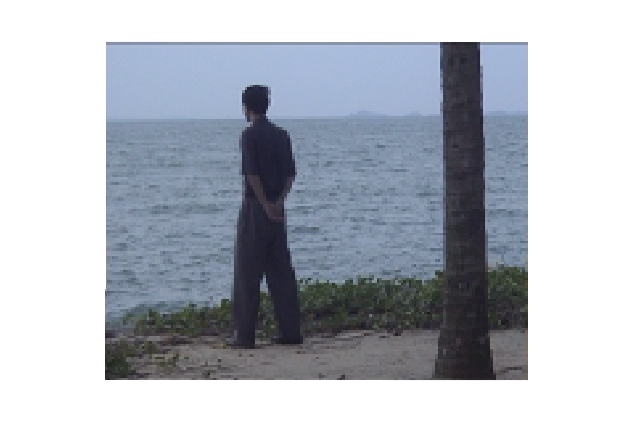}
\end{tabular}
\captionsetup{width=0.9\textwidth}
\caption{\centering \small Video frames 108, 487, 491 and 564}
\label{fig:video}
\end{figure*}

The second data set is comprised of a surveillance video of a beach, consisting of 633 frames of 160×128 pixels in the RGB color model. The data was previously studied in the context of outlier detection for functional data analysis \citep{DirOut} and originates from \cite{beach}. In the video we see a beach scenery, where in frame 483 a man comes into view, in frame 489 he disappears behind a tree and in frames 493 to 633 he remains in view, see for example Figure \ref{fig:video}. Hence from a robustness perspective, we consider frames 489 to 633 as outliers, while frames 1 to 488 are considered normal observations.

\begin{figure*}[h!]
\centering
\begin{tabular}{ccc}
\includegraphics[trim={0cm 0.5cm 1.05cm 0.7cm},clip,width=0.31\textwidth]{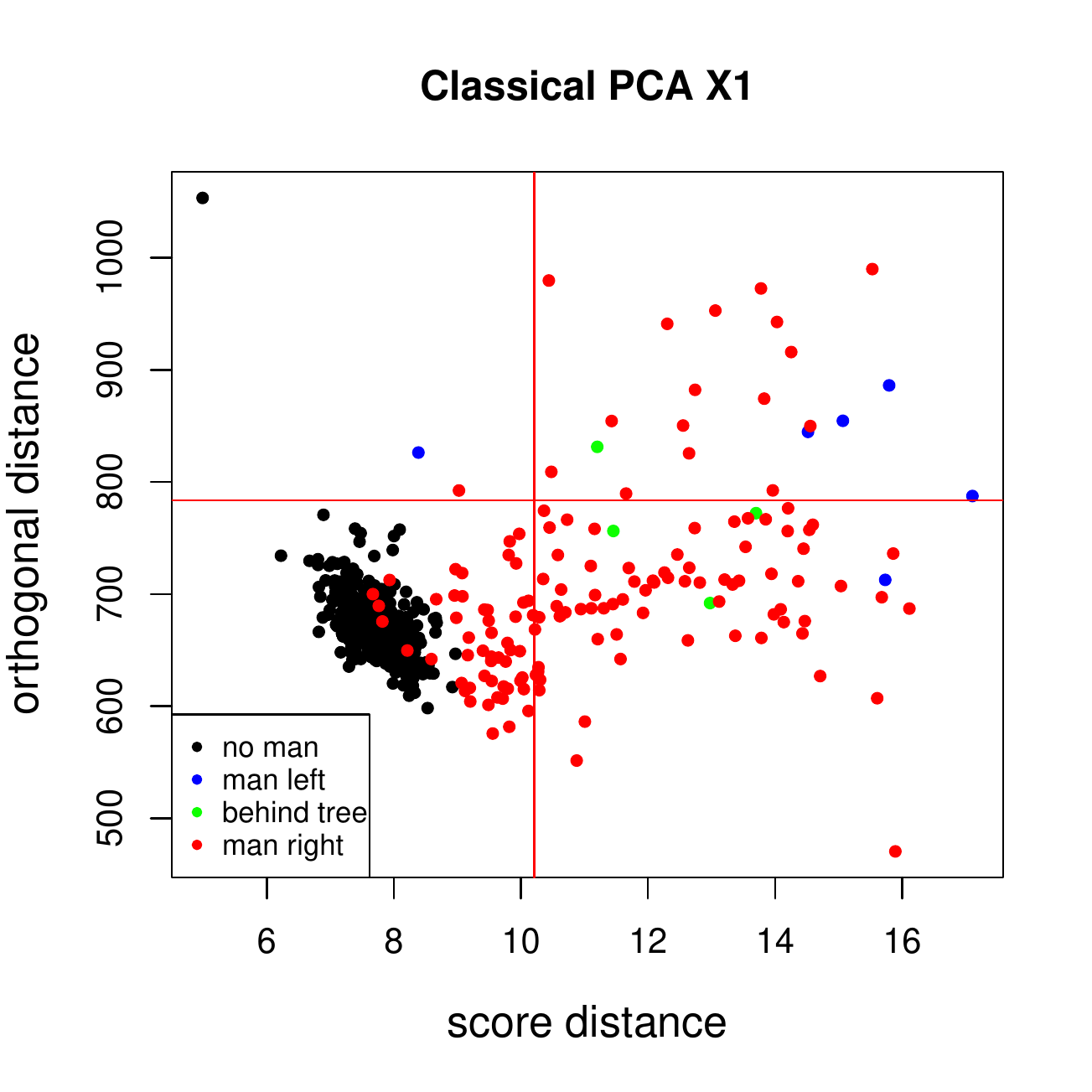} &
\includegraphics[trim={0cm 0.5cm 1.05cm 0.7cm},clip,width=0.31\textwidth]{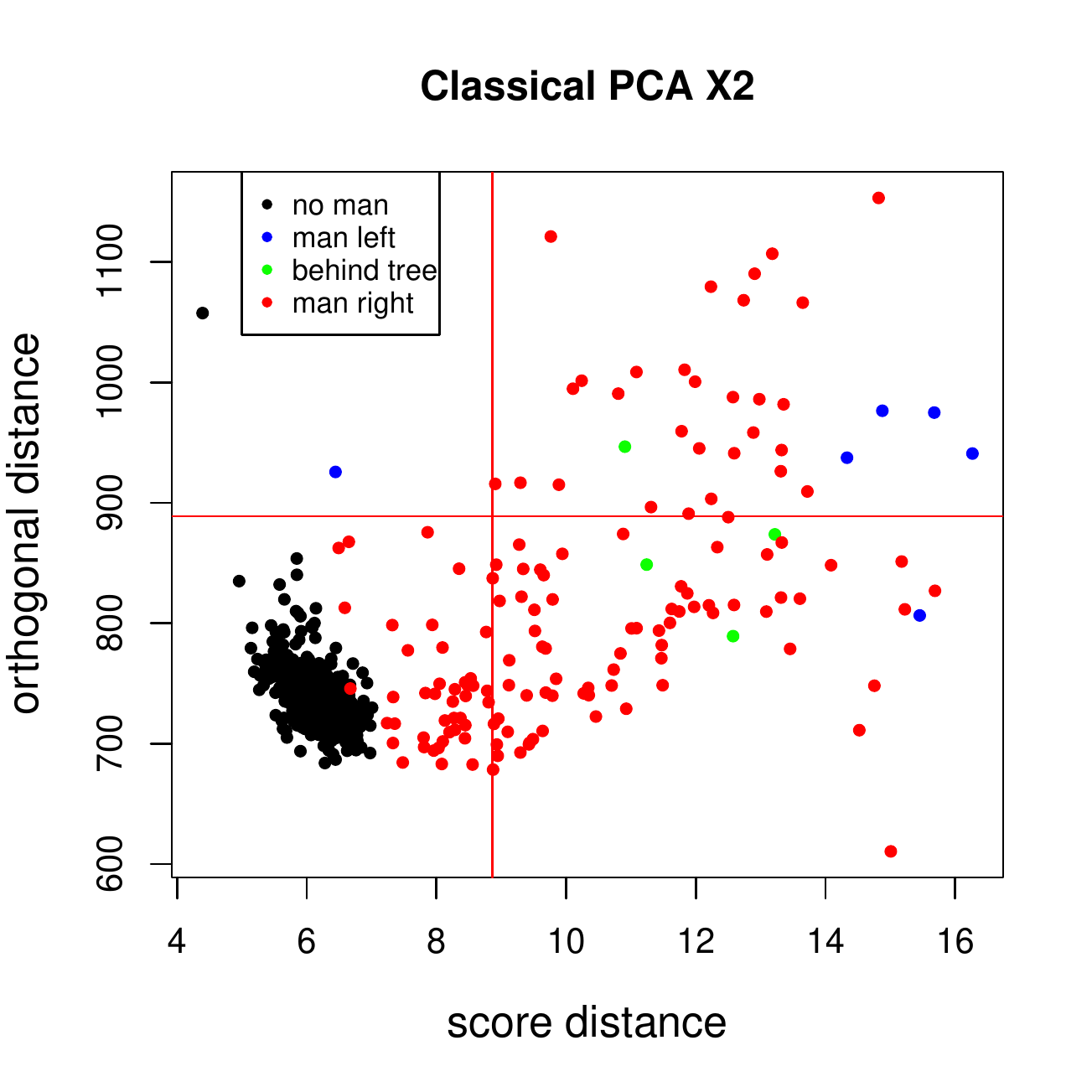} &
\includegraphics[trim={0cm 0.5cm 1.05cm 0.7cm},clip,width=0.31\textwidth]{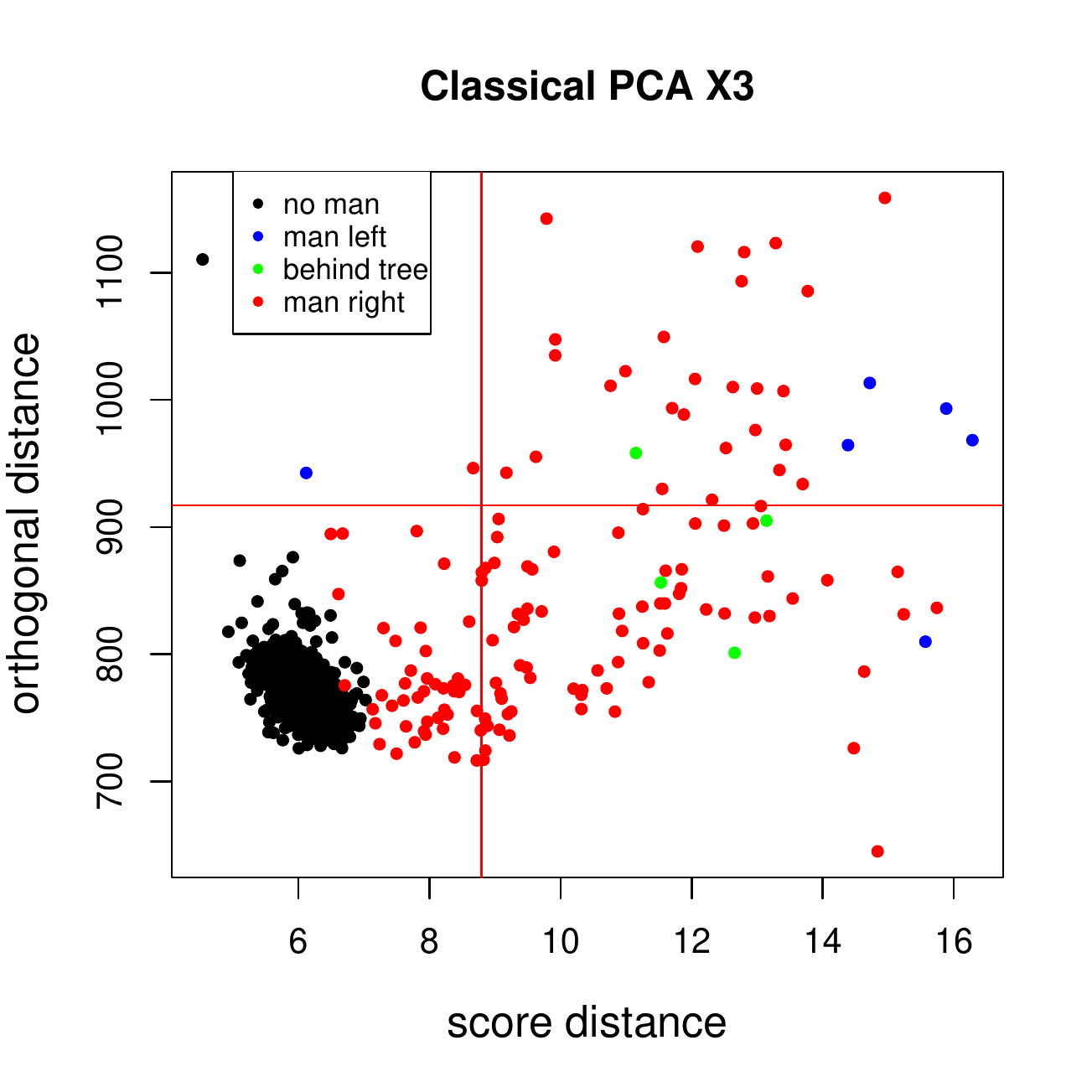}\\
\end{tabular}
\begin{tabular}{ccc}
\includegraphics[trim={0cm 0.5cm 1.05cm 0.7cm},clip,width=0.31\textwidth]{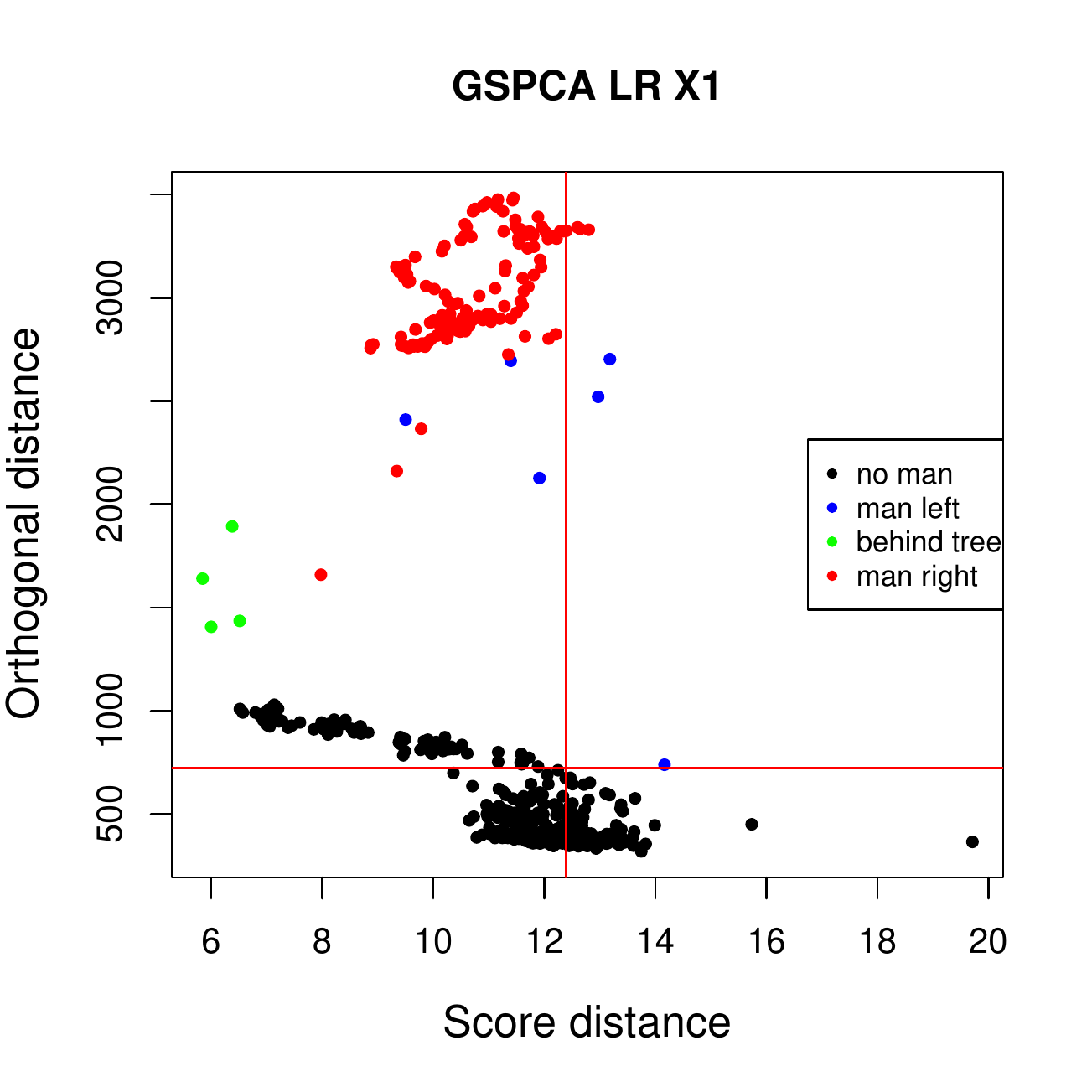} &
\includegraphics[trim={0cm 0.5cm 1.05cm 0.7cm},clip,width=0.31\textwidth]{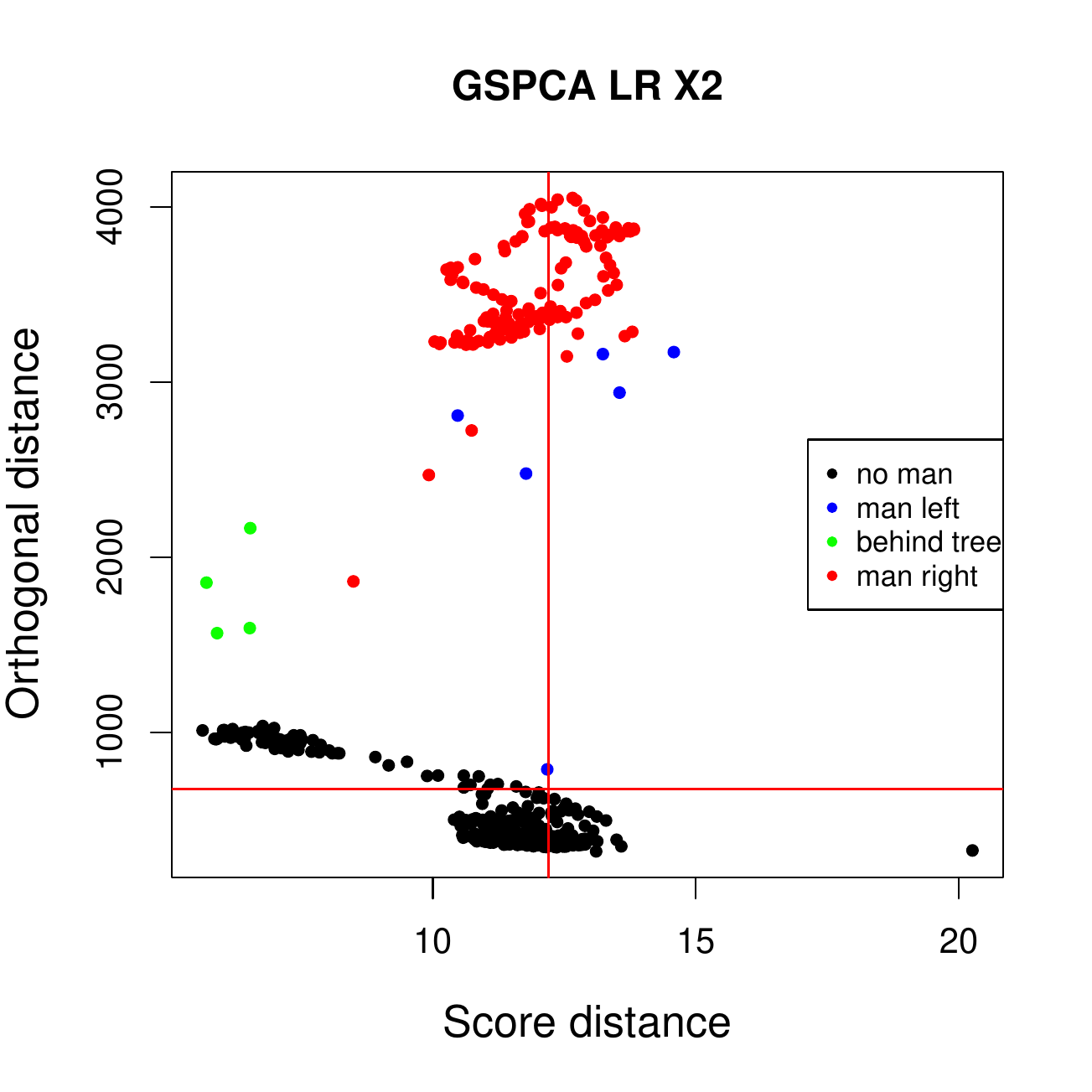} &
\includegraphics[trim={0cm 0.5cm 1.05cm 0.7cm},clip,width=0.31\textwidth]{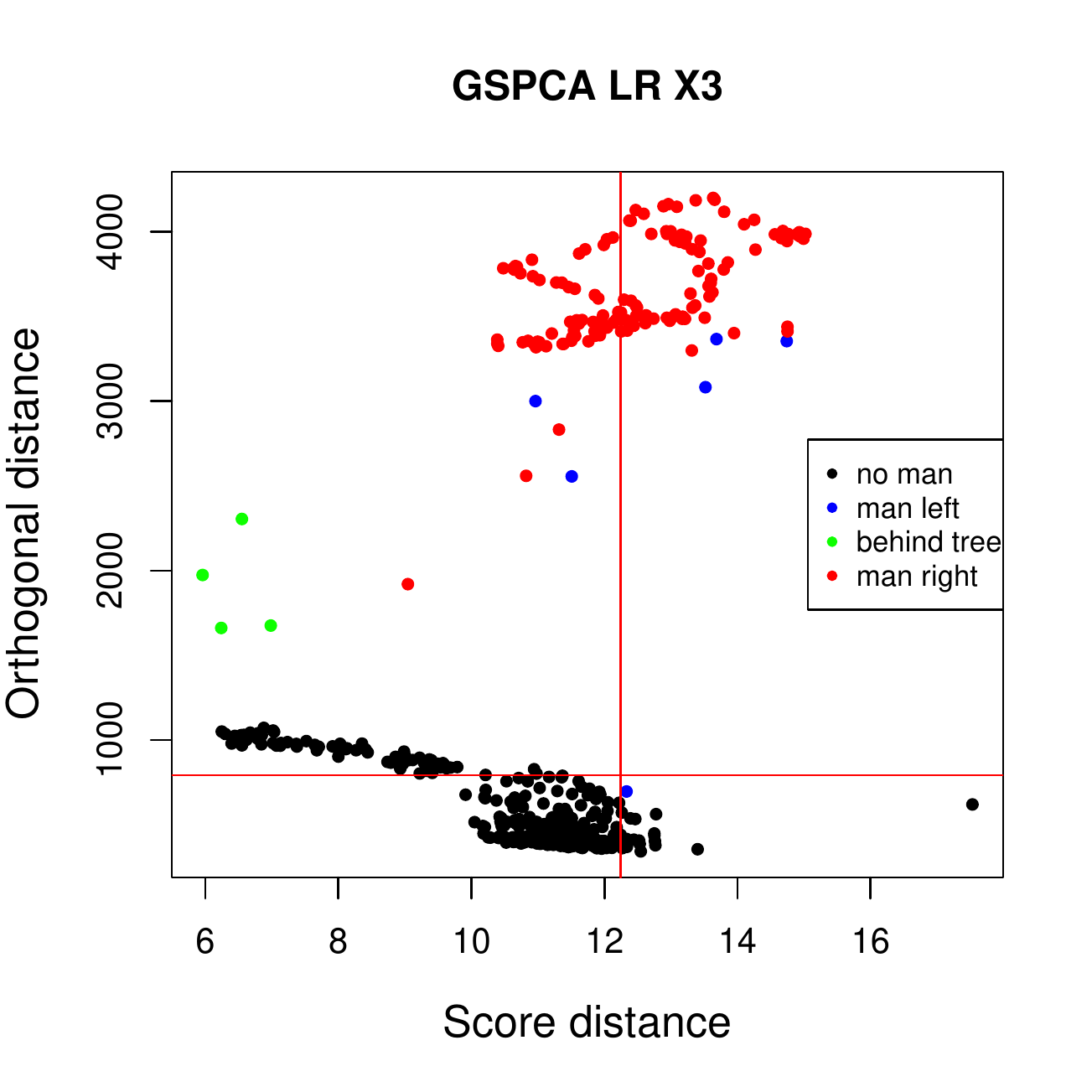}\\
\end{tabular}
\captionsetup{width=0.9\textwidth}
\caption{\centering \small Diagnostic plots for CPCA and GSPCA on the video data in RGB color}
\label{fig:videodiagplot}
\end{figure*}

To test GSPCA, we run our new method using the LR radial function and the classical method on the high dimensional, video data. As the video is color coded in the RGB model, we have 3 data sets ($\mathbf{X1}$, $\mathbf{X2}$ and $\mathbf{X3}$) of 633 observations with 20480 variables. We select the number of principal components to explain at least 85\% of the variance. The resulting diagnostic plots of the principal component analyses are shown in Figure \ref{fig:videodiagplot} for the 3 colors.

Figure \ref{fig:videodiagplot} shows that GSPCA distinguishes the normal observations (no man, black points) from the outliers (man in view, colored points). The method flags all outliers as orthogonal outliers or bad leverage points. It also separates the frames where the man is behind the tree (green). For CPCA, the distinction between the outliers and the normal observations is less clear. Many of the frames containing the man are flagged as good leverage points, indicating that CPCA was strongly influenced by these points. CPCA also doesn't separate the frames where the man is behind the tree.

Also worth noting, GSPCA distinguishes one point with a large score distance of 20. This observation corresponds to the first frame which is a bit lighter compared to the others, probably due to starting the video recording. CPCA also separates this point, but as a point with a large orthogonal distance around 1000.

\begin{figure}[h!]
\centering
\begin{tabular}{l}

\includegraphics[trim={0.4cm 6cm 0.3cm 3.9cm},clip,width=7.5cm]{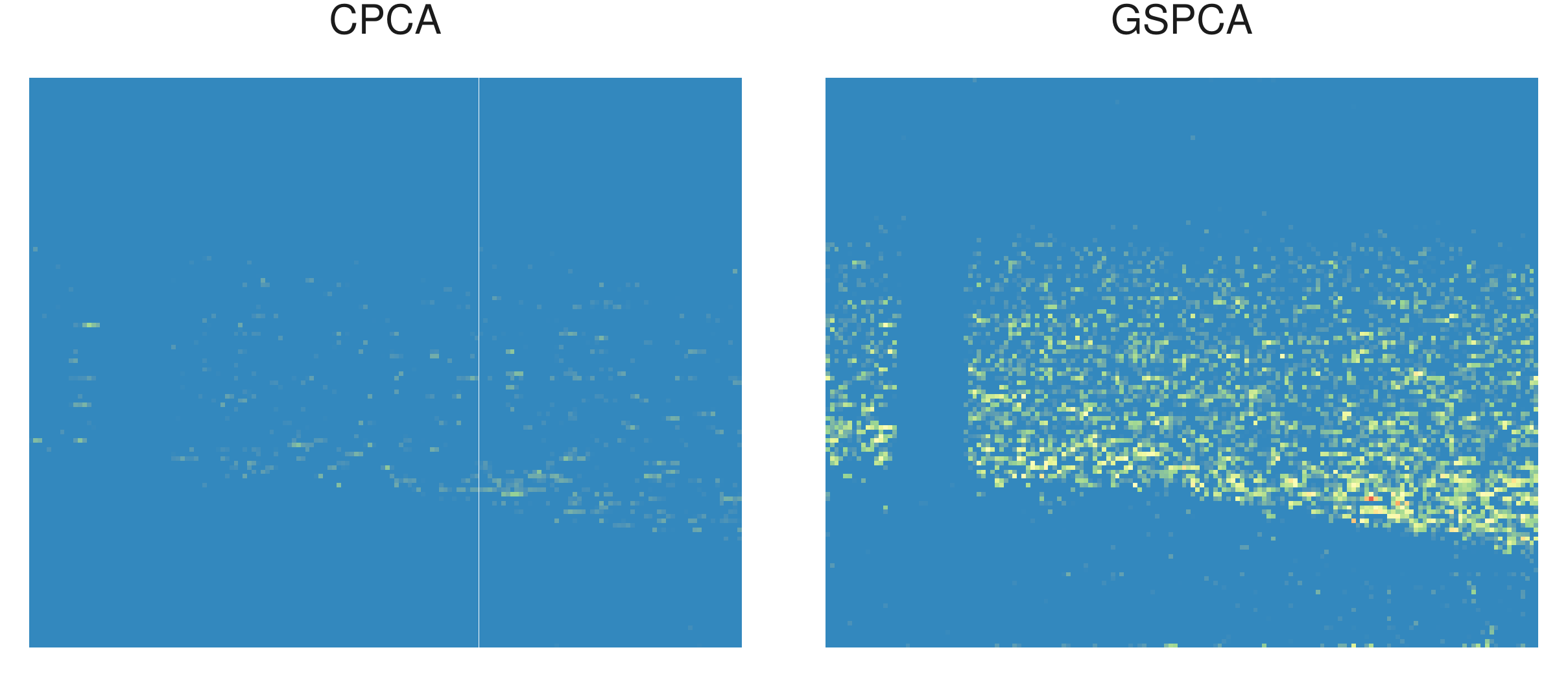} \\
\includegraphics[trim={0.4cm 6cm 0.3cm 3.9cm},clip,width=7.5cm]{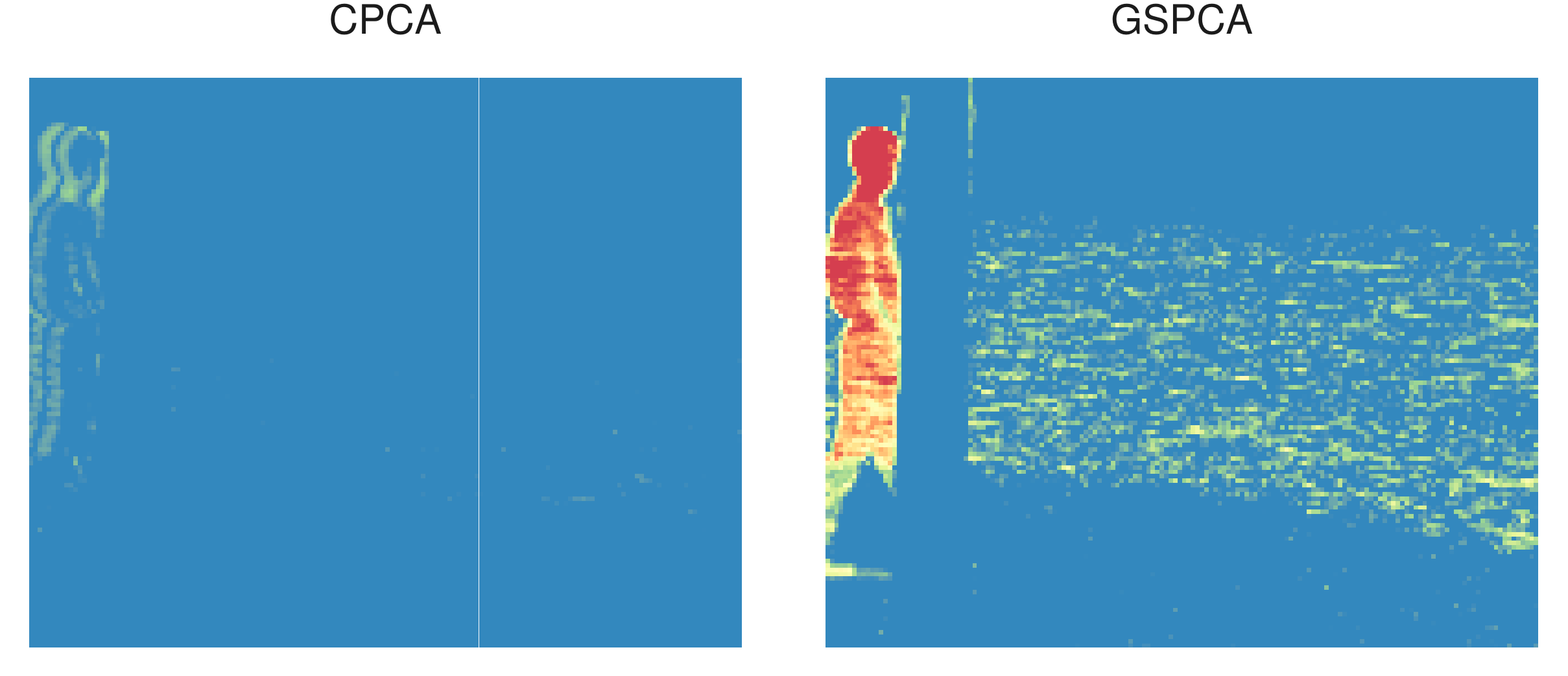} \\ 
\includegraphics[trim={0.4cm 6cm 0.3cm 3.9cm},clip,width=7.5cm]{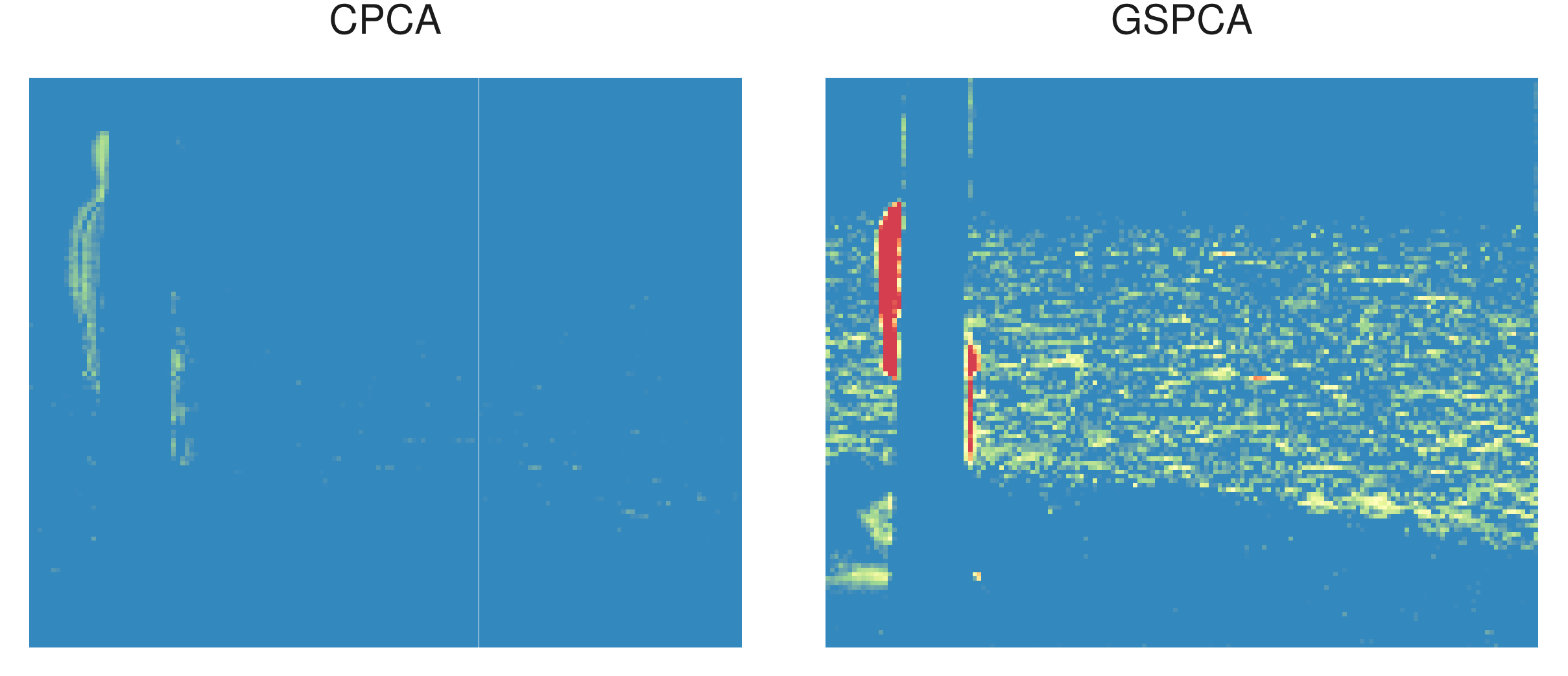} \\
\includegraphics[trim={0.4cm 3cm 0.3cm 3.9cm},clip,width=7.5cm]{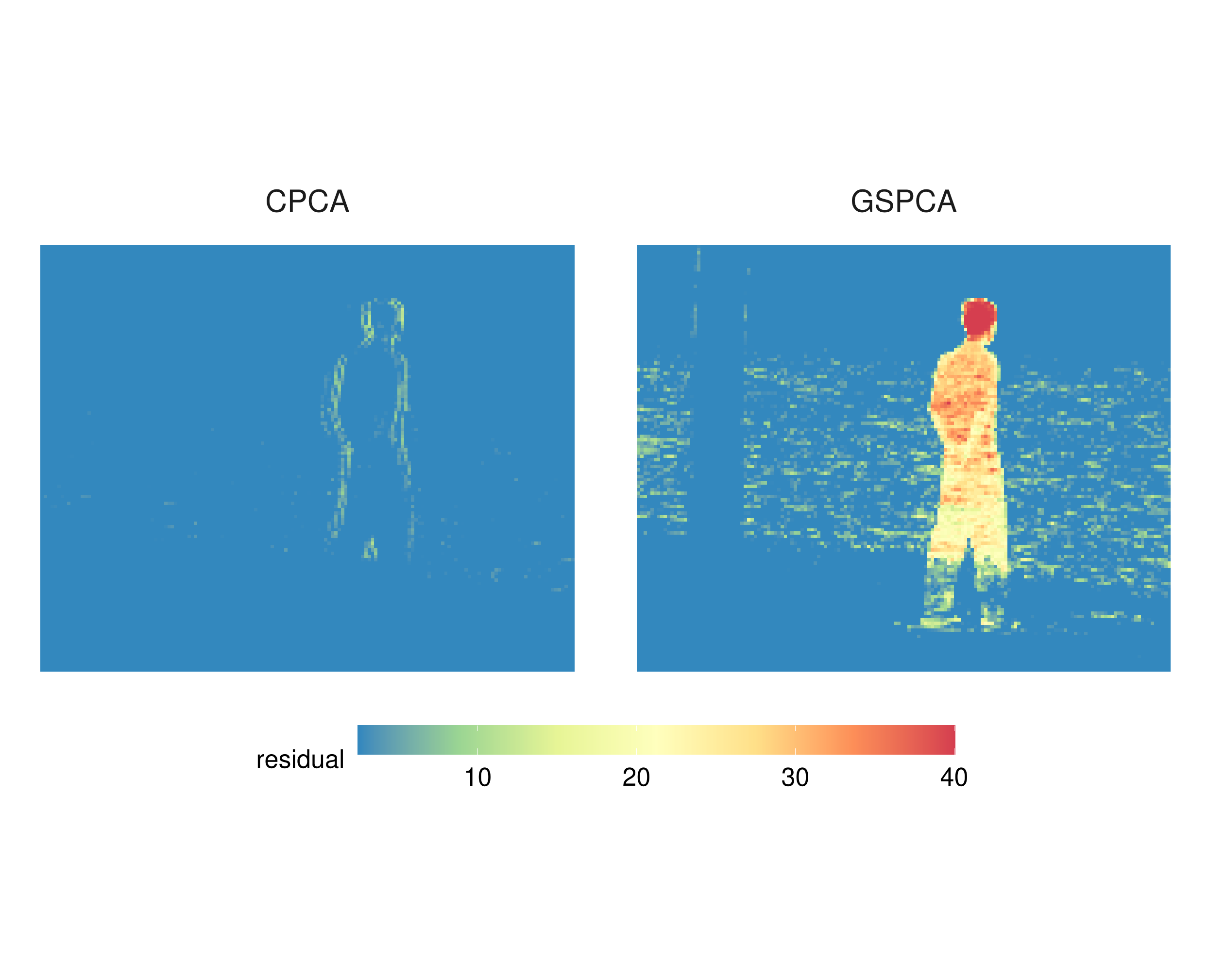} \\
\end{tabular}
\captionsetup{width=0.45\textwidth}
\caption{\centering \small Scaled residuals  of data set $\mathbf{X2}$ for frames 108, 487, 491 and 564}
\label{fig:residualframes}
\end{figure}

To illustrate the effect of the outliers on the analysis, we calculate the predicted values $\mathbf{\hat{X}}_{(k)}$ of the principal component analysis:
\begin{equation*}
    \mathbf{\hat{X}}_{(k)} = (\mathbf{X}_{n,p} - 1_n T(\mathbf{X})^T) V_{p,k} V_{p,k}^T + 1_n T(\mathbf{X})^T 
\end{equation*}
with $k$ the number of principal components to explain 85\% of the variance. From these, we compute the residuals from CPCA and GSPCA to detect outlying pixels and to study the difference between the original frames and the predictions:
\begin{equation*}
    r = \mathbf{X} - \mathbf{\hat{X}}_{(k)}
\end{equation*}
We also standardize the residuals per frame using the mean and standard deviation for CPCA, and the median and MAD for GSPCA. In Figure \ref{fig:residualframes} we show them for frames 108 (no man), 487 (man left), 491 (man behind tree) and 564 (man right). Here all scaled residuals below outlier cutoff 2.5 are given the same color.

From the residual plot we observe that CPCA only has few mild residuals, consisting mostly of noise around the man. In GSPCA, on the contrary, the entire man has very high residuals and is clearly detected by the analysis. The waves of the sea also have significant residuals as they move throughout the video.

\section{Conclusion}

In this paper we developed a new robust version of PCA by calculating the loading vectors on the generalized spatial sign covariance matrix (GSSCM) instead of on the classical covariance matrix. The GSSCM was subsequently combined with different radial functions, five were discussed in this work: Winsor, Quad, LR, Ball and Shell. This resulted in a new method we named \textit{generalized spherical principal component analysis} (GSPCA).

To evaluate GSPCA, three theoretical properties were studied. First, we calculated a breakdown value of $\left \lfloor{(n+1)/2}\right \rfloor /n$, meaning that GSPCA can resist upto 50\% contamination in a data set. Second, analytical forms for the influence functions of the eigenvectors of the GSSCM were derived and studied for the bivariate normal distribution. We demonstrated that all radial functions except Winsor, had bounded influence functions which redescended to zero, implying robustness for GSPCA. Last, the asymptotic relative efficiencies at bivariate normal distribution were computed, it was shown that GSPCA combined with Winsor is more efficient than SPCA. Ball, LR and Shell on the contrary had lower efficiencies.

To further examine GSPCA, a simulation study was conducted where the maxsub measure was analyzed for the different radial functions. First we studied data sets without contamination to examine efficiency. For the normal distribution and the Student's $t$-distribution with five degrees of freedom, classical PCA and Winsor performed the best. However, for distributions with heavier tails such as the Student's $t$-distribution with one or two degrees of freedom, classical PCA failed. GSPCA with the Winsor radial function, on the contrary, kept achieving great results.

Subsequently, contamination was introduced to the data sets to investigate robustness. In the presence of this contamination, Winsor and Quad performed bad. The other three radial functions LR, Shell and Ball achieved good results. Their outcomes were comparable to those of ROBPCA, the state-of-the-art method. However, GSPCA requires significantly less computational time.

Thereafter, our method was tested on real life data. The first data set was the well-studied, moderately sized Top Gear data set. The second data set consisted of high-dimensional video data. For both examples, GSPCA demonstrated favorable robustness properties, whereas CPCA was clearly influenced by outliers.

We also proposed an adjustment to the GSSCM to achieve Fisher consistency for the eigenvalues and the covariance estimate, this was done using projections.

In summary, we conclude that the newly developed GSPCA method can be considered a very interesting alternative for robust PCA based on the findings stated above. It is however important to combine GSPCA with the appropriate radial function. When aiming for efficiency, using GSPCA with the Winsor radial function can be a good alternative to classical PCA. When robustness is the main priority, GSPCA combined with the LR or Ball radial function are the most reliable.


\section*{Acknowledgments}
First author is supported by Fonds Wetenschappelijk onderzoek - Vlaanderen (FWO) as a PhD fellow Fundamental Research (PhD fellowship 11K5523N).

\bibliographystyle{abbrvnat}
\setcitestyle{authoryear,open={(},close={)}} 
\bibliography{arxiv_10032023}

\clearpage
\appendix

\section{ }
\subsection{Proof of Theorem \ref{thm:breakdown}}\label{app:thm1}

We follow the train of thought of the proof in \cite{GSSCM}:

\begin{proof}
\noindent
\underline{Part 1}: $\varepsilon \geq \left \lfloor{(n+1)/2}\right \rfloor /n$:

Given an $m < \left \lfloor{(n+1)/2}\right \rfloor$, change $m$ of the observations in $\mathbf{X}$ obtaining the contaminated data set $\mathbf{X_m^*}$. As location estimates we have $T(\mathbf{X})$ and $T(\mathbf{X_m^*})$ and we define $c_1 = \max _i \norm{x_i - T(\mathbf{X})} < \infty$. Since $m/n$ is smaller than the breakdown value of the location estimator, we have that there exist a constant $c_2$ such that $\norm{T(\mathbf{X}) - T(\mathbf{X_m^*})} \leq c_2 < \infty$. Hence by the triangle inequality, we get $d_i^* \coloneqq \norm{x_i - T(\mathbf{X_m^*})} \leq c_1 + c_2 < \infty$. Therefore we get $\text{med}_i(d_i^*) \leq c_1 + c_2$ and hence $\text{med}_i(d_i^*) + 1.4826 \cdot \text{MAD}_i(d_i^*) \leq 2.4826 \cdot \text{med}_i(d_i^*) \leq 2.4826 \cdot (c_1 + c_2)$. Using condition 3, this yields $\norm{g(t)} \leq 2.4826 \cdot (c_1 + c_2)$.
Now we compute:
\begin{align*}
    \lambda&_{\text{max}} = \underset{\norm{u}=1}{\sup} u^T S_{g}(\mathbf{X_m^*})u \\ & = \underset{\norm{u}=1}{\sup} \frac{1}{n} \sum_{i=1}^n u^T g(x_i^* - T(\mathbf{X_m^*})) g(x_i^* - T(\mathbf{X_m^*}))^T u\\
    &= \underset{\norm{u}=1}{\sup} \frac{1}{n} \sum_{i=1}^n \left[u^T g(x_i^* - T(\mathbf{X_m^*})) \right]^2 \\ &\leq \underset{\norm{u}=1}{\sup} \frac{1}{n} \sum_{i=1}^n \norm{u}^2 \norm{g(x_i^* - T(\mathbf{X_m^*}))}^2\\
    &\leq (2.4826 \cdot (c_1 + c_2))^2 < \infty
\end{align*}
Hence we have shown that the largest eigenvalue of $S_g(\mathbf{X_m^*})$ is bounded.\\
\\
\noindent
\underline{Part 2}: $\varepsilon \leq \left \lfloor{(n+1)/2}\right \rfloor /n$:

Given an $m >  \left \lfloor{(n+1)/2}\right \rfloor$, replace the last $m$ observations from $\mathbf{X}$ yielding the contaminated data set $\mathbf{X_m^*} = \{x_1,...,x_{n-m},x_{n-m+1}^*,...,x_n^*\}$. WLOG, by using location equivariance, we can assume that the mean of $x_1,...,x_{n-m}$ is zero. For the other data points, $j \in \{n-m+1,...,n\}$, we put $x_j^* = \lambda a_j$, with $a_j$ such that $\min_{i \in \{n-m+1,...,n\}} \norm{a_j - a_i} \geq 1$ and such that for all $\lambda > 1$: $\min_{i \in \{1,...,n-m\}} \norm{\lambda a_j - x_i} \geq \lambda$. We can do this by placing the $a_j$'s outside of the convex hull of $\mathbf{X}$, far enough apart from each other.

Further, we consider an increasing sequence $(\lambda_k)_k > 1$. Then for every $\lambda_k$ the set $\{x_{n-m+1}^*,...,x_n^*\}$ must contain at least one point such that $\xi(\norm{x_i^* - T(\mathbf{X_m^*})})=1$ by condition 2, say $x_b^*$. The set $\mathbf{X_m^*}$ contains other points with weight 1, take one arbitrarily and call it $x_c^*$. By the previous paragraph, we then have $\norm{x_b^* - x_c^*} \geq \lambda$ and hence $\norm{x_b^* - T(\mathbf{X_m^*})} + \norm{x_c^* - T(\mathbf{X_m^*})} \geq \lambda$ and further $\norm{x_b^* - T(\mathbf{X_m^*})}^2 + \norm{x_c^* - T(\mathbf{X_m^*})}^2 \geq \lambda^2/2$. We can then compute:
\begin{align*}
    \sum&_{j=1}^p \lambda_j(S_g(\mathbf{X_m^*})) = \text{trace}(S_g(\mathbf{X_m^*})) \\ &= \frac{1}{n} \sum_{i=1}^n \text{trace}\left[ g(x_i^* - T(\mathbf{X_m^*}))g(x_i^* - T(\mathbf{X_m^*}))^T \right]\\
    &= \frac{1}{n} \sum_{i=1}^n \norm{g(x_i^* - T(\mathbf{X_m^*}))}^2 \\
    &\geq \frac{1}{n} \left(\norm{x_b^* - T(\mathbf{X_m^*})}^2 + \norm{x_c* - T(\mathbf{X_m^*})}^2\right) \\ &\geq \lambda^2/(2n)
\end{align*}
For an unbounded increasing sequence of $\lambda$'s this becomes arbitrarily large and hence $\varepsilon \leq \left \lfloor{(n+1)/2}\right \rfloor /n$.
\end{proof}

\subsection{Proof of Theorem \ref{thm:IF1}}\label{app:thm2}

To proof the theorem, we use the following expression for the influence function of the GSSCM found in the paper of \citet{GSSCM}:
\begin{align}
    \text{IF}(x,\Sigma_g&,F) = g(x)g(x)^T - \Sigma_g(F) \nonumber \\ &+ \frac{\partial}{\partial \varepsilon} \int g_{\varepsilon}(X) g_{\varepsilon}(X)^TdF(X) \Big\vert_{\varepsilon=0} 
    \label{IFGSSCM}
\end{align}
Here the last term can be expanded as:
{\small
\begin{align*}
    &\frac{\partial}{\partial \varepsilon} \int g_{\varepsilon}(X) g_{\varepsilon}(X)^TdF(X) \Big\vert_{\varepsilon=0} = \\ &\int \Big( \left\{\frac{\partial}{\partial \varepsilon}  g_{\varepsilon}(X) \Big\vert_{\varepsilon=0} \right\}  g(X)^T  + g(X) \left\{\frac{\partial}{\partial \varepsilon}  g_{\varepsilon}(X)^T  \Big\vert_{\varepsilon=0} \right\} \Big)dF(X)
\end{align*}}
To ease the notational burden, we write $dge(x)$ for the derivative to $\varepsilon$ of $g_{\varepsilon}(x)$ in $\varepsilon=0$, for which we have:
\begin{align}
    dge(x) &= \frac{\partial}{\partial \varepsilon}  g_{\varepsilon}(X) \Big\vert_{\varepsilon=0}  \nonumber \\ &= X \frac{\partial}{\partial \varepsilon} \xi_{\varepsilon} (\norm{X}) \Big\vert_{\varepsilon=0} 
    \label{dge}
\end{align}

Explicit expressions for Equation \eqref{dge} per radial function can be found in appendix A.3 of the GSSCM paper \citep{GSSCM}.\\
Additionally, we make use of following lemma published in \citet{Croux2000} to compute the influence functions of the eigenvectors of the GSSCM:
\begin{lemma}[Croux \& Haesbroeck 2000]
\label{lemmaCenH2000}
Given $\Sigma:F \to \text{SPD}(p)$ a statistical functional such that $\text{IF}(x,\Sigma,F)$ exists. Let $v_{g,1},...,v_{g,p}$ and $\lambda_{g,1},...,\lambda_{g,p}$ be the eigenvectors and eigenvalues of $\Sigma(F)$. Then the influence function of $v_{g,j}$ is given by:
\begin{equation*}
    \text{IF}(x,v_{g,j},F) = \sum_{k=1,k \neq j}^{p} \frac{1}{\lambda_{g,j} - \lambda_{g,k}} \left( v_{g,k}^T \text{IF}(x,\Sigma,F) v_{g,j} \right) v_{g,k}
\end{equation*}
\end{lemma}
\noindent
The proof for this lemma follows from Lemma 2.1 in \citet{Sibson1979}. We are now ready to proof the theorem.

\begin{proof}
Making use of Lemma \ref{lemmaCenH2000} and Equation \eqref{IFGSSCM} for the influence function of the GSSCM, we note:
{\small
\begin{align*}
    &\text{IF}(x,v_{g,j}, F) \\
    &= \sum_{k=1, k \neq j}^{p} \frac{1}{\lambda_{g,j} - \lambda_{g,k}} \left( v_{g,k}^T \text{IF}(x,\Sigma_g, F) v_{g,j} \right) v_{g,k} \\
    &= \sum_{k=1, k \neq j}^{p} \frac{1}{\lambda_{g,j} - \lambda_{g,k}} \Big( v_{g,k}^T g(x)g(x)^T v_{g,j} - v_{g,k}^T \Sigma_g(F) v_{g,j} \\ 
    & \qquad \quad + v_{g,k}^T \frac{\partial}{\partial \varepsilon} \int g_{\varepsilon}(X) g_{\varepsilon}(X)^TdF(X) \Big\vert_{\varepsilon=0} v_{g,j} \Big) v_{g,k}\\ 
    &= \sum_{k=1, k \neq j}^{p} \frac{1}{\lambda_{g,j} - \lambda_{g,k}} \Big( (v_{g,k}^T g(x)) (v_{g,j}^T g(x)) - v_{g,k}^T \lambda_{g,j} v_{g,j}\\ 
    &\qquad \quad + v_{g,k}^T \frac{\partial}{\partial \varepsilon} \int g_{\varepsilon}(X) g_{\varepsilon}(X)^TdF(X) \Big\vert_{\varepsilon=0} v_{g,j} \Big) v_{g,k}\\
    &= \sum_{k=1, k \neq j}^{p} \frac{1}{\lambda_{g,j} - \lambda_{g,k}} \Bigg( (v_{g,k}^T g(x)) (v_{g,j}^T g(x)) \ - \ 0\\ 
    &\qquad \quad + v_{g,k}^T \int   \Big( \left\{\frac{\partial}{\partial \varepsilon}  g_{\varepsilon}(X) \Big\vert_{\varepsilon=0}  \right\} g(X)^T \\ 
    &\qquad \quad +  g(X) \left\{\frac{\partial}{\partial \varepsilon}  g_{\varepsilon}(X)^T  \Big\vert_{\varepsilon=0} \right\}  \Big)  dF(X)\ v_{g,j} \Bigg)   v_{g,k}\\
    &= \sum_{k=1, k \neq j}^{p} \frac{1}{\lambda_{g,j} - \lambda_{g,k}} \Big( (v_{g,k}^T g(x)) (v_{g,j}^T g(x)) \\ 
    &\qquad \quad + v_{g,k}^T \int   \big( dge(X) g(X)^T \\ 
    &\qquad \quad +  g(X) dge(X)^T  \big)  dF(X)\ v_{g,j} \Big)   v_{g,k}
\end{align*}
}%
\end{proof}

\subsection{Proof of Corollary \ref{col:IF1}}\label{app:col1}

\begin{proof}
When we assume that $F$ is a centered elliptically symmetric distribution with diagonal covariance matrix, we have that $v_{g,j} = v_j = e_j$. Hence Equation \eqref{IFeigenvectorsGSSCM} simplifies to:
\begin{align}
    \label{IFeigenvectorssimplified1}
    &\text{IF}(x,v_{g,j}, F) = \sum_{k=1, k \neq j}^{p} \frac{1}{\lambda_{g,j} - \lambda_{g,k}} \Big( g(x)_k g(x)_j \nonumber \\ &+ \int  \{ dge(X)_k \ g(X)_j   +   g(X)_k \ dge(X)_j \}  dF(X)  \Big)   v_k
\end{align}
Next we proof that the above integral in the second term equals zero. For this, we first know that $g(x) = x \cdot \xi(\norm{x})$, hence $g(x)_i = x_i \cdot \xi(\norm{x})$. Second, if we study appendix A.3 of the GSSCM paper \citep{GSSCM}, one can find that for every suggested radial function $dge(x)$ is of the form $dge(x) = \text{scalar} \cdot x \cdot f(\norm{x})$. Therefore, we have that $dge(x)_i = \text{scalar} \cdot x_i \cdot f(\norm{x})$. Last, the density function of a centered elliptically symmetric distribution has the following form: $f_X(x) = \text{det}(\Sigma)^{-1/2} \cdot h(x^T \Sigma^{-1} x)$ with $\Sigma$ diagonally assumed here, hence it is an even function in each of its variables $x_i$. Putting all this together, the integral in the second term of Equation \eqref{IFeigenvectorssimplified1} is of following form
\begin{align*}
    &\int_{-\infty}^{+\infty} \dots \int_{-\infty}^{+\infty} x_k x_j \cdot f(\norm{x}) \xi(\norm{x}) \cdot \omega(x) \cdot dx_1 dx_2 \dots dx_p\\& \text{for } k \neq j
\end{align*}
where $\omega(x)$ is an even function in each variable. Then one has that this integral is equal to zero because of symmetry reasons. Hence the second term in Equation \eqref{IFeigenvectorssimplified1} becomes zero for centered elliptically symmetric distributions.
\end{proof}

\subsection{Proof of Theorem \ref{thm:IF2}}\label{app:thm3}

Using Lemma \ref{lem:proj}, we can find an expression for the influence functions of our new eigenvalues:

\begin{proof}
\begin{align*}
    &\lambda_{S,k}(H) = S^2(v_{g,k}^T X) = S^2(H^{v_{g,k}})\\
    \implies &\text{IF}(x,\lambda_{S,k},H) = \frac{d}{d \varepsilon} \left( S^2\left(H_{\varepsilon,x}^{v_{g,k}(H_{\varepsilon,x})} \right) \right) \Big\vert_{\varepsilon=0}\\
    &= \frac{d}{d \varepsilon} \left( S^2(H_{\varepsilon,x}^{v_{k}}) \right)\Big\vert_{\varepsilon=0} \\ 
    &\qquad + \left( \frac{d}{da} \left( S^2(H^a) \right) \Big\vert_{a=v_{g,k}}
    \right) ^T \cdot \text{IF}(x,v_{g,k},H)
\end{align*}
We can use that $S^2(H^a) = a^T \Sigma a$, as $S$ is equivariant, which yields:
\begin{align*}
    \frac{d}{da} \left( S^2(H^a) \right) \Big\vert_{a=v_{g,k}}  &= \left( \frac{d}{da} \left( a^T \Sigma a \right) \Big\vert_{a=v_{g,k}} \right)\\ &= 2 \Sigma v_{g,k} = 2 \Sigma v_{k} = 2 \lambda_k v_k
\end{align*}
We can then continue our calculation:
\begin{align*}
    \text{IF}(x,\lambda_{S,k},H) &= \frac{d}{d \varepsilon} \left( S^2(H_{\varepsilon,x}^{v_{k}}) \right) \Big\vert_{\varepsilon=0} + 2 \lambda_k v_k^T \cdot \text{IF}(x,v_{g,k},H)\\
    &= \text{IF}(v_k^T x, \text{S}^2,H^{v_k}) + 2 \lambda_k v_k^T \cdot \text{IF}(x,v_{g,k},H)
\end{align*}
When we calculated the influence function of the eigenvector $v_{g,k}$ of GSSCM, we saw that this function had no component in the direction of $v_k$. Therefore we can remove the second term. Hence we obtain:
\begin{align*}
    \text{IF}(x,\lambda_{S,k},H) &= \text{IF}(v_k^T x, S^2,H^{v_k}) \\ &= v_k^T \Sigma v_k \ \text{IF}\left(\frac{v_k^T x}{\sqrt{v_k^T \Sigma v_k}}, S^2,F_0 \right) \\ &= 2 \lambda_k \ \text{IF}\left( \frac{v_k^T x}{\sqrt{\lambda_k}}, S,F_0 \right)
\end{align*}
\end{proof}

\subsection{Proof of Theorem \ref{thm:IF3}}\label{app:thm4}
\begin{proof}
{\small
\begin{align*}
    \text{IF}&(x,\Sigma_{\text{comb}},H) \\ &=  \frac{d}{d \varepsilon} \left( 
    \sum_{k=1}^p \lambda_{S,k}(H_{\varepsilon,x}) v_{g,k}(H_{\varepsilon,x}) v_{g,k}(H_{\varepsilon,x})^T \right) \Big\vert_{\varepsilon=0}\\
    &= \sum_{k=1}^p \left[  \text{IF}(x, \lambda_{S,k},H) v_k v_k^T \right. \\  & \left.  \qquad +  \ \lambda_{k} \ \text{IF}(x, v_{g,k},H) \ v_k^T +  \lambda_{k} v_k \ \text{IF}(x, v_{g,k},H)^T
    \right]
\end{align*}
}

Assuming that the original covariance matrix is diagonal and that $H$ is elliptically symmetric, we can use Corollary \ref{corr2IFeigenvectors} and Equation \eqref{IF_new_eigenvalue} to obtain:
\begin{align*}
    \text{IF}(&x,\Sigma_{\text{comb}},H) \\
    & = \sum_{k=1}^p \big[  \text{IF}(x, \lambda_{S,k},H) v_k v_k^T + \lambda_{k} \ \text{IF}(x, v_{g,k},H) \ v_k^T \\ 
    &\quad +  \lambda_{k} v_k \ \text{IF}(x, v_{g,k},H)^T \big] \\
    &= \sum_{k=1}^p 2 \lambda_k \ \text{IF}\left( \frac{v_k^T x}{\sqrt{\lambda_k}}, S, F_0 \right) v_k v_k^T\\
    & \quad + \sum_{k=1}^p \lambda_k \Big[
    \sum_{j=1, j \neq k}^{p} \frac{1}{\lambda_{g,k} - \lambda_{g,j}} (g(x)_j g(x)_k) v_j v_k^T \\
    & \quad + \sum_{j=1, j \neq k}^{p} \frac{1}{\lambda_{g,k} - \lambda_{g,j}} (g(x)_j g(x)_k) v_k v_j^T \Big] \\
    &= 2  \sum_{k=1}^p \lambda_k \ \text{IF}\left( \frac{v_k^T x}{\sqrt{\lambda_k}}, S,F_0 \right) v_k v_k^T  \\
    & \quad + \sum_{k=1}^p \lambda_k 
    \sum_{j=1, j \neq k}^{p} \frac{1}{\lambda_{g,k} - \lambda_{g,j}} (g(x)_j g(x)_k) (v_j v_k^T + v_k v_j^T)
\end{align*}
\end{proof}

\end{document}